\documentclass{article}
\usepackage[normalem]{ulem}% Pour distinguer les contribs
\usepackage{graphicx}
\usepackage{latexsym,amssymb,amsmath,amssymb,xspace,theorem}
\usepackage{multicol}
\usepackage{t1enc}
\usepackage[T1]{fontenc}
\usepackage{epic,eepic}
\usepackage{subfigure}
\theoremstyle{plain} \theoremheaderfont{\scshape}

\newtheorem{Thm}{\bf Theorem}

\newtheorem{Lem}[Thm]{\bf Lemma}
\newtheorem{Clm}{\bf Claim}[Thm]

\newtheorem{Claimsansnum}{\bf Claim}
\newtheorem{Conj}[Thm]{{\bf Conjecture}}
\newtheorem{Pb}[Thm]{{\bf Problem}}

\newtheorem{Prop}[Thm]{\bf Proposition}
\newtheorem{Cor}[Thm]{ \bf Corollary}
{\theorembodyfont{\rmfamily}
 \newtheorem{Def}[Thm]{\bf Definition}

}

\newenvironment{Prf}{{\bf \noindent Proof } }{\hfill$\square$\\}
\newenvironment{PrfClaim}{{\bf Proof }}{{\hfill\tiny{$\blacksquare$\\}}}

\newcommand{\ignore}[1]{}

\newcommand{\trou}{\vspace{5mm} \noindent}

\newcommand{\cqfd}{\unskip\kern 6pt\penalty 500
\raise -2pt\hbox{\vrule\vbox to 10pt{\hrule width 4pt
\vfill\hrule}\vrule}\par}

\newcommand{\stp}{$\mathcal T=\{T_1,T_2 \ldots, T_k\}\ $}
\newcommand{\np}{$\mathcal T=\{T_1,T_2 \ldots, T_{\frac{n}{2}}\}\ $}
\newcommand{\npprime}{$\mathcal T'=\{T'_1,T'_2 \ldots, T'_{\frac{n}{2}}\}\ $}
\newcommand{\npsecond}{$\mathcal T''=\{T''_1,T''_2 \ldots, T''_{\frac{n}{2}}\}\ $}

\newcommand{\cubthree}{cubic $3$-edge-colorable graph\xspace}

\newcommand{\cubthreev}{cubic $3$-edge-colorable graph \xspace}
\newcommand{\cubthreesv}{cubic $3$-edge-colorable graphs}
\newcommand{\threenc}{$\mathcal T$, $\mathcal T'$ and $\mathcal T''$\xspace}

\newcommand{\TroisPNICsv}{three compatible normal odd partitions}
\newcommand{\TroisPNIC}{three compatible normal odd partitions\xspace}
\let\oldmarginpar\marginpar
\renewcommand\marginpar[1]{\-\oldmarginpar[\raggedleft\footnotesize #1]{\raggedright\footnotesize #1}}
\input{epsf.sty}
\begin{document}
%\begin{frontmatter}


\title{On Compatible Normal Odd Partitions in Cubic Graphs}
\author{J.L. Fouquet\\
 L.I.F.O., Facult\'e des Sciences, B.P. 6759 \\
 Universit\'e d'Orl\'eans, 45067 Orl\'eans Cedex 2, FR \thanks{email: jean-luc.fouquet@univ-orleans.fr}
\and J.M. Vanherpe \\
L.I.F.O., Facult\'e des Sciences, B.P. 6759 \\
Universit\'e d'Orl\'eans, 45067 Orl\'eans Cedex 2, FR\thanks{email: jean-marie.vanherpe@univ-orleans.fr}
}

%\address{L.I.F.O., Facult\'e des Sciences, B.P. 6759 \\
%Universit\'e d'Orl\'eans, 45067 Orl\'eans Cedex 2, FR}
\maketitle
\begin{abstract}
A {\em normal odd  partition} $\mathcal T$ of the edges of a cubic
graph is a partition into {\em trails} of odd length (no repeated
edge) such that each vertex is the end vertex of exactly one trail
of the partition and internal in some trail. For each vertex $v$,
we can distinguish the edge for which this
vertex is pending. Three normal odd  partitions are compatible
whenever these distinguished edges are distinct for each vertex. We examine this notion and show that a \cubthree can always be
provided with \TroisPNIC. The Petersen graph has this property and
we can construct other cubic graphs with chromatic index four with
the same property. Finally, we propose a new conjecture which, if
true, would imply the well known Fan and Raspaud Conjecture.\\
\\
Keywords:Cubic graph;  Edge-partition
\end{abstract}

%\begin{keyword}
%Cubic graph;  Edge-partition;
%\end{keyword}
%\end{frontmatter}

\section{Introduction}

For basic graph-theoretic terms, we refer the reader to Bondy and
Murty \cite{BonMur08}.  A {\em walk} in a
graph $G$ is a sequence $W = v_0e_1v_1 \ldots e_kv_k$,
 where $v_0, v_1, \ldots , v_k$ are vertices of $G$, and $e_1, e_2 \ldots , e_k$ are edges
 of $G$ and $v_{i-1}$ and $v_i$ are the ends of $e_i$, $1 \leq i \leq k$. The vertices $v_0$ and $v_k$ are the {\em end
 vertices} and $e_1$ and $e_k$ are the {\em end edges} of this walk while $v_1, \ldots, v_{k-1}$ are the
 {\em internal vertices} and $e_2, \ldots, e_{k-1}$ are the {\em internal edges}. The {\em length} $l(W)$ of $W$ is
 the number of edges (namely $k$). The walk $W$ is {\em odd} whenever $k$ is odd, {\em even} otherwise. The
walk
 $W$ is a {\em trail} if its edges
 $e_1, e_2, \ldots, e_k$ are distinct and a {\em path} if its vertices
 $v_0, v_1, \ldots, v_k$ are distinct. If $W = v_0e_1v_1 \ldots e_kv_k$ is a walk of $G$, then $W'=v_ie_{i+1} \ldots e_jv_j$
 ($0 \leq i \leq j \leq k$) is a {\em subwalk}  of $W$ ({\em  subtrails} and
 {\em subpaths} are defined analogously) .

 If $v$ is an internal vertex of a walk $W$ with ends $x$ and $y$, then
 $W(x,v)$ and $W(v,y)$ are the subwalks of $W$ obtained by cutting $W$ at
 $v$. Conversely if $W_1$ and $W_2$ have precisely one common end $v$, then the
 {\em concatenation} of these two walks {\em at $v$} gives rise to a new
 walk (denoted by $W_1+W_2$) with $v$ as an internal vertex.
 When there is no possible confusion as to the edges being used, it would be convenient to omit the edges in the description of a walk,
 i.e., $W = v_0e_1v_1 \ldots e_kv_k$ can be shortened to $W = v_0v_1 \ldots
 v_k$.

 In what follows, $G$ is a cubic graph on $n$ vertices where loops and multiple edges are allowed.
\begin{Def}\label{Definition:NormalPartition}
A partition   of $E(G)$ into trails \stp  is {\em normal} when every
vertex is an internal vertex of some trail of  $\mathcal T$, say $T_i$, $i\in\{1,\ldots k\}$ and
an end vertex in $T_j \in \mathcal T$, $j\in\{1,\ldots k\}$. The \emph{length} of a normal partition
is the maximum length of the trails in the partition, that is ${\rm
max}\{l(T_i) | T_i \in \mathcal T\}$.
\end{Def}
If \stp is a normal partition,  then $k=\frac{n}{2}$. We can associate to each vertex $v$ the unique edge with end
$v$ that is the end edge of a trail of $\mathcal T$. We shall
denote this edge by $e_{\mathcal T}(v)$ and it will be convenient to
say that $e_{\mathcal T}(v)$ is the {\em marked} edge associated to
$v$. When it is  necessary to illustrate our purpose by a figure,
we represent the marked edge associated to a vertex by a
$\vdash$ close to this vertex. 

Let $v$ be a vertex such that $v$ is an
internal vertex in $T_i \in \mathcal T$ and an end vertex in $T_j
\in \mathcal T$ (as an end of $e_{\mathcal T}(v)$). We can associate to $v$ the set $E_{\mathcal{T}}(v)$ containing the end vertices of $T_i$. Note that
 $T_i$ and $T_j$ are not necessarily distinct, in this case we have $v\in E_{\mathcal{T}}(v)$. When $x$ and $y$ are the ends of $T_i$, one of these two vertices is
certainly different from $v$. Let us transform $\mathcal T$ into a new normal
partition $\mathcal T'$ by the so called {\em switching} operation (see Definition \ref{Def:switchingOperation}).

\begin{figure}
\setlength{\unitlength}{0.00052493in}
\begingroup\makeatletter\ifx\SetFigFont\undefined
% extract first six characters in \fmtname
\def\x#1#2#3#4#5#6#7\relax{\def\x{#1#2#3#4#5#6}}%
\expandafter\x\fmtname xxxxxx\relax \def\y{splain}%
\ifx\x\y   % LaTeX or SliTeX?
\gdef\SetFigFont#1#2#3{%
  \ifnum #1<17\tiny\else \ifnum #1<20\small\else
  \ifnum #1<24\normalsize\else \ifnum #1<29\large\else
  \ifnum #1<34\Large\else \ifnum #1<41\LARGE\else
     \huge\fi\fi\fi\fi\fi\fi
  \csname #3\endcsname}%
\else
\gdef\SetFigFont#1#2#3{\begingroup
  \count@#1\relax \ifnum 25<\count@\count@25\fi
  \def\x{\endgroup\@setsize\SetFigFont{#2pt}}%
  \expandafter\x
    \csname \romannumeral\the\count@ pt\expandafter\endcsname
    \csname @\romannumeral\the\count@ pt\endcsname
  \csname #3\endcsname}%
\fi
\fi\endgroup
{\renewcommand{\dashlinestretch}{30}
\begin{picture}(8870,2060)(0,-10)
\thicklines
\put(8032.031,863.000){\arc{1628.476}{4.8145}{7.7519}}
\put(7800.000,1695.500){\arc{3344.865}{1.3813}{2.3086}}
\put(3442.031,863.000){\arc{1628.476}{4.8145}{7.7519}}
\put(3210.000,1695.500){\arc{3344.865}{1.3813}{2.3086}}
\thinlines
\put(4875,1673){\blacken\ellipse{202}{202}}
\put(4875,1673){\ellipse{202}{202}}
\put(6675,1673){\blacken\ellipse{202}{202}}
\put(6675,1673){\ellipse{202}{202}}
\put(7350,1673){\blacken\ellipse{202}{202}}
\put(7350,1673){\ellipse{202}{202}}
\put(5460,1673){\blacken\ellipse{202}{202}}
\put(5460,1673){\ellipse{202}{202}}
\put(6675,458){\blacken\ellipse{202}{202}}
\put(6675,458){\ellipse{202}{202}}
\put(285,1673){\blacken\ellipse{202}{202}}
\put(285,1673){\ellipse{202}{202}}
\put(2085,1673){\blacken\ellipse{202}{202}}
\put(2085,1673){\ellipse{202}{202}}
\put(2760,1673){\blacken\ellipse{202}{202}}
\put(2760,1673){\ellipse{202}{202}}
\put(870,1673){\blacken\ellipse{202}{202}}
\put(870,1673){\ellipse{202}{202}}
\put(2085,458){\blacken\ellipse{202}{202}}
\put(2085,458){\ellipse{202}{202}}
\thicklines
\path(6675,458)(6675,1673)
\path(4875,1673)(6675,1673)
\path(8115,1673)(6900,1673)
\path(6900,1763)(6900,1583)
\path(285,1673)(870,1673)(2130,1673)
	(2760,1673)(3525,1673)
\path(2085,458)(2085,1493)
\path(1995,1493)(2175,1493)
\put(4605,1898){\makebox(0,0)[lb]{\smash{{\SetFigFont{7}{8.4}{\rmdefault}{}{}$x_0$}}}}
\put(5280,1898){\makebox(0,0)[lb]{\smash{{\SetFigFont{7}{8.4}{\rmdefault}{}{}$x_1$}}}}
\put(6450,1853){\makebox(0,0)[lb]{\smash{{\SetFigFont{7}{8.4}{\rmdefault}{}{}$x_r$}}}}
\put(7170,1853){\makebox(0,0)[lb]{\smash{{\SetFigFont{7}{8.4}{\rmdefault}{}{}$x_{r+1}$}}}}
\put(6765,818){\makebox(0,0)[lb]{\smash{{\SetFigFont{7}{8.4}{\rmdefault}{}{}$e_s$}}}}
\put(6990,1403){\makebox(0,0)[lb]{\smash{{\SetFigFont{7}{8.4}{\rmdefault}{}{}$e_r$}}}}
\put(15,1898){\makebox(0,0)[lb]{\smash{{\SetFigFont{7}{8.4}{\rmdefault}{}{}$x_0$}}}}
\put(690,1898){\makebox(0,0)[lb]{\smash{{\SetFigFont{7}{8.4}{\rmdefault}{}{}$x_1$}}}}
\put(1860,1853){\makebox(0,0)[lb]{\smash{{\SetFigFont{7}{8.4}{\rmdefault}{}{}$x_r$}}}}
\put(2580,1853){\makebox(0,0)[lb]{\smash{{\SetFigFont{7}{8.4}{\rmdefault}{}{}$x_{r+1}$}}}}
\put(2175,818){\makebox(0,0)[lb]{\smash{{\SetFigFont{7}{8.4}{\rmdefault}{}{}$e_s$}}}}
\put(3660,773){\makebox(0,0)[lb]{\smash{{\SetFigFont{7}{8.4}{\rmdefault}{}{}$T_i$}}}}
\put(6495,413){\makebox(0,0)[rb]{\smash{{\SetFigFont{7}{8.4}{\rmdefault}{}{}$x_s$}}}}
\put(2355,1448){\makebox(0,0)[lb]{\smash{{\SetFigFont{7}{8.4}{\rmdefault}{}{}$e_r$}}}}
\put(1950,413){\makebox(0,0)[rb]{\smash{{\SetFigFont{7}{8.4}{\rmdefault}{}{}$x_s$}}}}
\put(4965,1448){\makebox(0,0)[lb]{\smash{{\SetFigFont{7}{8.4}{\rmdefault}{}{}$e_0$}}}}
\put(330,1448){\makebox(0,0)[lb]{\smash{{\SetFigFont{7}{8.4}{\rmdefault}{}{}$e_0$}}}}
\end{picture}
}
 \caption{The switching  operation when $T_i=T_j$.}\label{Fig:SwitchingOp}
\end{figure}
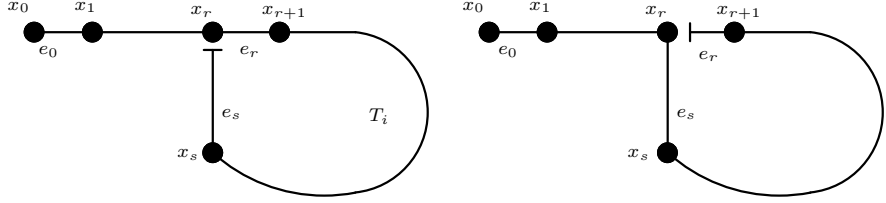

\begin{Def}\label{Def:switchingOperation}
Let $\mathcal T$ be a normal partition and $v$ be a vertex of the graph such that $v$ is an internal vertex in $T_i \in \mathcal T$ and an end vertex in $T_j \in \mathcal T$ . 
Let $x$  and $y$ be the ends of $T_i$, ($x\neq v$).
\begin {itemize}
\item When $T_i\neq T_j$, let $T'_i=T_i(x,v)+T_j$, $T'_j=T_i(y,v)$ and $\mathcal T' = \mathcal T -\{T_i,T_j\} \cup\{T'_i,T'_j\}$.
\item When $T_i=T_j$, let us write $T_i=x_0e_0x_1e_1\ldots x_re_rx_{r+1}\ldots x_se_sx_{s+1}$ where $x_0=x$, $x_r=v$, $e_s=e_{\mathcal T}(v)$ and $x_{s+1}=v$. \\
We set $T'_i=x_0e_0x_1e_1\ldots x_re_sx_s\ldots x_{r+1}e_rx_r$ and $\mathcal T' = \mathcal T -\{T_i\} \cup\{T'_i\}$ (see Figure \ref{Fig:SwitchingOp}). 
\end{itemize}
The normal partition $\mathcal T'$ is the result of the {\em switch of }$\mathcal T$ {\em on} $v$.
\end{Def}

\begin{Def}\label{Definition:OddPartition}
A normal partition \np  of $E(G)$ into trails is {\em odd} when
every trail in $\mathcal T$ is odd.  For each trail of odd length $T_i \in
\mathcal T$,  let us say that an edge $e$ of $T_i$ is {\em odd} 
whenever  the subtrails of $T_i$ obtained by deleting $e$ have odd 
lengths. The edges of $T_i$ that are not odd are said to be {\em even}.

\end{Def}

Given two normal partitions \np and \npprime, $A_{\mathcal T
\mathcal T'}$  is the set of vertices such that $e_{\mathcal
T}(v)=e_{\mathcal T'}(v)$. It must be clear that two normal partitions $\mathcal{T}$ and $\mathcal{T'}$ are identical whenever $A_{\mathcal T\mathcal T'}=V(G)$

\begin{Def}\label{Definition:CompatiblePartition}
Two normal partitions $\mathcal T$ and $\mathcal T'$ of $E(G)$ into
trails are {\em compatible} when $e_{\mathcal T}(v) \not=
e_{\mathcal T'}(v)$ for every vertex $v$ of $G$ (in other words
$A_{\mathcal T \mathcal T'}=\emptyset$).
\end{Def}

Given three normal partitions \np, \npprime and \npsecond we let
$A(\mathcal{T},\mathcal{T'},\mathcal{T''})=A_{\mathcal T \mathcal T'} \cup A_{\mathcal T' \mathcal  T''} \cup
A_{\mathcal T' \mathcal T''}$. We  say that $G$ has \TroisPNIC
\threenc whenever these partitions are pairwise compatible, that is  $A(\mathcal{T},\mathcal{T'},\mathcal{T''})=\emptyset$.

It is shown in \cite{FouVan07a} (see Theorem \ref{Theorem:ThreeCompatibleGeneral}) that a cubic graph without loops can
always be provided with three compatible normal partitions.

\begin{Thm} \cite{FouVan07a} \label{Theorem:ThreeCompatibleGeneral} A
cubic graph  $G$ has three compatible normal partitions if and only
if $G$ has no loop.
\end{Thm}

Normal odd partitions  are directly associated to
perfect matchings and it is  natural to ask whether  the
problem of finding \TroisPNIC is connected to the edge-coloring problem. We show that cubic
graphs with chromatic index 3 can be provided with \TroisPNICsv .
It turns out that the Petersen graph, the Flower snarks, and the Goldberg snarks have also three such partitions.

\section{Preliminary results}

\subsection{Switching equivalence}
In \cite{FouVan07a} we proved that if $\mathcal T$ and $\mathcal T'$ are two normal partitions of a cubic graph then we can transform  $\mathcal T$ into  $\mathcal T'$ 
by a sequence of at most $2n$ switchings. In other words $\mathcal T$ and $\mathcal T'$ are {\em switching equivalent}.

When $\mathcal T$ is a normal odd partition, a switching leading to a new odd partition $\mathcal T'$ is said to be an {\em odd switching}.
When we can transform a normal odd partition $\mathcal T$  in $\mathcal T'$ by a sequence of odd switching operations, $\mathcal T$ and $\mathcal T'$ are said to be {\em odd switching equivalent}.

\begin{Thm}\label{Thm:UneSeuleClasseDeNormaOddPartition}
Any two normal odd partitions of a cubic graph $G$ are odd switching equivalent.
\end{Thm}
\begin{Prf}
Let $M_{\mathcal T}(v)$ denote the set of edges $x$ incident with a vertex $v$ for which there exists a normal odd partition $\mathcal{T}$ of $G$ odd switching equivalent with $\mathcal{T}$, 
such that $x = e_{\mathcal{T}}(v)$ and $e_{\mathcal{T}}(u) = e_{\mathcal{T'}}(u)$ for all $u$, $v$ such that $u \neq v$.

If $\mathcal{T}$ is a normal odd partition of a cubic graph $G$, then for every vertex $v$ of $G$ there exists a normal odd partition $\mathcal{T'}$ of $G$ such that 
$e_{\mathcal{T}}(v)\neq  e_{\mathcal{T'}}(v)$ and $e_{\mathcal{T}}(u) = e_{\mathcal{T'}}(u)$ for all $u$ and $v$ such that $u \neq v$, $\mathcal{T}$, and $\mathcal{T'}$ are odd switching equivalent. 
Therefore, $|M_{\mathcal{T}}(v)| \geq 2$ 
for every $v$.

Assume that $\mathcal{T}$ and $\mathcal{T'}$ are normal odd partitions of $G$ that are not odd switching equivalent and such that $A_{\mathcal{T}\mathcal{T'}}$ has maximum cardinality. 
Then there is a vertex $v \notin A_{\mathcal{T}\mathcal{T'}}$. 
Since $|M_{\mathcal{T}}(v)| \geq 2$ and $|M_{\mathcal{T'}}(v)| \geq 2$, we have $M_{\mathcal{T}}(v)\cap M_{\mathcal{T'}}(v)\neq\emptyset$. 
Therefore, there exist two normal odd partitions $\mathcal{S}$ and $\mathcal{S'}$ of $G$ that are not odd switching equivalent and $A_{\mathcal{T}\mathcal{T'}}\subsetneq A_{\mathcal{S}\mathcal{S'}}$, a contradiction.

\end{Prf}

\begin{Thm} \label{Theorem:PerfectmatchingOddNormalPartition} Let $G$ be a cubic graph. 
Then $G$ has an odd normal partition if and only if $G$ has a perfect matching.
\end{Thm}
\begin{Prf}
If $M$ is a perfect matching of $G$, then $G-M$ is a $2-$factor of
$G$. Let us give an orientation to this $2-$factor and for each
vertex $v$ let us denote the outgoing edge $o(v)$. For each edge $e$ such that 
$e=uv \in M$, let $P_{uv}$ be the path of length $3$ obtained by
concatenating $o(u)$, $uv$ and $o(v)$. Then $T=\{P_{uv} | uv \in M
\}$ is a normal odd partition (of length $3$) of $G$. Conversely let
\np be a normal odd partition of $G$. A vertex $v \in V(G)$ is
internal in exactly one trail of $\mathcal T$. The edges of this
trail being alternatly odd and even, $v$ is incident to exactly
one odd edge. Hence the odd edges defined above induce a perfect
matching of $G$.
\end{Prf}

\begin{figure}[htb] 
\begin{center}
\setlength{\unitlength}{0.00030621in}
\begingroup\makeatletter\ifx\SetFigFont\undefined
% extract first six characters in \fmtname
\def\x#1#2#3#4#5#6#7\relax{\def\x{#1#2#3#4#5#6}}%
\expandafter\x\fmtname xxxxxx\relax \def\y{splain}%
\ifx\x\y   % LaTeX or SliTeX?
\gdef\SetFigFont#1#2#3{%
  \ifnum #1<17\tiny\else \ifnum #1<20\small\else
  \ifnum #1<24\normalsize\else \ifnum #1<29\large\else
  \ifnum #1<34\Large\else \ifnum #1<41\LARGE\else
     \huge\fi\fi\fi\fi\fi\fi
  \csname #3\endcsname}%
\else
\gdef\SetFigFont#1#2#3{\begingroup
  \count@#1\relax \ifnum 25<\count@\count@25\fi
  \def\x{\endgroup\@setsize\SetFigFont{#2pt}}%
  \expandafter\x
    \csname \romannumeral\the\count@ pt\expandafter\endcsname
    \csname @\romannumeral\the\count@ pt\endcsname
  \csname #3\endcsname}%
\fi
\fi\endgroup
{\renewcommand{\dashlinestretch}{30}
\begin{picture}(7738,6792)(0,-10)
\put(6364,2566){\blacken\ellipse{336}{336}}
\put(6364,2566){\ellipse{336}{336}}
\put(3612,2566){\blacken\ellipse{336}{336}}
\put(3612,2566){\ellipse{336}{336}}
\put(972,2582){\blacken\ellipse{336}{336}}
\put(972,2582){\ellipse{336}{336}}
\put(3644,5734){\blacken\ellipse{336}{336}}
\put(3644,5734){\ellipse{336}{336}}
\thicklines
\path(6344,2584)(3644,5734)
\path(3644,2584)(3644,5284)
\path(944,2584)(943,2582)(940,2579)
	(935,2572)(926,2562)(914,2547)
	(898,2527)(879,2502)(855,2472)
	(827,2436)(795,2396)(761,2351)
	(723,2303)(684,2251)(643,2196)
	(601,2140)(558,2083)(516,2025)
	(474,1968)(433,1911)(394,1854)
	(356,1800)(321,1746)(287,1695)
	(256,1645)(227,1598)(200,1552)
	(175,1508)(153,1466)(133,1425)
	(115,1387)(99,1349)(86,1313)
	(74,1278)(65,1244)(57,1211)
	(51,1178)(47,1146)(45,1115)
	(44,1084)(45,1052)(47,1019)
	(51,987)(57,955)(65,923)
	(74,891)(85,859)(97,827)
	(111,795)(127,764)(144,732)
	(163,700)(184,669)(206,638)
	(229,607)(254,576)(280,546)
	(308,517)(337,488)(366,460)
	(397,432)(429,406)(461,380)
	(494,355)(527,331)(561,308)
	(596,285)(630,264)(665,244)
	(700,225)(735,207)(771,190)
	(806,174)(841,159)(876,145)
	(911,132)(946,120)(982,109)
	(1015,99)(1049,90)(1083,81)
	(1117,74)(1152,67)(1188,61)
	(1223,56)(1260,51)(1296,48)
	(1334,45)(1372,44)(1410,44)
	(1448,44)(1487,46)(1527,49)
	(1566,52)(1606,57)(1646,64)
	(1686,71)(1726,80)(1766,89)
	(1806,100)(1846,112)(1885,126)
	(1924,140)(1963,156)(2001,172)
	(2038,190)(2075,209)(2112,229)
	(2147,250)(2183,272)(2217,295)
	(2251,319)(2285,344)(2317,369)
	(2350,396)(2382,425)(2413,454)
	(2444,484)(2470,510)(2495,537)
	(2521,566)(2546,595)(2572,626)
	(2597,658)(2623,691)(2649,726)
	(2675,763)(2701,802)(2728,842)
	(2756,885)(2784,929)(2813,976)
	(2842,1025)(2872,1077)(2903,1130)
	(2935,1187)(2967,1245)(3001,1306)
	(3035,1369)(3070,1434)(3105,1501)
	(3141,1570)(3177,1640)(3214,1711)
	(3250,1783)(3286,1854)(3322,1926)
	(3357,1996)(3391,2064)(3424,2130)
	(3455,2193)(3484,2252)(3511,2308)
	(3535,2358)(3557,2404)(3577,2444)
	(3593,2478)(3607,2507)(3619,2531)
	(3628,2550)(3634,2564)(3639,2573)
	(3642,2579)(3643,2583)(3644,2584)
\path(6344,2584)(7694,1909)
\path(6344,2584)(7694,1909)
\put(3644,6409){\makebox(0,0)[lb]{\smash{{{\SetFigFont{7}{8.4}{rm}V}}}}}
\path(3507,5294)(3782,5294)
\dashline{120.000}(3632,5757)(907,2519)(907,2532)
\put(5607,4044){\makebox(0,0)[lb]{\smash{{{\SetFigFont{7}{8.4}{rm}T}}}}}
\thinlines
\drawline(3582,5294)(3582,5294)
\end{picture}
}
\caption{Conformal switching on $v$ not allowed (the dashed edge is odd).}
\label{Figure:SwitchingNotAllowed}
\end{center}
\end{figure}
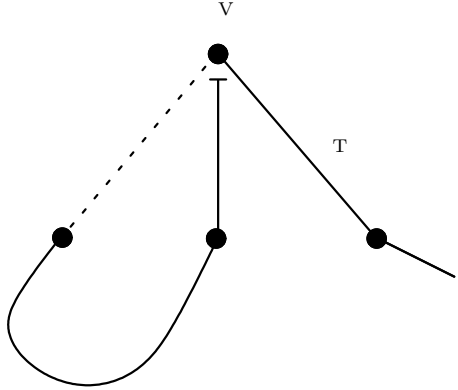
Given an odd normal partition $\mathcal T$  of $G$, we can define
the  {\em associated} perfect matching as the set of odd edges of
$\mathcal T$. Conversely, given a perfect matching $M$, we can say
that a normal odd partition $\mathcal T$ is {\em conformal} to $M$
whenever $M$ is the set of odd edges of $\mathcal T$. Let  $\mathcal T'$ be a normal odd partition obtained from a normal odd partition $\mathcal T$ by one operation of switching.
If $\mathcal T$ and $\mathcal T'$ are conformal to a perfect matching $M$, then we can say that we have performed a {\em conformal (to $M$) switching}. This
operation of conformal switching is  not always possible on a
vertex. Indeed, assuming that $v$ is an internal vertex in $T \in
\mathcal T$ and an end vertex of this trail, then the conformal switching is
not allowed since we would obtain a cycle in the transformation (see
Figure \ref{Figure:SwitchingNotAllowed}, the edge of $M$ is the
dashed edge).

\begin{Thm} \label{Theorem:PerfectmatchingConformalNormalPartition}If $G$ is a cubic graph of order at least four
and $M$ is a perfect matching in $G$, then any two normal odd partitions
$\mathcal T$ and $\mathcal T'$  conformal to $M$ are conformal switching equivalent.
\end{Thm}
\begin{Prf}
Assume that  $\mathcal T$ and $\mathcal T'$ are two normal odd partitions conformal to a perfect matching $M$ that does not belong to the same equivalence class.
Suppose that $|A_{\mathcal T \mathcal T'}|$ is maximum. In particular, we have  $A_{\mathcal T \mathcal T'}\not =V(G)$.

Let $v \not \in A_{\mathcal T \mathcal T'}$ and let $u_1,u_2$, and
$u_3$ be its neighbors. Put $vu_1=e_1$, $vu_2=e_2$, and $vu_3=e_3$. Without loss of generality we may assume that $vu_1$ is an edge of $M$ and that $e_{\mathcal T}(v)=vu_2$, while $e_{\mathcal
T'}(v)=vu_3$. Since a conformal switching of $\mathcal T$ on $v$  leads
to a conformal normal partition $\mathcal T''$ where $e_{\mathcal
T"}(v)=e_{\mathcal T'}(v)$ while nothing is changed elsewhere, we
can suppose that this conformal switching is not allowed on $v$. In the
same way a conformal switching of $\mathcal T'$ on $v$ is not allowed
as well. Hence $v$ is an internal vertex of $T \in \mathcal T$ and
an end vertex of this trail. Symmetrically, $v$ is an internal
vertex of $T' \in \mathcal T'$ and an end vertex of this trail (see
Figure \ref{Figure:SwitchingEquivalentProof}). We suppose that $y$
is the second end vertex of $T$ and $y'$ the second end vertex of
$T'$.

\begin{figure}[htb]
\begin{center}
\setlength{\unitlength}{0.00030621in}
\begingroup\makeatletter\ifx\SetFigFont\undefined%
\gdef\SetFigFont#1#2#3#4#5{%
  \reset@font\fontsize{#1}{#2pt}%
  \fontfamily{#3}\fontseries{#4}\fontshape{#5}%
  \selectfont}%
\fi\endgroup%
{\renewcommand{\dashlinestretch}{30}
\begin{picture}(15922,6631)(0,-10)
\put(3644,5734){\blacken\ellipse{336}{336}}
\put(3644,5734){\ellipse{336}{336}}
\put(892,2518){\blacken\ellipse{336}{336}}
\put(892,2518){\ellipse{336}{336}}
\put(3628,2582){\blacken\ellipse{336}{336}}
\put(3628,2582){\ellipse{336}{336}}
\put(6332,2582){\blacken\ellipse{336}{336}}
\put(6332,2582){\ellipse{336}{336}}
\put(12876,5750){\blacken\ellipse{336}{336}}
\put(12876,5750){\ellipse{336}{336}}
\put(10172,2550){\blacken\ellipse{336}{336}}
\put(10172,2550){\ellipse{336}{336}}
\put(12908,2566){\blacken\ellipse{336}{336}}
\put(12908,2566){\ellipse{336}{336}}
\put(15644,2566){\blacken\ellipse{336}{336}}
\put(15644,2566){\ellipse{336}{336}}
\dashline{96.000}(14459,1832)(15099,1524)
\dashline{96.000}(7868,1832)(8508,1524)
\thicklines
\path(6344,2584)(3644,5734)
\path(3644,2584)(3644,5284)
\path(944,2584)(943,2582)(940,2579)
	(935,2572)(926,2562)(914,2547)
	(898,2527)(879,2502)(855,2472)
	(827,2436)(795,2396)(761,2351)
	(723,2303)(684,2251)(643,2196)
	(601,2140)(558,2083)(516,2025)
	(474,1968)(433,1911)(394,1854)
	(356,1800)(321,1746)(287,1695)
	(256,1645)(227,1598)(200,1552)
	(175,1508)(153,1466)(133,1425)
	(115,1387)(99,1349)(86,1313)
	(74,1278)(65,1244)(57,1211)
	(51,1178)(47,1146)(45,1115)
	(44,1084)(45,1052)(47,1019)
	(51,987)(57,955)(65,923)
	(74,891)(85,859)(97,827)
	(111,795)(127,764)(144,732)
	(163,700)(184,669)(206,638)
	(229,607)(254,576)(280,546)
	(308,517)(337,488)(366,460)
	(397,432)(429,406)(461,380)
	(494,355)(527,331)(561,308)
	(596,285)(630,264)(665,244)
	(700,225)(735,207)(771,190)
	(806,174)(841,159)(876,145)
	(911,132)(946,120)(982,109)
	(1015,99)(1049,90)(1083,81)
	(1117,74)(1152,67)(1188,61)
	(1223,56)(1260,51)(1296,48)
	(1334,45)(1372,44)(1410,44)
	(1448,44)(1487,46)(1527,49)
	(1566,52)(1606,57)(1646,64)
	(1686,71)(1726,80)(1766,89)
	(1806,100)(1846,112)(1885,126)
	(1924,140)(1963,156)(2001,172)
	(2038,190)(2075,209)(2112,229)
	(2147,250)(2183,272)(2217,295)
	(2251,319)(2285,344)(2317,369)
	(2350,396)(2382,425)(2413,454)
	(2444,484)(2470,510)(2495,537)
	(2521,566)(2546,595)(2572,626)
	(2597,658)(2623,691)(2649,726)
	(2675,763)(2701,802)(2728,842)
	(2756,885)(2784,929)(2813,976)
	(2842,1025)(2872,1077)(2903,1130)
	(2935,1187)(2967,1245)(3001,1306)
	(3035,1369)(3070,1434)(3105,1501)
	(3141,1570)(3177,1640)(3214,1711)
	(3250,1783)(3286,1854)(3322,1926)
	(3357,1996)(3391,2064)(3424,2130)
	(3455,2193)(3484,2252)(3511,2308)
	(3535,2358)(3557,2404)(3577,2444)
	(3593,2478)(3607,2507)(3619,2531)
	(3628,2550)(3634,2564)(3639,2573)
	(3642,2579)(3643,2583)(3644,2584)
\path(6344,2584)(7694,1909)
\path(6344,2584)(7694,1909)
\put(14909,1785){\makebox(0,0)[lb]{\smash{{\SetFigFont{6}{7.2}{\rmdefault}{}{}$y'$}}}}
\put(8626,1761){\makebox(0,0)[lb]{\smash{{\SetFigFont{6}{7.2}{\rmdefault}{}{}$y$}}}}
\put(3644,6409){\makebox(0,0)[lb]{\smash{{\SetFigFont{6}{7.2}{\rmdefault}{}{}$V$}}}}
\put(12894,6396){\makebox(0,0)[lb]{\smash{{\SetFigFont{6}{7.2}{\rmdefault}{}{}$V$}}}}
\put(12907.000,2803.000){\arc{5491.930}{0.0853}{3.0654}}
\drawline(3582,5294)(3582,5294)
\path(3507,5294)(3782,5294)
\drawline(12832,5281)(12832,5281)
\path(13207,5557)(13007,5407)
\path(12919,5682)(12919,2557)
\path(13119,5482)(15619,2544)
\dashline{150.000}(12882,5744)(10157,2506)(10157,2519)
\dashline{150.000}(3632,5757)(907,2519)(907,2532)
\path(12908,2563)(14258,1888)
\put(5607,4044){\makebox(0,0)[lb]{\smash{{\SetFigFont{6}{7.2}{\rmdefault}{}{}$T$}}}}
\put(14857,4031){\makebox(0,0)[lb]{\smash{{\SetFigFont{6}{7.2}{\rmdefault}{}{}$T'$}}}}
\put(13282,2482){\makebox(0,0)[lb]{\smash{{\SetFigFont{6}{7.2}{\rmdefault}{}{}$u_2$}}}}
\put(10494,2469){\makebox(0,0)[lb]{\smash{{\SetFigFont{6}{7.2}{\rmdefault}{}{}$u_1$}}}}
\put(15907,2469){\makebox(0,0)[lb]{\smash{{\SetFigFont{6}{7.2}{\rmdefault}{}{}$u_3$}}}}
\put(6704,2562){\makebox(0,0)[lb]{\smash{{\SetFigFont{6}{7.2}{\rmdefault}{}{}$u_3$}}}}
\put(3869,2472){\makebox(0,0)[lb]{\smash{{\SetFigFont{6}{7.2}{\rmdefault}{}{}$u_2$}}}}
\put(1169,2427){\makebox(0,0)[lb]{\smash{{\SetFigFont{6}{7.2}{\rmdefault}{}{}$u_1$}}}}
\put(5039,3237){\makebox(0,0)[lb]{\smash{{\SetFigFont{6}{7.2}{\rmdefault}{}{}$e_3$}}}}
\put(3014,3417){\makebox(0,0)[lb]{\smash{{\SetFigFont{6}{7.2}{\rmdefault}{}{}$e_2$}}}}
\put(14399,3192){\makebox(0,0)[lb]{\smash{{\SetFigFont{6}{7.2}{\rmdefault}{}{}$e_3$}}}}
\put(12329,3642){\makebox(0,0)[lb]{\smash{{\SetFigFont{6}{7.2}{\rmdefault}{}{}$e_2$}}}}
\put(11159,4272){\makebox(0,0)[lb]{\smash{{\SetFigFont{6}{7.2}{\rmdefault}{}{}$e_1$}}}}
\put(2024,4452){\makebox(0,0)[lb]{\smash{{\SetFigFont{6}{7.2}{\rmdefault}{}{}$e_1$}}}}
\end{picture}
} 
\caption{Conformal switchings on $v$ are not allowed in $\mathcal T$ and $\mathcal T'$ (the dashed edges are odd).}
\label{Figure:SwitchingEquivalentProof}
\end{center}
\end{figure}
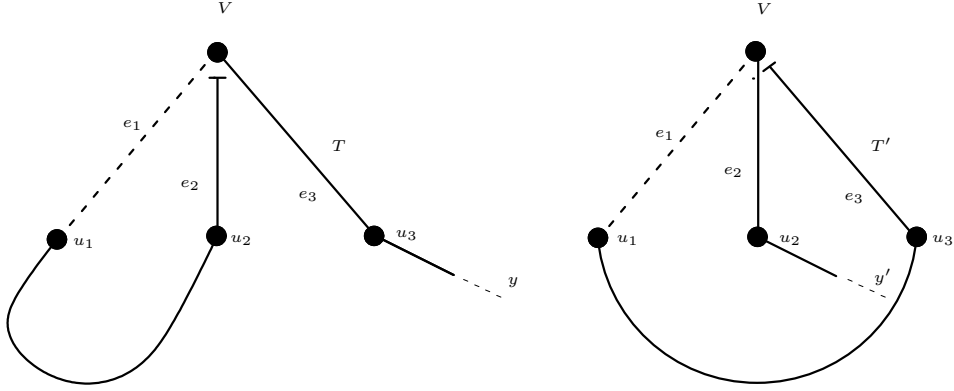
\begin{Clm}The vertices $u_2$ and $u_3$ are distinct.
\label{Claim:PerfectmatchingOddNormalPartitionClaim1}
\end{Clm}
\begin{PrfClaim}
 If $u_2=u_3$, then we have $e_{\mathcal T}(u_3)=e_3$
(formally we need to  distinguish between $u_2$ and $u_3$) and
$e_{\mathcal T'}(u_2)=e_2$. Hence $u_2 \not \in A_{\mathcal T
\mathcal T'}$. Let us denote by $u_2u_4$ the edge of $M$ incident to $u_2$ ($=u_3$) on
the subtrail of $\mathcal T$ joining $u_1$ to $u_2$.

We may assume that $u_4\neq u_1$, otherwise $G$ would be a graph on two vertices, a contradiction.

But now, conformal switchings of $\mathcal T$ on
$u_4$, $u_2$, and $v$ lead to a normal partition $\mathcal T''$ conformal switching equivalent to $\mathcal T$. Whether $u_4$ belongs or not to $A_{\mathcal T \mathcal T'}$,
$A_{\mathcal T'' \mathcal T'}$ has more vertices than $A_{\mathcal T \mathcal T'}$, a contradiction.
\end{PrfClaim}

\begin{Clm} The vertices $u_1$ and  $u_2$ are distinct.
\label{Claim:PerfectmatchingOddNormalPartitionClaim2}
\end{Clm}
\begin{PrfClaim}
Assume not: thus $u_1\notin A_{\mathcal T\mathcal T'}$. From Claim \ref{Claim:PerfectmatchingOddNormalPartitionClaim1}, we then have $u_1\neq u_3$.

We have $y=u_1$, otherwise we could transform $\mathcal T$ to $\mathcal  T''$ by using a conformal switching on $u_1$ followed by a conformal switching on $v$ 
and we would obtain $|A_{\mathcal  T'' \mathcal T'}|>|A_{\mathcal T \mathcal T'}|$, a contradiction.

But now, conformal switchings of $\mathcal T$ on
$u_3$, $u_1$, and $v$ lead to a normal partition $\mathcal T''$ conformal switching equivalent to $\mathcal T$. Whether $u_3$ belongs or not to $A_{\mathcal T \mathcal T'}$,
$A_{\mathcal T'' \mathcal T'}$ has more vertices than $A_{\mathcal T \mathcal T'}$, a contradiction.
\end{PrfClaim}

Similarly $u_1\neq u_3$.  Consider the subtrails
$T(u_1,u_2)$.  There is a certainly a vertex on that trail for which
the associated marked edge $e_{\mathcal T}(w) \not = e_{\mathcal
T'}(w)$. Assume that $w$ is the first such vertex when
running from $u_1$ to $u_2$ on $T$ (let us remark that
$T(u_1,w)=T'(u_1,w)$). Hence $w \not \in A_{\mathcal T \mathcal
T'}$. Let $x$ be the neighboring vertex of $w$ on $T(v,w)$ (it may happen that $u_1=w$, in which case $x=v$) and let $Q$ be the trail
of $\mathcal T$ ending in $w$ with the marked edge $e_{\mathcal
T}(w)$.

Since $e_{\mathcal{T}}(w)\notin M$ and $e_{\mathcal{T'}}(w)\notin M$, we have $xw\in M$.

\begin{Clm}\label{Claim:PerfectmatchingOddNormalPartitionClaim4}$w=y$.
\end{Clm}
\begin{PrfClaim}
Assume that $w \not = y$. Since $xw \in M$, we can perform a
conformal switching of $\mathcal{T}$ on $w$ leading to the conformal
partition $\mathcal T''$:
$$\mathcal T'' =\mathcal T -\{T,Q\}+\{T(y,w)+Q,T(w,v)\}$$
But now, $|A_{\mathcal{T}\mathcal{T''}}|>|A_{\mathcal{T}\mathcal{T'}}|$, a contradiction.
\end{PrfClaim}

In the same way, we certainly have $w=y'$ (take $\mathcal T'$
instead of $\mathcal T$). Since by Claim
\ref{Claim:PerfectmatchingOddNormalPartitionClaim1} $u_2 \not =
u_3$, we must have either $u_2 \not = w$ or $u_3 \not = w$. By
considering $\mathcal T$, we can decide without loss of generality that $u_2 \not =w$ (if not, we consider $\mathcal T'$ where the roles of $u_2$ and
$u_3$ are exchanged).

In $\mathcal T$, the vertex $u_2$ is an internal vertex of $T$ and an end
vertex of a trail $S$ with $S \not = T$. A conformal switching is
allowed  on $u_2$ and this switching leads to the conformal partition
$$\mathcal Q=\mathcal T -\{T,S\}+\{T(y,u_2)+S,u_2v\}.$$
We have $A_{\mathcal Q \mathcal T'}=A_{\mathcal T \mathcal T'}$
or  $A_{\mathcal Q \mathcal T'}=A_{\mathcal T \mathcal T'}-u_2$. But
now we can perform a conformal switching on $v$ followed by a conformal
switching on $w$. The first switching on $v$ leads to $\mathcal R$ defined as follows:
$$\mathcal R=\mathcal Q -\{T(y,u_2)+S,vu_2\}+\{S+T(u_2,v)+vu_2,T(v,w)\}.$$
The second switching on $w$ leads to $\mathcal S$ defined by
$$\mathcal S=\mathcal R -\{S+T(u_2,v)+vu_2,T(v,w)\}+\{u_2v+T(v,w)+T(w,v),T(w,u_2)+S\}.$$
We have now $v$ and $w$  in $A_{\mathcal S \mathcal T'}$. Since this
set has at least one vertex more than $A_{\mathcal T \mathcal T'}$,
we have a contradiction.
\end{Prf}
\begin{figure}[t]
\begin{center}
\includegraphics[scale=0.35]{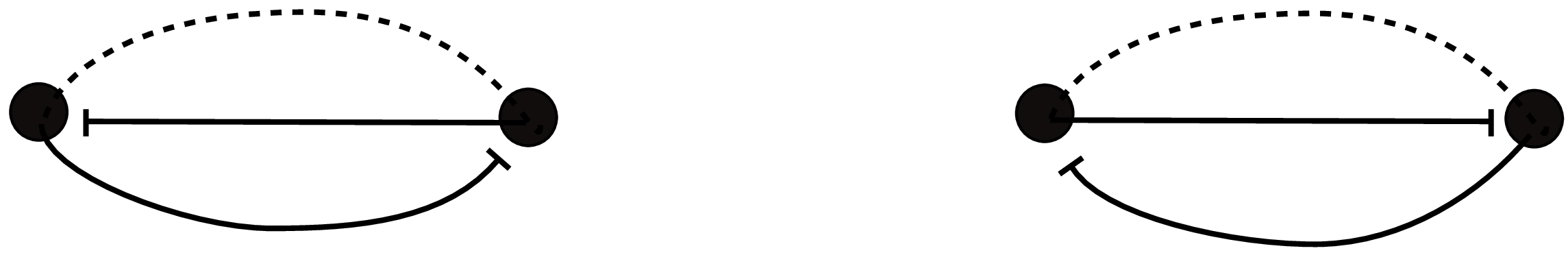}
\caption{Two non-equivalent conformal partitions for the cubic graph on two vertices (the dashed edges are odd).}\label{Figure:ConformalPartitionOnThetaGraph}
\end{center}
\end{figure}
It turns out that the cubic graph on two vertices depicted in Figure \ref{Figure:ConformalPartitionOnThetaGraph} has two non-equivalent conformal  partitions (with respect to the dashed edge).

\subsection{Miscellaneous}
The following proposition will be essential in the next section.
\begin{Prop} \label{Proposition:EdgeStructure}Let $G$ be a cubic graph
having three normal partitions $\mathcal T$, $\mathcal T'$ and
$\mathcal T''$. If $e=xy$ is an edge of $G$ such that $x$ and $y$ are not in
$A(\mathcal T,\mathcal T',\mathcal T'')$, then one of the followings is true:

\begin{itemize}
  \item  $e$ is an internal edge in exactly one partition,
  \item $e$ is an internal edge in exactly two partitions.
\end{itemize}
Moreover, in the second case, the edge $e$ itself is a trail of the third partition.
\end{Prop}

\begin{Prf}
 Assume that $e$ is an end edge in $\mathcal T$, in $\mathcal T'$, and in $\mathcal T''$. Then in $x$ or $y$ we
would have two partitions (say $\mathcal T$ and $\mathcal T'$) for
which $e_{\mathcal T}(x)=e_{\mathcal T'}(x)$ ($e_{\mathcal
T}(y)=e_{\mathcal T'}(y)$ respectively), a contradiction. 

If $e$ is an internal edge in $\mathcal T$, $\mathcal T'$ and $\mathcal T''$, then let $a$ and $b$ be the two other neighbors of $x$. We would then have 
\begin{itemize}
  \item $e_{\mathcal T}(x)=xa$ or $xb$
  \item $e_{\mathcal T'}(x)=xa$ or $xb$
  \item $e_{\mathcal T''}(x)=xa$ or $xb$,
\end{itemize}
which is impossible. 

Assume now that $e$ is an internal edge of a
trail in $\mathcal T$ and in $\mathcal T'$ and let $a$ and $b$ be the
two other neighbors of $x$. Up to the names of the vertices, we have
\begin{itemize}
  \item $e_{\mathcal T}(x)=xa$
  \item $e_{\mathcal T'}(x)=xb$.
\end{itemize}
>From the third partition $T\mathcal "$, we must have $e_{\mathcal
T"}(x)=xy$. In the same way we would obtain $e_{\mathcal
T"}(y)=yx$. Hence the trail containing $e=xy$ is reduced to $e$, as
claimed.\end{Prf}

Given a normal partition $\mathcal T$, the average length of the trails in $\mathcal T$ is denoted $\mu(\mathcal T)$ while $n_{\mathcal T}(i)$ is the
number of trails of length $i$.

\begin{Prop} \cite{FouVan07a}\label{Proposition:Distribution}
Let  $\mathcal T$ be a normal partition of a cubic graph $G$ on $n$
vertices. It follows that
\begin{itemize}
    \item $\mu(\mathcal T)=3$,
    \item $\sum_{i=1}^{i=n+1}(3-i)n_{\mathcal T}(i)=0$.
\end{itemize}
\end{Prop}
Hence a normal partition whose average length is $3$ has all its trails
of length $3$.

\begin{Prop} \label{Proposition:Bridgeless3compatibleNormalOdd}If $G$ is
a cubic graph with \TroisPNIC, then  $G$ is bridgeless.
\end{Prop}
\begin{Prf}
Assume that $xy$ is a bridge of $G$ and let $C$ be the component of $G-xy$ containing $x$. Since $G$ has \TroisPNICsv , one
of these partitions, say $\mathcal T$, is such that $e_{\mathcal
T}(x)=xy$. Thus the edges of $C$ are partitioned into odd trails.
We have $$m=|E(C)| =\frac{3(|C|-1)+2}{2}$$ and $m$ is even whenever $|C| \equiv 3 \ mod
\ 4$ while $m$ is odd whenever $|C| \equiv 1 \ mod \ 4$. The trace
of $\mathcal T$ on $C$ is a set of $\frac{|C|-1}{2}$ trails and this
number is odd when $|C| \equiv 3 \ mod \ 4$ and even otherwise.
Hence when $|C| \equiv 3 \ mod \ 4$ we must have an odd number of
odd trails partitioning $E(C)$, but in that case $m$ is even, a contradiction.
When $|C| \equiv 1 \ mod \ 4$, we must have an even number of odd
trails partitioning $E(C)$, but in that case $m$ is odd,
contradiction.
\end{Prf}

 In Figure \ref{Figure:K4}, we show $K_4$ provided with \TroisPNICsv. Let us remark that,
 following Theorem \ref{Theorem:Bipartite}, we need to have trails
 of length $5$ in at least one partition.

\begin{figure}[htb] 
\begin{center}
\setlength{\unitlength}{0.00026247in}
\begingroup\makeatletter\ifx\SetFigFont\undefined
% extract first six characters in \fmtname
\def\x#1#2#3#4#5#6#7\relax{\def\x{#1#2#3#4#5#6}}%
\expandafter\x\fmtname xxxxxx\relax \def\y{splain}%
\ifx\x\y   % LaTeX or SliTeX?
\gdef\SetFigFont#1#2#3{%
  \ifnum #1<17\tiny\else \ifnum #1<20\small\else
  \ifnum #1<24\normalsize\else \ifnum #1<29\large\else
  \ifnum #1<34\Large\else \ifnum #1<41\LARGE\else
     \huge\fi\fi\fi\fi\fi\fi
  \csname #3\endcsname}%
\else
\gdef\SetFigFont#1#2#3{\begingroup
  \count@#1\relax \ifnum 25<\count@\count@25\fi
  \def\x{\endgroup\@setsize\SetFigFont{#2pt}}%
  \expandafter\x
    \csname \romannumeral\the\count@ pt\expandafter\endcsname
    \csname @\romannumeral\the\count@ pt\endcsname
  \csname #3\endcsname}%
\fi
\fi\endgroup
{\renewcommand{\dashlinestretch}{30}
\begin{picture}(15157,4039)(0,-10)
\thicklines
\path(2055,3576)(263,504)
\thinlines
\put(176,200){\blacken\ellipse{336}{336}}
\put(176,200){\ellipse{336}{336}}
\put(4376,200){\blacken\ellipse{336}{336}}
\put(4376,200){\ellipse{336}{336}}
\put(14981,212){\blacken\ellipse{336}{336}}
\put(14981,212){\ellipse{336}{336}}
\put(12868,3849){\blacken\ellipse{336}{336}}
\put(12868,3849){\ellipse{336}{336}}
\put(12881,1462){\blacken\ellipse{336}{336}}
\put(12881,1462){\ellipse{336}{336}}
\put(10781,212){\blacken\ellipse{336}{336}}
\put(10781,212){\ellipse{336}{336}}
\put(9576,175){\blacken\ellipse{336}{336}}
\put(9576,175){\ellipse{336}{336}}
\put(7463,3812){\blacken\ellipse{336}{336}}
\put(7463,3812){\ellipse{336}{336}}
\put(7476,1425){\blacken\ellipse{336}{336}}
\put(7476,1425){\ellipse{336}{336}}
\put(5376,175){\blacken\ellipse{336}{336}}
\put(5376,175){\ellipse{336}{336}}
\put(2276,1450){\blacken\ellipse{336}{336}}
\put(2276,1450){\ellipse{336}{336}}
\thicklines
\path(2276,1712)(2276,3862)(4388,212)
	(151,212)(2301,1462)(4163,324)
\path(11056,161)(14706,161)
\path(14719,249)(14719,61)
\path(11044,249)(11044,86)
\path(12615,1320)(10822,206)(12897,3906)
	(15035,231)(12860,1468)(12855,3560)
\path(12797,3544)(12935,3544)
\path(12606,1399)(12731,1261)
\path(7651,3599)(9451,412)
\path(9351,349)(9551,474)
\path(7563,3537)(7751,3637)
\path(7688,1262)(9551,187)(5313,187)
	(7451,3812)(7451,1449)(5588,324)
\path(7738,1337)(7651,1174)
\path(5563,399)(5651,274)
\path(189,551)(352,451)
\path(2200,1675)(2400,1675)
\path(4126,295)(4201,408)
\path(1952,3597)(2115,3497)
\thinlines
\put(2267,3839){\blacken\ellipse{328}{332}}
\put(2267,3839){\ellipse{328}{332}}
\end{picture}
}
\caption{$K_4$ with \TroisPNIC.}
\label{Figure:K4}
\end{center}
\end{figure}

\section{On cubic graphs with chromatic index three}
In this section the existence of \TroisPNIC in \cubthreesv\xspace is considered.
\begin{Thm} \label{Theorem:Bipartite}If $G$ is a cubic graph, then
the following are equivalent:
\begin{itemize}
  \item []\makebox[20 pt][l]{\rm{i)}} $G$ has \TroisPNIC of length $3$
  \item []\makebox[20 pt][l]{\rm{ii)}} $G$ has \TroisPNIC, where each  edge is an  internal edge in exactly one partition
  \item []\makebox[20 pt][l]{\rm {iii)}} $G$ is bipartite.
\end{itemize}
\end{Thm}

\begin{Prf} Assume first that $G$ can be provided with  \TroisPNIC of length $3$, say 
$\mathcal T$, $\mathcal T'$, and $\mathcal T''$ . Since the average
length of each partition is $3$ (Proposition
\ref{Proposition:Distribution}), each trail of each partition has
length exactly $3$. Thus $\mathcal T$, $\mathcal T'$, and $\mathcal T''$
are  three normal odd partitions and from Proposition
\ref{Proposition:EdgeStructure}, each edge is the internal edge of
one trail in exactly one partition. Conversely suppose that $G$ can
be provided with \TroisPNIC  where each edge is an internal edge in
exactly one partition,  the edge of each trail of length $1$ must be an internal edge of two partitions, thus there is no trail of length $1$ in any 
of these partitions.
 Since the average length of each partition is
$3$, that means that each trail in each partition has length exactly
$3$. Hence $i) \equiv ii)$.

We prove now that $i) \equiv iii)$. Let $\mathcal T$, $\mathcal T'$
and $\mathcal T''$ be \TroisPNIC of length $3$. Following the proof of
Theorem \ref{Theorem:PerfectmatchingOddNormalPartition}, the set of internal
edges of trails of $\mathcal T$ ($\mathcal T'$ and $\mathcal T''$
respectively) is a perfect matching, say $M$ ($M'$ and $M"$
respectively).

Let $a_0a_1a_2a_3$ be a trail of $\mathcal T$ and let $b_1$ and
$b_2$ be the third neighbors of $a_1$ and $a_2$ respectively. By
definition, we have $e_{\mathcal T}(a_1)=a_1b_1$ and $e_{\mathcal
T}(a_2)=a_2b_2$.

Assume without loss of generality that $a_0a_1$ is an internal edge of a trail $T_1'$ of $\mathcal T'$.
The trail $T_1'$ does not use $a_1a_2$: otherwise $e_{\mathcal
T'}(a_1)=a_1b_1$, a contradiction to $e_{\mathcal T}(a_1)=a_1b_1$
since $\mathcal T$ and $\mathcal T'$ are compatible. Hence $T_1'$ uses
$a_1b_1$ and $e_{\mathcal T'}(a_1)=a_1a_2$.

Assume now that $a_2a_3$ is an internal edge of a trail $T'_2$ of
$\mathcal T'$. Reasoning in the same way, we get that $e_{\mathcal
T'}(a_2)=a_2a_1$.
 These two results lead to the fact that $a_2a_3$ must be a trail
 in $\mathcal T'$, which is impossible since each trail has length
 exactly $3$.

 Hence $a_2a_3$ is an internal edge
 in a trail of $\mathcal T''$. Thus the two internal vertices of
 $a_0a_1a_2a_3$ can be distinguished, as follows from the fact that
 the end edge to which they are incident is internal in
 $\mathcal T'$ (say {\em white} vertices) or $\mathcal T''$ (say {\em black} vertices).
 The same holds for each trail in $\mathcal T$ (and incidently for each partition $\mathcal T'$ and
$\mathcal T''$). We can now remark that $a_1b_1$ is an end edge of a
trail in $\mathcal T$. This end edge cannot be an internal edge in
$\mathcal T'$ since the trail of length $3$ going through $a_0a_1$ ends with
$a_1b_1$. Hence $a_1b_1$ is an internal edge in $\mathcal T''$ and
$b_1$ is a black vertex. Considering now $a_0$, this vertex is the
internal vertex of a trail of length $3$ of $\mathcal T$. Since
$a_0a_1 \in M'$ and $M'$ is a perfect matching, $a_0$ cannot be
incident to an other internal edge of a trail in $\mathcal T'$ and
$a_0$ must be a black vertex. Hence $a_1$ is a white vertex and its
neighbors are all black vertices. Since we can perform this reasoning
for each vertex, $G$ is bipartite as claimed.

\trou Conversely, suppose that $G$ is bipartite and let
$V(G)=\{W,B\}$ be the bipartition of its vertex set. In the
following, a vertex in $W$ will be represented by a circle ($\circ$)
while a vertex in $B$ will be represented by a bullet ($\bullet$).
>From K\"{o}nig's Theorem \cite{Kon}, $G$ is a \cubthree. Let us
consider a coloring of its edge set with three colors $ \{\alpha,
\beta, \gamma\}$. A trail of length $3$ that is obtained by considering an
edge $uv$ ($u \in B$ and $v \in W$) colored with $\beta$ together
with the edge colored $\alpha$ incident with $u$ and the edge
colored with $\gamma$ incident with $v$ will be said to have the type $\alpha \bullet \beta \circ\gamma$. 

It can be easily checked
that the set $\mathcal T$ of trails of type $\alpha \bullet \beta \circ \gamma$
 is a normal odd partition of length $3$. We can
define in the same way $\mathcal T'$ as the set of trails of type $\beta \bullet
\gamma \circ \alpha$  and $\mathcal T''$ as the
set of trails of type $\gamma \bullet \alpha \circ \beta$ .

Hence $\mathcal T$, $\mathcal T'$, and $\mathcal T''$ are
three normal odd partitions of length $3$. We claim that these
partitions are compatible. Indeed, let $v \in W$ be a vertex and
$u_1,u_2$ and $u_3$ be its neighbors. Assume that $u_1v$ is colored
with $\alpha$, $u_2v$ is colored with $\beta$ and $u_3v$ is colored
with $\gamma$. Hence $u_1v$ is internal in a trail of $\mathcal T''$ and $e_{\mathcal
T"}(v)=vu_3$. The edge $u_2v$ is internal in a  trail of $\mathcal T$ and $e_{\mathcal
T}(v)=vu_1$. The edge $u_3v$ is internal in a trail of $\mathcal T'$ and $e_{\mathcal T'}(v)=vu_2$.
Since the same reasoning can be performed in each vertex of $G$, the
three normal partitions $\mathcal T$, $\mathcal T'$ and $\mathcal T''$ are compatible.
\end{Prf}

\begin{Thm} \label{Theorem:1Length3PlusTwoOdd}Let $G$ be a cubic graph with \TroisPNIC $\mathcal{T}$, $\mathcal T'$, and $\mathcal T''$.
If $\mathcal T$ has length $3$ then $G$ is a \cubthreev.
\end{Thm}
\begin{Prf}  Since $\mathcal T$ has length $3$, every trail of
$\mathcal T$  has length $3$ (see Proposition
\ref{Proposition:Distribution}). Hence there is no edge which can be
an internal edge of a trail of $\mathcal T'$ and a trail of
$\mathcal T''$, since by Proposition \ref{Proposition:EdgeStructure}
such an edge would be a trail of length $1$ in $\mathcal T$. Thus the
perfect matchings associated to $\mathcal T'$ and $\mathcal T''$ (see
Theorem \ref{Theorem:PerfectmatchingOddNormalPartition}) would then be 
disjoint and induce an even $2$-factor of $G$, which means that $G$
is a \cubthreev as claimed.
\end{Prf}

\begin{figure}
\begin{center}
\hspace{-2cm}\input{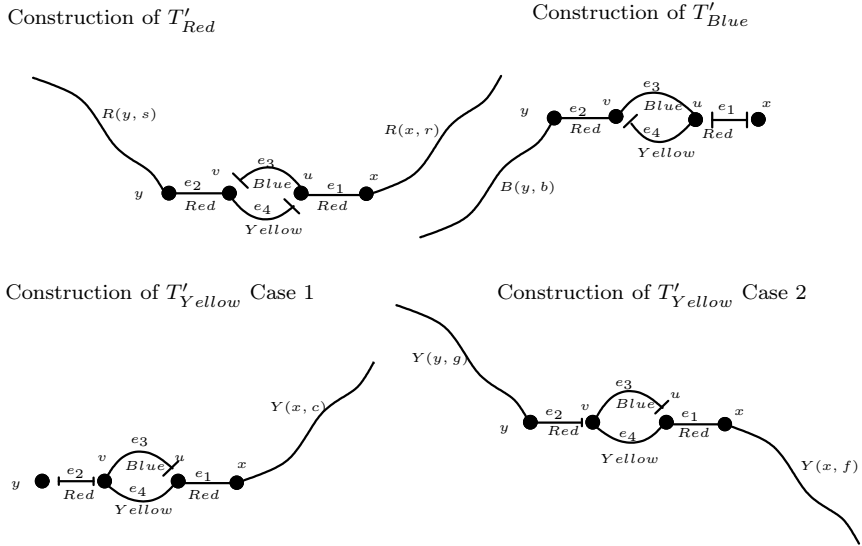}
\caption{The different cases in Lemma \ref{Lemma:AretesMultiples3PNIC}.}\label{Fig:ExtensionAretesMultiples}
\end{center}
\end{figure}
Lemmas \ref{Lemma:AretesMultiples3PNIC}, \ref{Lemma:Triangle3PNIC}, and \ref{Lemma:SommetxDansTiTj}, below, together with Theorem \ref{Theoremm:3EdgeCoulouredA3PNICAvecOddProperty}, 
will be useful in proving that a cubic graph with a $3$-edge-coloring has three compatible normal odd partitions which are conformal with respect to this coloring (see Corollary \ref{Cor:3EdgeCoulouredA3PNICAvecOddProperty}).
\begin{Lem}\label{Lemma:AretesMultiples3PNIC}
Let $G$ be a \cubthree. Assume that $G$ has a proper 3-edge-coloring $\{Red, Blue, Yellow\}$ together with \TroisPNIC $\mathcal
T_{Red}$, $\mathcal T_{Blue}$, $\mathcal T_{Yellow}$ which are,
respectively, conformal to $Red$, $Blue$ and $Yellow$. Then the graph $G'$ obtained from $G$ by subdividing an edge $e$ such that $e=xy$ with two vertices $u$
and $v$ ($u$ adjacent to $x$ and $v$ adjacent to $y$) and joining
these two vertices by an additional edge has also this property.
\end{Lem}
\begin{Prf} 
Assume that $e=xy$ is colored with $Red$.

We get a proper $3$-edge-coloring of $G'$ as follows: let the
edges $xu$ and $vy$ be colored $Red$ while the two other edges
incident to $u$ and $v$ are  colored with the two remaining colors. 

Since $xy$ is colored $Red$, this edge is internal in $\mathcal T_{Red}$. Moreover, by Proposition \ref{Proposition:EdgeStructure}, we know that $xy$ is an end edge in some other partition, 
say $\mathcal T_{Blue}$. For $\mu\in\{Red,Blue,Yellow\}$, we are going to transform the normal odd partition
$\mathcal T_{\mu}$ of $G$, conformal to $\mu$, into the normal odd
partition $\mathcal T'_{\mu}$ of $G'$, conformal to $\mu$.

Let us put $xu=e_1$ and $vy=e_2$, while the edge incident to $u$ and $v$ colored with $Blue$ is denoted by $e_3$ and the edge incident to $u$ and $v$ colored $Yellow$ is denoted by $e_4$.

Let $R \in \mathcal T_{Red}$ be the trail containing
$e$. We write $R=R(r,x)+xey+R(y,s)$ where $r$ and $s$ are end vertices of $R$. In order to get the normal odd partition $\mathcal T'_{Red}$
the subtrail $xey$ of $R$ is split into two subtrails, namely $xe_1ue_3v$ and $ye_2ve_4u$ (see Figure \ref{Fig:ExtensionAretesMultiples}). Thus   $$\mathcal T'_{Red}=\mathcal T_{Red}-\{R\}\cup\{R(r,x)+xe_1ue_3v,R(y,s)+ye_2ve_4u\}.$$ Obviously all the trails of $\mathcal T'_{Red}$ have odd edges of color $Red$.

The edge $e$ is an end edge of some trail $B$ of {$\mathcal T_{Blue}$}. This trail has $x$ and some other vertex $b$ as end vertices. We replace the subtrail $yex$ of
$B$ with  $ye_2ve_3ue_4v$ and we
consider the trail of length $1$, $xe_1u$. Hence we get a
normal odd partition $\mathcal T'_{Blue}$ of $G'$ conformal with $Blue$ as follows:
$$\mathcal T'_{Blue}=\mathcal T_{Blue}-\{B\}\cup\{xe_1u,ve_4ue_3ve_2y+B(y,b)\}.$$

We have now  two cases to consider.

 \noindent {\bf Case 1:}  $e$ is an end edge of some trail $Y$ of $\mathcal T_{Yellow}$.

Hence $y$ is one end vertex of $Y$ while $c$ is the other one.
  We replace the subtrail
$xey$ of $Y$ with $xe_1ue_4ve_3u$ and
we add the trail of length $1$ $ye_2v$. In other words:
$$\mathcal T'_{Yellow}=\mathcal T_{Yellow}-\{Y\}\cup\{ue_3ve_4ue_1x+Y(x,c), ye_2v\}.$$

\noindent {\bf Case 2:} $e$ is an internal edge of some trail $Y$  of  $\mathcal T_{Yellow}$.

We write $Y=Y(f,x)+xey+Y(y,g)$, where $f$ and $g$ are end vertices of $Y$. We replace the subtrail $xey$ of $Y$ with  $xe_1ue_4ve_3u$ and we
add the trail  $ye_2v$ of length $1$. Thus

$$\mathcal T'_{Yellow}=\mathcal T_{Yellow}-\{Y\}\cup\{Y(f,x)+xe_1ue_4ve_3u, ve_2y+T_{Yellow}(y,g)\}.$$

In all cases, we get a normal odd partition $T'_{Yellow}$ of $G'$
conformal with $Yellow$ and we can check that these three normal odd
partitions $\mathcal T'_{Red}, T'_{Blue}$ and $T'_{Yellow}$ of
$G'$ are compatible, as expected.
\end{Prf}
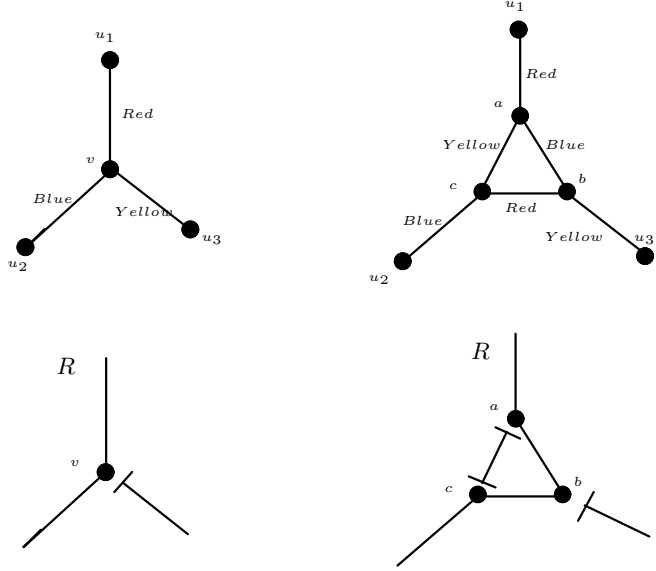
\begin{figure}
\begin{center}
\setlength{\unitlength}{0.00026247in}
\begingroup\makeatletter\ifx\SetFigFont\undefined%
\gdef\SetFigFont#1#2#3#4#5{%
  \reset@font\fontsize{#1}{#2pt}%
  \fontfamily{#3}\fontseries{#4}\fontshape{#5}%
  \selectfont}%
\fi\endgroup%
{\renewcommand{\dashlinestretch}{30}
\begin{picture}(13061,11585)(0,-10)
\put(2085,8054){\blacken\ellipse{336}{336}}
\put(2085,8054){\ellipse{336}{336}}
\put(10365,9134){\blacken\ellipse{328}{332}}
\put(10365,9134){\ellipse{328}{332}}
\put(11310,7604){\blacken\ellipse{328}{332}}
\put(11310,7604){\ellipse{328}{332}}
\put(9600,7604){\blacken\ellipse{328}{332}}
\put(9600,7604){\ellipse{328}{332}}
\put(1995,1934){\blacken\ellipse{336}{336}}
\put(1995,1934){\ellipse{336}{336}}
\put(10275,3014){\blacken\ellipse{328}{332}}
\put(10275,3014){\ellipse{328}{332}}
\put(11220,1484){\blacken\ellipse{328}{332}}
\put(11220,1484){\ellipse{328}{332}}
\put(9510,1484){\blacken\ellipse{328}{332}}
\put(9510,1484){\ellipse{328}{332}}
\put(3705,6839){\blacken\ellipse{336}{336}}
\put(3705,6839){\ellipse{336}{336}}
\put(375,6479){\blacken\ellipse{336}{336}}
\put(375,6479){\ellipse{336}{336}}
\put(2085,10259){\blacken\ellipse{336}{336}}
\put(2085,10259){\ellipse{336}{336}}
\put(10332,10877){\blacken\ellipse{336}{336}}
\put(10332,10877){\ellipse{336}{336}}
\put(7992,6197){\blacken\ellipse{336}{336}}
\put(7992,6197){\ellipse{336}{336}}
\put(12885,6299){\blacken\ellipse{336}{336}}
\put(12885,6299){\ellipse{336}{336}}
\thicklines
\path(2130,8054)(3750,6794)
\path(10365,9134)(9600,7649)
\path(2355,1709)(3660,674)
\path(2175,1529)(2535,1934)
\path(10275,4724)(10275,3014)(11265,1439)
	(9510,1439)(7890,44)
\path(11670,1259)(12975,629)
\path(10095,2744)(9600,1709)
\path(11535,944)(11850,1529)
\path(9330,1844)(9870,1619)
\path(9870,2834)(10365,2609)
\path(2085,10304)(2085,8009)(405,6478)(765,6838)
\path(2010,4230)(2010,1935)(330,404)(690,764)
\path(10365,10844)(10365,9134)(11355,7559)
	(9600,7559)(7980,6164)
\path(11312,7562)(12932,6302)
\put(9825,9314){\makebox(0,0)[lb]{\smash{{\SetFigFont{5}{6.0}{}{}{}$a$}}}}
\put(8925,7649){\makebox(0,0)[lb]{\smash{{\SetFigFont{5}{6.0}{}{}{}$c$}}}}
\put(10455,9899){\makebox(0,0)[lb]{\smash{{\SetFigFont{5}{6.0}{}{}{}$Red$}}}}
\put(11535,7784){\makebox(0,0)[lb]{\smash{{\SetFigFont{5}{6.0}{}{}{}$b$}}}}
\put(9735,3194){\makebox(0,0)[lb]{\smash{{\SetFigFont{5}{6.0}{}{}{}$a$}}}}
\put(8835,1529){\makebox(0,0)[lb]{\smash{{\SetFigFont{5}{6.0}{}{}{}$c$}}}}
\put(11445,1664){\makebox(0,0)[lb]{\smash{{\SetFigFont{5}{6.0}{}{}{}$b$}}}}
\put(2310,9089){\makebox(0,0)[lb]{\smash{{\SetFigFont{5}{6.0}{}{}{}$Red$}}}}
\put(510,7379){\makebox(0,0)[lb]{\smash{{\SetFigFont{5}{6.0}{}{}{}$Blue$}}}}
\put(2175,7109){\makebox(0,0)[lb]{\smash{{\SetFigFont{5}{6.0}{}{}{}$Yellow$}}}}
\put(1005,3914){\makebox(0,0)[lb]{\smash{{\SetFigFont{9}{10.8}{}{}{}$R$}}}}
\put(9375,4229){\makebox(0,0)[lb]{\smash{{\SetFigFont{9}{10.8}{}{}{}$R$}}}}
\put(1590,8189){\makebox(0,0)[lb]{\smash{{\SetFigFont{5}{6.0}{}{}{}$v$}}}}
\put(3930,6614){\makebox(0,0)[lb]{\smash{{\SetFigFont{5}{6.0}{}{}{}$u_3$}}}}
\put(15,6074){\makebox(0,0)[lb]{\smash{{\SetFigFont{5}{6.0}{}{}{}$u_2$}}}}
\put(1275,2069){\makebox(0,0)[lb]{\smash{{\SetFigFont{5}{6.0}{}{}{}$v$}}}}
\put(10050,11339){\makebox(0,0)[lb]{\smash{{\SetFigFont{5}{6.0}{}{}{}$u_1$}}}}
\put(7305,5804){\makebox(0,0)[lb]{\smash{{\SetFigFont{5}{6.0}{}{}{}$u_2$}}}}
\put(8790,8459){\makebox(0,0)[lb]{\smash{{\SetFigFont{5}{6.0}{}{}{}$Yellow$}}}}
\put(7980,6929){\makebox(0,0)[lb]{\smash{{\SetFigFont{5}{6.0}{}{}{}$Blue$}}}}
\put(10860,8459){\makebox(0,0)[lb]{\smash{{\SetFigFont{5}{6.0}{}{}{}$Blue$}}}}
\put(10050,7199){\makebox(0,0)[lb]{\smash{{\SetFigFont{5}{6.0}{}{}{}$Red$}}}}
\put(1770,10709){\makebox(0,0)[lb]{\smash{{\SetFigFont{5}{6.0}{}{}{}$u_1$}}}}
\put(10860,6614){\makebox(0,0)[lb]{\smash{{\SetFigFont{5}{6.0}{}{}{}$Yellow$}}}}
\put(12660,6614){\makebox(0,0)[lb]{\smash{{\SetFigFont{5}{6.0}{}{}{}$u_3$}}}}
\end{picture}
}
\caption{Situation in Lemma \ref{Lemma:Triangle3PNIC}.}\label{Fig:ExtensionTriangle}
\end{center}
\end{figure}

\begin{Lem}\label{Lemma:Triangle3PNIC}
Let $G$ be a  3-edge-colorable  cubic graph with the  proper 3-edge-coloring $\{Red, Blue, Yellow\}$. 
If $\mathcal T_{Red}$, $\mathcal T_{Blue}$, and $\mathcal T_{Yellow}$
are three compatible normal odd partitions conformal, respectively, to $Red$, $Blue$, and $Yellow$, then the graph $G'$
obtained from $G$ by expanding a vertex by a triangle also has this property.
\end{Lem}
\begin{Prf}
Let $v$ be a vertex of $G$ with neighbors $u_1$, $u_2$, $u_3$. Let us expand the vertex $v$ by a triangle, say $abc$. We color
the edges $ab$, $ac$, and $bc$ in order to get a proper $3$-edge-coloring in $G'$\ (see Figure \ref{Fig:ExtensionTriangle}). Assume without loss of generality that the edge $ab$ (respectively, $bc$,
$ac$) is colored $Blue$ (respectively, $Red$, $Yellow$).

We suppose  $e_{\mathcal T_{Red}}(v)=vu_3$. The
vertex $v$ is an internal vertex of some trail in
$\mathcal T_{Red}$, say $R$, we replace $v$ in
$R$ with $abc$ and we add the trail of
length $1$, $ac$. Thus we get a normal odd partition of
the edge set of $G'$ all of whose trails have odd edges of color
$Red$.

We proceed in a similar way for $\mathcal T_{Blue}$ and for $\mathcal T_{Yellow}$ and we get \TroisPNIC with the desired
property.
\end{Prf}

\begin{figure}
\begin{center}
\hspace{-2cm}\setlength{\unitlength}{0.00021872in}
\begingroup\makeatletter\ifx\SetFigFont\undefined%
\gdef\SetFigFont#1#2#3#4#5{%
  \reset@font\fontsize{#1}{#2pt}%
  \fontfamily{#3}\fontseries{#4}\fontshape{#5}%
  \selectfont}%
\fi\endgroup%
{\renewcommand{\dashlinestretch}{30}
\begin{picture}(25372,11889)(0,-10)
\put(12498,10905){\blacken\ellipse{336}{336}}
\put(12498,10905){\ellipse{336}{336}}
\put(12498,8115){\blacken\ellipse{328}{332}}
\put(12498,8115){\ellipse{328}{332}}
\put(14973,7530){\blacken\ellipse{328}{332}}
\put(14973,7530){\ellipse{328}{332}}
\put(14928,11490){\blacken\ellipse{328}{332}}
\put(14928,11490){\ellipse{328}{332}}
\put(10023,11535){\blacken\ellipse{328}{332}}
\put(10023,11535){\ellipse{328}{332}}
\put(10113,7530){\blacken\ellipse{328}{332}}
\put(10113,7530){\ellipse{328}{332}}
\put(3588,4560){\blacken\ellipse{336}{336}}
\put(3588,4560){\ellipse{336}{336}}
\put(3588,1770){\blacken\ellipse{328}{332}}
\put(3588,1770){\ellipse{328}{332}}
\put(6063,1185){\blacken\ellipse{328}{332}}
\put(6063,1185){\ellipse{328}{332}}
\put(6018,5145){\blacken\ellipse{328}{332}}
\put(6018,5145){\ellipse{328}{332}}
\put(1113,5190){\blacken\ellipse{328}{332}}
\put(1113,5190){\ellipse{328}{332}}
\put(1203,1185){\blacken\ellipse{328}{332}}
\put(1203,1185){\ellipse{328}{332}}
\put(12498,4560){\blacken\ellipse{336}{336}}
\put(12498,4560){\ellipse{336}{336}}
\put(12498,1770){\blacken\ellipse{328}{332}}
\put(12498,1770){\ellipse{328}{332}}
\put(14973,1185){\blacken\ellipse{328}{332}}
\put(14973,1185){\ellipse{328}{332}}
\put(14928,5145){\blacken\ellipse{328}{332}}
\put(14928,5145){\ellipse{328}{332}}
\put(10023,5190){\blacken\ellipse{328}{332}}
\put(10023,5190){\ellipse{328}{332}}
\put(10113,1185){\blacken\ellipse{328}{332}}
\put(10113,1185){\ellipse{328}{332}}
\put(21408,4740){\blacken\ellipse{336}{336}}
\put(21408,4740){\ellipse{336}{336}}
\put(23883,1365){\blacken\ellipse{328}{332}}
\put(23883,1365){\ellipse{328}{332}}
\put(23838,5325){\blacken\ellipse{328}{332}}
\put(23838,5325){\ellipse{328}{332}}
\put(18933,5370){\blacken\ellipse{328}{332}}
\put(18933,5370){\ellipse{328}{332}}
\put(19023,1365){\blacken\ellipse{328}{332}}
\put(19023,1365){\ellipse{328}{332}}
\put(21408,1995){\blacken\ellipse{328}{332}}
\put(21408,1995){\ellipse{328}{332}}
\thicklines
\path(12498,10905)(12498,8115)
\path(3588,4560)(3588,1770)
\path(1203,5550)(3633,4965)(5928,5550)
\path(3273,4155)(3273,2715)
\path(3003,4155)(3498,4155)
\dottedline{120}(6108,5595)(7233,5910)
\dottedline{120}(1068,5595)(33,5865)
\dottedline{120}(3273,2625)(3273,1860)
\path(12498,4560)(12498,1770)
\path(10113,5550)(12543,4965)(14838,5550)
\path(12183,4155)(12183,2715)
\path(11913,4155)(12408,4155)
\dottedline{120}(9978,5595)(8943,5865)
\dottedline{120}(15018,5595)(16143,5910)
\dottedline{120}(12183,2625)(12183,1860)
\path(21408,4740)(21408,1950)
\path(21048,4110)(21183,4605)
\path(21093,4380)(19428,4785)
\dottedline{120}(19303,4837)(18628,4972)
\path(21784,2740)(21784,4450)(24079,5035)
\dottedline{120}(24214,5069)(25339,5384)
\dottedline{120}(21768,2715)(21768,2085)
\path(12498,10905)(10068,11490)
\path(14962,7533)(12532,8118)
\path(14965,11500)(12535,10915)
\path(12499,8134)(10069,7549)
\path(3588,4560)(1158,5145)
\path(6052,1188)(3622,1773)
\path(6055,5155)(3625,4570)
\path(3589,1789)(1159,1204)
\path(12498,4560)(10068,5145)
\path(14962,1188)(12532,1773)
\path(14965,5155)(12535,4570)
\path(12499,1789)(10069,1204)
\path(21408,4740)(18978,5325)
\path(23872,1368)(21442,1953)
\path(23875,5335)(21445,4750)
\path(21409,1969)(18979,1384)
\put(11463,3345){\makebox(0,0)[lb]{\smash{{\SetFigFont{6}{7.2}{\rmdefault}{}{}$B'_v$}}}}
\put(3588,150){\makebox(0,0)[b]{\smash{{\SetFigFont{8}{9.6}{\rmdefault}{}{}Partition $\mathcal T_{Red}$ on $v$}}}}
\put(12588,285){\makebox(0,0)[b]{\smash{{\SetFigFont{8}{9.6}{\rmdefault}{}{}Partition $\mathcal T_{Blue}$ on $v$}}}}
\put(22038,285){\makebox(0,0)[b]{\smash{{\SetFigFont{8}{9.6}{\rmdefault}{}{}Partition $\mathcal T_{Yellow}$ on $v$}}}}
\put(10878,11490){\makebox(0,0)[lb]{\smash{{\SetFigFont{8}{9.6}{\rmdefault}{}{}$Red$}}}}
\put(13398,11535){\makebox(0,0)[lb]{\smash{{\SetFigFont{8}{9.6}{\rmdefault}{}{}$Blue$}}}}
\put(9033,10950){\makebox(0,0)[lb]{\smash{{\SetFigFont{8}{9.6}{\rmdefault}{}{}$v_1$}}}}
\put(9303,7845){\makebox(0,0)[lb]{\smash{{\SetFigFont{8}{9.6}{\rmdefault}{}{}$w_1$}}}}
\put(10608,7215){\makebox(0,0)[lb]{\smash{{\SetFigFont{8}{9.6}{\rmdefault}{}{}$Red$}}}}
\put(12633,9600){\makebox(0,0)[lb]{\smash{{\SetFigFont{8}{9.6}{\rmdefault}{}{}$Yellow$}}}}
\put(12768,10590){\makebox(0,0)[lb]{\smash{{\SetFigFont{8}{9.6}{\rmdefault}{}{}$v$}}}}
\put(12858,8205){\makebox(0,0)[lb]{\smash{{\SetFigFont{8}{9.6}{\rmdefault}{}{}$w$}}}}
\put(15153,7845){\makebox(0,0)[lb]{\smash{{\SetFigFont{8}{9.6}{\rmdefault}{}{}$w_2$}}}}
\put(13353,7260){\makebox(0,0)[lb]{\smash{{\SetFigFont{8}{9.6}{\rmdefault}{}{}$Blue$}}}}
\put(14703,10950){\makebox(0,0)[lb]{\smash{{\SetFigFont{8}{9.6}{\rmdefault}{}{}$v_2$}}}}
\put(438,5910){\makebox(0,0)[lb]{\smash{{\SetFigFont{8}{9.6}{\rmdefault}{}{}$r_v$}}}}
\put(2373,5325){\makebox(0,0)[lb]{\smash{{\SetFigFont{8}{9.6}{\rmdefault}{}{}$R_v$}}}}
\put(6288,5955){\makebox(0,0)[lb]{\smash{{\SetFigFont{8}{9.6}{\rmdefault}{}{}$s_v$}}}}
\put(6468,5010){\makebox(0,0)[lb]{\smash{{\SetFigFont{8}{9.6}{\rmdefault}{}{}$v_2$}}}}
\put(3858,4155){\makebox(0,0)[lb]{\smash{{\SetFigFont{8}{9.6}{\rmdefault}{}{}$v$}}}}
\put(438,4650){\makebox(0,0)[lb]{\smash{{\SetFigFont{8}{9.6}{\rmdefault}{}{}$v_1$}}}}
\put(2553,3390){\makebox(0,0)[lb]{\smash{{\SetFigFont{8}{9.6}{\rmdefault}{}{}$R'_v$}}}}
\put(303,1545){\makebox(0,0)[lb]{\smash{{\SetFigFont{8}{9.6}{\rmdefault}{}{}$w_1$}}}}
\put(2643,2175){\makebox(0,0)[lb]{\smash{{\SetFigFont{8}{9.6}{\rmdefault}{}{}$r'_v$}}}}
\put(3813,1905){\makebox(0,0)[lb]{\smash{{\SetFigFont{8}{9.6}{\rmdefault}{}{}$w$}}}}
\put(6063,1545){\makebox(0,0)[lb]{\smash{{\SetFigFont{8}{9.6}{\rmdefault}{}{}$w_2$}}}}
\put(8403,6000){\makebox(0,0)[lb]{\smash{{\SetFigFont{8}{9.6}{\rmdefault}{}{}$c_v$}}}}
\put(11283,5325){\makebox(0,0)[lb]{\smash{{\SetFigFont{8}{9.6}{\rmdefault}{}{}$B_v$}}}}
\put(9033,5100){\makebox(0,0)[lb]{\smash{{\SetFigFont{8}{9.6}{\rmdefault}{}{}$v_1$}}}}
\put(15918,6045){\makebox(0,0)[lb]{\smash{{\SetFigFont{8}{9.6}{\rmdefault}{}{}$b_v$}}}}
\put(15378,5010){\makebox(0,0)[lb]{\smash{{\SetFigFont{8}{9.6}{\rmdefault}{}{}$v_2$}}}}
\put(12768,4155){\makebox(0,0)[lb]{\smash{{\SetFigFont{8}{9.6}{\rmdefault}{}{}$v$}}}}
\put(9213,1545){\makebox(0,0)[lb]{\smash{{\SetFigFont{8}{9.6}{\rmdefault}{}{}$w_1$}}}}
\put(11463,2085){\makebox(0,0)[lb]{\smash{{\SetFigFont{8}{9.6}{\rmdefault}{}{}$b'_v$}}}}
\put(12723,1905){\makebox(0,0)[lb]{\smash{{\SetFigFont{8}{9.6}{\rmdefault}{}{}$w$}}}}
\put(14973,1545){\makebox(0,0)[lb]{\smash{{\SetFigFont{8}{9.6}{\rmdefault}{}{}$w_2$}}}}
\put(18663,5775){\makebox(0,0)[lb]{\smash{{\SetFigFont{8}{9.6}{\rmdefault}{}{}$v_1$}}}}
\put(20958,5145){\makebox(0,0)[lb]{\smash{{\SetFigFont{8}{9.6}{\rmdefault}{}{}$v$}}}}
\put(22983,5640){\makebox(0,0)[lb]{\smash{{\SetFigFont{8}{9.6}{\rmdefault}{}{}$v_2$}}}}
\put(24378,4740){\makebox(0,0)[lb]{\smash{{\SetFigFont{8}{9.6}{\rmdefault}{}{}$z_v$}}}}
\put(22263,4155){\makebox(0,0)[lb]{\smash{{\SetFigFont{8}{9.6}{\rmdefault}{}{}$Y_v$}}}}
\put(19113,4200){\makebox(0,0)[lb]{\smash{{\SetFigFont{8}{9.6}{\rmdefault}{}{}$Y'_v$}}}}
\put(17538,4875){\makebox(0,0)[lb]{\smash{{\SetFigFont{8}{9.6}{\rmdefault}{}{}$y'_v$}}}}
\put(18123,1725){\makebox(0,0)[lb]{\smash{{\SetFigFont{8}{9.6}{\rmdefault}{}{}$w_1$}}}}
\put(21858,2175){\makebox(0,0)[lb]{\smash{{\SetFigFont{8}{9.6}{\rmdefault}{}{}$y_v$}}}}
\put(20958,1320){\makebox(0,0)[lb]{\smash{{\SetFigFont{8}{9.6}{\rmdefault}{}{}$w$}}}}
\put(23883,1725){\makebox(0,0)[lb]{\smash{{\SetFigFont{8}{9.6}{\rmdefault}{}{}$w_2$}}}}
\end{picture}
}
\caption{Initial situation in Lemma \ref{Lemma:SommetxDansTiTj}.}\label{Fig:SituationDebutLemma17}
\end{center}
\end{figure}
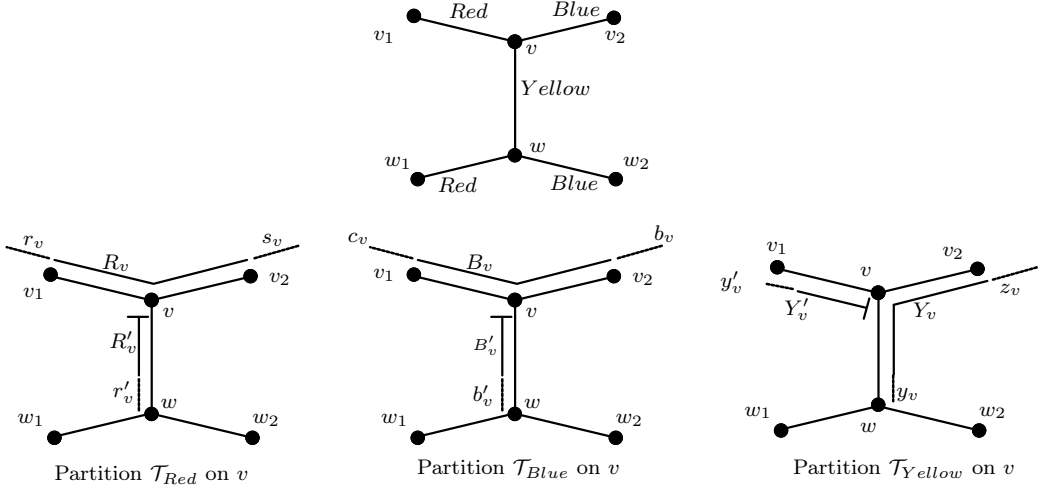
%\begin{Nots}\label{Nots:TrailsRBY}
Let $G$ be a simple \cubthree without triangles and with  a proper 3-edge-coloring using
colors in $\{Red, Blue, Yellow\}$. Let  $\mathcal T_{Red}$, $\mathcal T_{Blue}$, and $\mathcal T_{Yellow}$ be three normal odd partitions conformal, respectively, to  $Red$, $Blue$, and $Yellow$. With these hypotheses, given a vertex $v$ of $G$, 
denote by $v_1$, $v_2$, and $w$ the neighbors of $v$ such that $vv_1\in Red$, $vv_2\in Blue$, and $vw\in Yellow$. In addition, let $w_1$ and $w_2$ be the neighbors of $w$ satisfying $ww_1\in Red$ and $ww_2\in Blue$. 
Let  $R_v$ (respectively, $B_v$, $Y_v$) be the trail of $\mathcal T_{Red}$ (respectively, of $\mathcal T_{Blue}$, $\mathcal T_{Yellow}$) that contains $v$ as an internal vertex and $R'_v$ (respectively, $B'_v$, $Y'_v$) be the trail 
of $\mathcal T_{Red}$ (respectively, $\mathcal T_{Blue}$, $\mathcal T_{Yellow}$) for which $v$ is an end vertex.\newline
Let $r_v$ and $s_v$ be the end vertices of $R_v$: more precisely, $r_v$ is the end vertex of the subtrail $R(v,r_v)$ having the end edge adjacent to $v$ colored with $Red$. 
The vertices $b_v$ and $c_v$ are defined in an analogous way for the trail $B_v$ as well as the vertices $y_v$ and $z_v$ for $Y_v$.\newline
Moreover, $r'_v$ denotes the end vertex of the trail $R'_v$ distinct from $v$, and the vertices $b'_v$ and $y'_v$ are defined similarly for the trails $B'_v$ and $Y'_v$.%\end{Nots}

\begin{Lem}\label{Lemma:SommetxDansTiTj}
Let $G$ be a simple \cubthree without triangles and with  the proper 3-edge-coloring  $\{Red, Blue, Yellow\}$. Let  $\mathcal T_{Red}$, $\mathcal T_{Blue}$, and $\mathcal T_{Yellow}$ be three normal odd partitions conformal to  $Red$, $Blue$, and $Yellow$ 
such that  $A(\mathcal{T}_{Red},\mathcal{T}_{Blue},\mathcal{T}_{Yellow})$ has minimum size. 

Let $v$ be a vertex of $G$. If $v\in A_{\mathcal T_{Red}\mathcal{T}_{Blue}}$, and, using the notations above, if  $vv_1$ is an end edge of $Y'_v$ (see Figure \ref{Fig:SituationDebutLemma17}), 
then, as shown in Figure \ref{Fig:SituationFinLemma17}:
\begin{enumerate}
\item\label{Lemma:SommetxDansTiTj:Item1}$r_v=v$ and  $R_v=R'_v$,
\item\label{Lemma:SommetxDansTiTj:Item2}$w\notin A(\mathcal{T}_{Red},\mathcal{T}_{Blue},\mathcal{T}_{Yellow})$,
\item\label{Lemma:SommetxDansTiTj:Item3} $B'_v=B'_w=vw$,
\item $Y_v=Y_w=Y'_v=Y'_w$ and the edges $vv_1$, $w_2w$, $wv$, $vv_2$ and $w_1w$ occur in that order on the trail.
\end{enumerate}
\end{Lem}

\begin{Prf} We prove successively items one to four.
\begin{figure}
\begin{center}
\hspace*{-2cm}\input{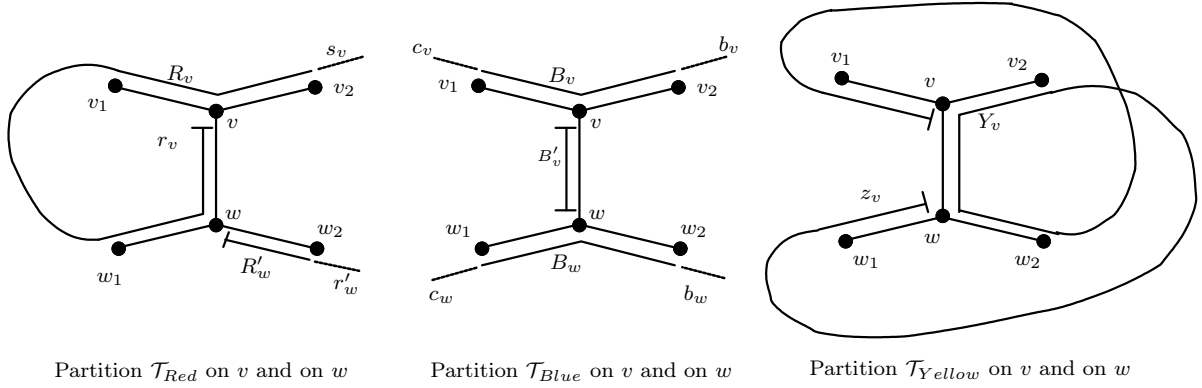}
\caption{Final situation in Lemma \ref{Lemma:SommetxDansTiTj}.}\label{Fig:SituationFinLemma17}
\end{center}
\end{figure}

{\bf Proof of item 1.}

Assume that $r_v\neq v$.  We use  a conformal switching of
$\mathcal T_{Red}$ on $v$ and we get a normal odd partition
$\mathcal{T'}_{Red}$ as
follows:\\

$$\mathcal T'_{Red}=\mathcal T_{Red}-\{R_v,R'_v\}\cup\{R'_v+R_v(v,r_v), R_v(v,s_v)\}.$$
Observe that the trails
$R'_v+R_v(v,r_v)$ and $R_v(v,s_v)$ are odd and
that $vv_1$ remains to be an odd edge of $R'_v+R_v(v,r_v)$.  Since we have
$e_{\mathcal T'_{Red}}(v)=vv_2$, $|A(\mathcal{T'}_{Red},\mathcal{T}_{Blue},\mathcal{T}_{Yellow})|= |A(\mathcal{T}_{Red},\mathcal{T}_{Blue},\mathcal{T}_{Yellow})|-1$, a contradiction. 
Thus $r_v=v$ and $R_v=R'_v$.

{\bf Proof of item 2.}

Assume that $w\in A(\mathcal{T}_{Red},\mathcal{T}_{Blue},\mathcal{T}_{Yellow})$. We know by item 1 that $e_{\mathcal T_{Red}}(w)=ww_2$. We get a normal odd
partition $\mathcal T'_{Red}$ from $\mathcal T_{Red}$  by
using conformal switches of $\mathcal T_{Red}$ on $w$ and $v$. More precisely, we write

$$\mathcal T'_{Red}=\mathcal T_{Red}-\{R_v,R'_w\}\cup\{wv+R_v(v,w)+R'_w,R_v(v,s_v)\}.$$ 
Once again, when performing those operations we
get three odd normal partitions which are compatible on $v$, a
contradiction since $|A(\mathcal{T'}_{Red},\mathcal{T}_{Blue},\mathcal{T}_{Yellow})|= |A(\mathcal{T}_{Red},\mathcal{T}_{Blue},\mathcal{T}_{Yellow})|-1$.

{\bf Proof of item 3.}

Since $R_v$ contains the edge $w_1w$, the vertex $w$ and
ends with $v$, we must have $e_{\mathcal T_{Red}}(w)=ww_2$. Moreover, since $w\notin A(\mathcal{T}_{Red},\mathcal{T}_{Blue},\mathcal{T}_{Yellow})$,  we have $e_{\mathcal T_{Yellow}}(w)\neq ww_2$ and the edge $wv$ must be an internal edge of $Y_w$. 
Hence $e_{\mathcal T_{Yellow}}(w)=ww_1$ and $Y_w=Y_v$,  
it follows that $wv$ is an end edge of $B'_v$, i.e., that the trail $B'_v$ has length $1$   and $B'_v=B'_w=vw$.

{\bf Proof of item 4.}

Now consider the trails of $\mathcal T_{Yellow}$. We already know that $Y_v=Y_w$, thus $z_v=y_w$ and $y_v=z_w$, and furthermore $vv_1$ is an end edge of $Y'_v$ while $ww_1$ is an end edge of $Y'_w$.
\newline Assume first that  $z_v\neq w$. We proceed successively to two conformal switchings (of $\mathcal T_{Blue}$ and $\mathcal T_{Yellow}$)
on $w$:
\begin{enumerate}
 \item a switching of $\mathcal T_{Blue}$ on $w$ which leads to
$$\mathcal T'_{Blue}=\mathcal T_{Blue}-\{B_w, B'_v\}\cup\{B_w(w,c_w),vw+B_w(w,b_w)\},$$
\item a switching $\mathcal T_{Yellow}$ on $w$:
$$\mathcal T'_{Yellow}=\mathcal T_{Yellow}-\{Y_v,Y'_w\}\cup\{Y'_w+wv+Y_v(v,z_v),Y_v(w,y_v)\}.$$
\end{enumerate}

When $R_v$ and $R'_w$ are distinct trails, we proceed to a switching of $\mathcal T_{Red}$ on $w$ and $v$:
$$\mathcal T'_{Red}=\mathcal T_{Red}-\{R_v, R'_w\}\cup\{wv+R_v(v,w)+R'_w,R_v(v,s_v)\}.$$

If, on the contrary, $R_v = R'_v=R'_w = R_w$, we set $\mathcal T'_{Red}$ in order to have $e_{\mathcal{T}_{Red}}(v)=vv_2$ and $e_{\mathcal{T}_{Red}}(w)=wv$, 
that is, we proceed successively to the four following conformal switchings:
\begin{description}
 \item a conformal switching of $\mathcal T_{Red}$ on $v_1$,
 \item a conformal switching of $\mathcal T_{Red}$ on $w$,
 \item a conformal switching of $\mathcal T_{Red}$ on $v$,
 \item a conformal switching of $\mathcal T_{Red}$ on $v_1$.
\end{description}
Hence we get
$$\mathcal T'_{Red}=\mathcal T_{Red}-\{R_v\}\cup\{wvv_1+R_v(v_1,w_1)+w_1ww_2+R'_w(w_2,v_2)+v_2v\}.$$
Moreover, a conformal switching of $\mathcal T_{Blue}$ on $w$ and a conformal switching of $\mathcal T_{Yellow}$ on $w$ lead us to 
$$\mathcal T'_{Blue}=\mathcal T_{Blue}-\{B_w, B'_v\}\cup\{B_w(w,c_w),vw+B_w(w,b_w)\},$$
$$\mathcal T'_{Yellow}=\mathcal T_{Yellow}-\{Y_v,Y'_w\}\cup\{Y'_w+wv+Y_v(v,z_v),Y_v(w,y_v)\}.$$

But now, in both cases, $A(\mathcal{T'}_{Red},\mathcal{T'}_{Blue},\mathcal{T'}_{Yellow})$  has fewer vertices
than $A(\mathcal{T}_{Red},\mathcal{T}_{Blue},\mathcal{T}_{Yellow})$, a contradiction.

>From now on we can suppose $z_v=w$ and therefore $Y_v=Y_w=Y'_w$.
 When $z_w\neq v$, we perform the conformal switching of
$\mathcal T_{Blue}$ and $\mathcal T_{Yellow}$ on $v$ and we get two
new normal odd partitions, which are

$$\mathcal T'_{Yellow}=\mathcal T_{Yellow}-\{Y_v,Y'_v\}\cup\{Y_v(v,z_v),Y'_v+vw+Y_w(w,z_w)\},$$
$$\mathcal T'_{Blue}=\mathcal T_{Blue}-\{B_v,B'_v\}\cup\{wv+B_v(v,b_v),B_v(v,c_v)\}.$$
It follows that $|A(\mathcal{T}_{Red},\mathcal{T'}_{Blue},\mathcal{T'}_{Yellow})|= |A(\mathcal{T}_{Red},\mathcal{T}_{Blue},\mathcal{T}_{Yellow})|-1$, a contradiction.  Consequently,
$z_w=v$ and $Y_v=Y_w=Y'_v=Y'_w$,
which proves the lemma.
\end{Prf}

\begin{Thm}\label{Theoremm:3EdgeCoulouredA3PNICAvecOddProperty}
Let $G$ be a simple \cubthree without triangles.  If $G$ has a proper 3-edge-coloring $\{Red, Blue, Yellow\}$, then $G$ has
\TroisPNIC $\mathcal T_{Red}$, $\mathcal T_{Blue}$ and $\mathcal T_{Yellow}$ which are conformal, respectively, to $Red$, $Blue$, and $Yellow$.
\end{Thm}

\begin{Prf}
Let us consider three odd normal
partitions $\mathcal T_{Red}$, $\mathcal T_{Blue}$, and $\mathcal T_{Yellow}$
such that the odd edges are colored $Red$, $Blue$, and $Yellow$, respectively. Assume that
$\mathcal T_{Red}$, $\mathcal T_{Blue}$, and $\mathcal T_{Yellow}$ are such that the size of $A(\mathcal{T}_{Red},\mathcal{T}_{Blue},\mathcal{T}_{Yellow})$ is minimum. We
suppose  $A(\mathcal{T}_{Red},\mathcal{T}_{Blue},\mathcal{T}_{Yellow})\neq\emptyset$: otherwise, $\mathcal T_{Red}$, $\mathcal T_{Blue}$ and $\mathcal T_{Yellow}$ are compatible and the proof is complete.

Let $v\in A(\mathcal{T}_{Red},\mathcal{T}_{Blue},\mathcal{T}_{Yellow})$. Without loss of generality, we suppose  $v\in A_{\mathcal T_{Red}\mathcal T_{Blue}}$.  
Using the  notations given above, $v_1$, $v_2$ and $w$ are the neighbors of $v$ such that $vv_1\in Red$, $vv_2\in Blue$, and $vw\in Yellow$, while $w_1$ and $w_2$ are neighbors of $w$ 
such that $ww_1\in Red$ and $ww_2\in Blue$. Hence $e_{\mathcal T_{Red}}(v)=vw=e_{\mathcal T_{Blue}}(v)$. 
Moreover, since $e_{\mathcal T_{Yellow}}(v)\neq vw$, we can assume that $e_{\mathcal T_{Yellow}}(v)=vv_1$.

We know, by Lemma \ref{Lemma:SommetxDansTiTj}, that $r_v=v$, $B'_v=B'_w=vw$,   $e_{\mathcal T_{Yellow}}(w)=w_1w$, $Y_v=Y_w$, $z_v=y_w=w$, and $y_v=z_w=v$. Furthermore, $w\notin A(\mathcal{T}_{Red},\mathcal{T}_{Blue},\mathcal{T}_{Yellow})$.

\begin{Claimsansnum}
 $w_2\notin A(\mathcal{T}_{Red},\mathcal{T}_{Blue},\mathcal{T}_{Yellow})$ and $r'_w=w_2$.
\end{Claimsansnum}

\begin{PrfClaim}
  Assume $w_2\in A(\mathcal{T}_{Red},\mathcal{T}_{Blue},\mathcal{T}_{Yellow})$. By using  the conformal switching of $\mathcal T_{Blue}$ on $v$, the conformal switching of $\mathcal T_{Yellow}$  on $w_2$, and  a final conformal switching on $v$, we obtain
\begin{enumerate}
 \item the conformal switching of $\mathcal T_{Blue}$ on $v$ leads to $$\mathcal T'_{Blue}=\mathcal T_{Blue}-\{B_v,B'_v\}\cup\{wv+B(v,b_v), B(v,c_v)\},$$
 \item after a conformal switching of $\mathcal T_{Yellow}$  on $w_2$, we have: $$\mathcal T'_{Yellow}=\mathcal T_{Yellow}-\{Y_v, Y'_{w_2}\}\cup\{Y'_v(v,w_2)+Y'_{w_2}, w_2w+wv+Y_v(v,w)\},$$
 \item we now perform a conformal switching of $\mathcal T'_{Yellow}$ on $v$, hence 
$$\mathcal  T''_{Yellow}=\mathcal T'_{Yellow}-\{Y_v(v,y'_{w_2}),Y_v(w,w_2)\}\cup\{Y_v(v,w), w_2w+wv+Y_v(v,y'_{w_2}\}.$$
\end{enumerate}

But we have $A(\mathcal{T}_{Red},\mathcal{T'}_{Blue},\mathcal{T''}_{Yellow})=A(\mathcal{T}_{Red},\mathcal{T}_{Blue},\mathcal{T}_{Yellow})-\{v\}$,
 a contradiction to the choice of  
$\mathcal T_{Red}$, $\mathcal T_{Blue}$ and $\mathcal T_{Yellow}$.

We can suppose now $w_2\notin A(\mathcal{T}_{Red},\mathcal{T}_{Blue},\mathcal{T}_{Yellow})$, the edge $ww_2$ being an internal edge of
$B_w\in \mathcal T_{Blue}$ and an internal edge of
$Y_v\in\mathcal T_{Yellow}$, and since $w, w_2\notin A(\mathcal{T}_{Red},\mathcal{T}_{Blue},\mathcal{T}_{Yellow})$, by Proposition \ref{Proposition:EdgeStructure}, the trail
$R'_w$ has length $1$. Hence we can write
$R'_w=ww_2$, that is, $r'_w=w_2$.
\end{PrfClaim}\\
Let us first use a conformal switching of $\mathcal T_{Red}$ on $w$ (recall that $r'_w=w_2$) followed by a conformal switching of the resulting $Red$ partition on  $w$ as
well as  a final conformal switching of $\mathcal T_{Blue}$ on $w$, in other
words, we get the odd normal partitions:
$$\mathcal T'_{Red}=\mathcal T_{Red}-\{R_v, R'_w\}\cup\{R_v(v,s_v), wv+R_v(v,w)+ww_2\}$$
and
$$\mathcal T'_{Blue}=\mathcal T_{Blue}-\{B'_v, B_w\}\cup\{vw+B_w(w,b_w),B_w(w,c_w)\}.$$

We have $v\notin A(\mathcal{T'}_{Red},\mathcal{T'}_{Blue},\mathcal{T}_{Yellow})$ and $w\in A(\mathcal{T'}_{Red},\mathcal{T'}_{Blue},\mathcal{T}_{Yellow})$. More precisely,
 $w\in A_{\mathcal T'_{Blue}\mathcal T_{Yellow}}$. It follows that $A(\mathcal{T'}_{Red},\mathcal{T'}_{Blue},\mathcal{T}_{Yellow})$ and $A(\mathcal{T}_{Red},\mathcal{T}_{Blue},\mathcal{T}_{Yellow})$ have the same size.
 \newline Observe that the trail $wv+R_v(v,w)+ww_2$ of $\mathcal T'_{Red}$  contains $w$ and $w_2$ as end vertices. Since $w\in A_{\mathcal T'_{Blue}\mathcal T_{Yellow}}$ and $A(\mathcal{T'}_{Red},\mathcal{T'}_{Blue},\mathcal{T}_{Yellow})$ 
has minimum size, we apply Lemma \ref{Lemma:SommetxDansTiTj} to the vertices $w$ and $w_1$. 
It follows that the trail of $\mathcal T'_{Red}$ having $wv$ as an end edge must have $w$ and $w_1$ as end vertices, a contradiction since this trail ends with $w$ and $w_2$.
\end{Prf}

Due to Lemmas \ref{Lemma:AretesMultiples3PNIC} and \ref{Lemma:Triangle3PNIC}, Theorem \ref{Theoremm:3EdgeCoulouredA3PNICAvecOddProperty} can be easily extended to 3-edge-colorable cubic graphs having multiple edges or triangles. 

\begin{Cor}\label{Cor:3EdgeCoulouredA3PNICAvecOddProperty}
If $G$ is a  \cubthree  with  a proper 3-edge-coloring  $\{Red, Blue, Yellow\}$, then $G$ has
\TroisPNIC $\mathcal T_{Red}$, $\mathcal T_{Blue}$ and $\mathcal T_{Yellow}$ which are conformal, respectively, to $Red$, $Blue$, $Yellow$.
\end{Cor}

\section{On cubic graphs with chromatic index four}
A snark is a bridgeless cubic graph with edge chromatic number four.
 By Proposition \ref{Proposition:Bridgeless3compatibleNormalOdd}, a cubic graph with \TroisPNIC must be bridgeless. Thus in this section we consider the problem of providing \TroisPNIC 
for some known snarks as the families of Flower snarks as well as  Goldberg snarks.

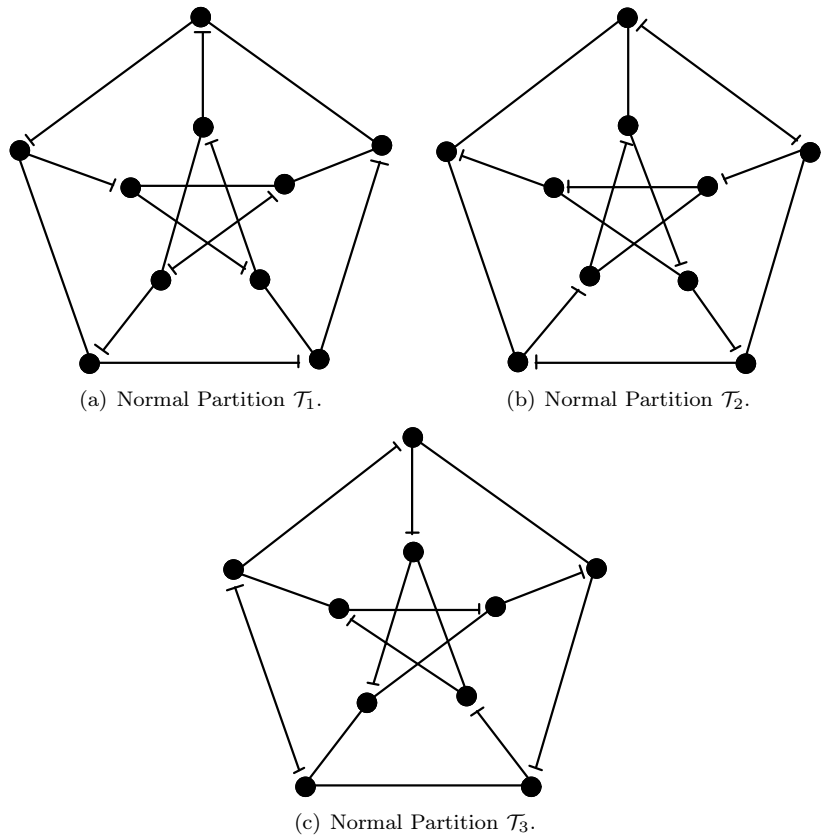
\begin{figure}
\begin{center}
\subfigure[Normal Partition $\mathcal T_1$.]{\label{Figure:PetersenPartitionT1}\setlength{\unitlength}{0.00030621in}
\begingroup\makeatletter\ifx\SetFigFont\undefined%
\gdef\SetFigFont#1#2#3#4#5{%
  \reset@font\fontsize{#1}{#2pt}%
  \fontfamily{#3}\fontseries{#4}\fontshape{#5}%
  \selectfont}%
\fi\endgroup%
{\renewcommand{\dashlinestretch}{30}
\begin{picture}(6565,6313)(0,-10)
\put(1375,175){\blacken\ellipse{336}{336}}
\put(1375,175){\ellipse{336}{336}}
\put(5311,248){\blacken\ellipse{336}{336}}
\put(5311,248){\ellipse{336}{336}}
\put(4294,1617){\blacken\ellipse{336}{336}}
\put(4294,1617){\ellipse{336}{336}}
\put(2598,1604){\blacken\ellipse{336}{336}}
\put(2598,1604){\ellipse{336}{336}}
\put(2077,3191){\blacken\ellipse{336}{336}}
\put(2077,3191){\ellipse{336}{336}}
\put(3325,4233){\blacken\ellipse{336}{336}}
\put(3325,4233){\ellipse{336}{336}}
\put(4718,3252){\blacken\ellipse{336}{336}}
\put(4718,3252){\ellipse{336}{336}}
\put(6389,3918){\blacken\ellipse{336}{336}}
\put(6389,3918){\ellipse{336}{336}}
\put(3277,6123){\blacken\ellipse{336}{336}}
\put(3277,6123){\ellipse{336}{336}}
\put(176,3833){\blacken\ellipse{336}{336}}
\put(176,3833){\ellipse{336}{336}}
\thicklines
\path(1614,409)(2589,1659)(3314,4209)(3314,5846)
\path(3451,3959)(4251,1659)(5289,234)(6351,3634)
\path(4964,184)(1401,184)(114,3834)(1776,3234)
\path(2789,1796)(4489,3034)
\path(3176,5859)(3439,5859)
\path(4439,3121)(4589,2959)
\path(3364,3934)(3564,4009)
\path(1814,3334)(1726,3134)
\path(2726,1871)(2851,1709)
\path(1464,496)(1689,334)
\path(4951,309)(4951,109)
\path(6201,3659)(6489,3584)
\path(276,4096)(414,3921)
\path(364,3996)(3264,6134)(6376,3896)
	(4664,3221)(1976,3221)(4039,1796)
\path(4064,1914)(3976,1714)
\end{picture}
}}
\subfigure[Normal Partition $\mathcal T_2$.]{\label{Figure:PetersenPartitionT2}\setlength{\unitlength}{0.00030621in}
\begingroup\makeatletter\ifx\SetFigFont\undefined%
\gdef\SetFigFont#1#2#3#4#5{%
  \reset@font\fontsize{#1}{#2pt}%
  \fontfamily{#3}\fontseries{#4}\fontshape{#5}%
  \selectfont}%
\fi\endgroup%
{\renewcommand{\dashlinestretch}{30}
\begin{picture}(6590,6300)(0,-10)
\put(4658,3215){\blacken\ellipse{336}{336}}
\put(4658,3215){\ellipse{336}{336}}
\put(1400,199){\blacken\ellipse{336}{336}}
\put(1400,199){\ellipse{336}{336}}
\put(2635,1677){\blacken\ellipse{336}{336}}
\put(2635,1677){\ellipse{336}{336}}
\put(2017,3191){\blacken\ellipse{336}{336}}
\put(2017,3191){\ellipse{336}{336}}
\put(4319,1592){\blacken\ellipse{336}{336}}
\put(4319,1592){\ellipse{336}{336}}
\put(5312,175){\blacken\ellipse{336}{336}}
\put(5312,175){\ellipse{336}{336}}
\put(6414,3797){\blacken\ellipse{336}{336}}
\put(6414,3797){\ellipse{336}{336}}
\put(3289,4257){\blacken\ellipse{336}{336}}
\put(3289,4257){\ellipse{336}{336}}
\put(176,3809){\blacken\ellipse{336}{336}}
\put(176,3809){\ellipse{336}{336}}
\put(3277,6110){\blacken\ellipse{336}{336}}
\put(3277,6110){\ellipse{336}{336}}
\thicklines
\path(3514,5973)(6189,3998)
\path(6239,4085)(6102,3935)
\path(3602,6073)(3439,5885)
\path(4940,3335)(6365,3872)(5277,185)(1702,185)
\path(1714,310)(1714,85)
\path(5127,423)(4277,1623)(2027,3160)(439,3735)
\path(464,3835)(402,3660)
\path(5214,473)(5052,385)
\path(4902,3423)(4977,3260)
\path(2252,3198)(4677,3198)(2577,1610)(3252,3935)
\path(2264,3310)(2264,3110)
\path(3164,3985)(3364,3923)
\path(4102,1848)(4289,1910)
\path(4177,1885)(3289,4210)(3289,6173)
	(164,3773)(1389,185)(2439,1435)
\path(2327,1485)(2552,1348)
\end{picture}
}}
\subfigure[Normal Partition $\mathcal T_3$.]{\label{Figure:PetersenPartitionT3}\setlength{\unitlength}{0.00030621in}
\begingroup\makeatletter\ifx\SetFigFont\undefined%
\gdef\SetFigFont#1#2#3#4#5{%
  \reset@font\fontsize{#1}{#2pt}%
  \fontfamily{#3}\fontseries{#4}\fontshape{#5}%
  \selectfont}%
\fi\endgroup%
{\renewcommand{\dashlinestretch}{30}
\begin{picture}(6576,6359)(0,-10)
\put(176,3903){\blacken\ellipse{336}{336}}
\put(176,3903){\ellipse{336}{336}}
\put(1984,3231){\blacken\ellipse{336}{336}}
\put(1984,3231){\ellipse{336}{336}}
\put(3264,4207){\blacken\ellipse{336}{336}}
\put(3264,4207){\ellipse{336}{336}}
\put(4672,3263){\blacken\ellipse{336}{336}}
\put(4672,3263){\ellipse{336}{336}}
\put(4176,1727){\blacken\ellipse{336}{336}}
\put(4176,1727){\ellipse{336}{336}}
\put(2464,1615){\blacken\ellipse{336}{336}}
\put(2464,1615){\ellipse{336}{336}}
\put(1408,175){\blacken\ellipse{336}{336}}
\put(1408,175){\ellipse{336}{336}}
\put(5280,175){\blacken\ellipse{336}{336}}
\put(5280,175){\ellipse{336}{336}}
\put(6400,3919){\blacken\ellipse{336}{336}}
\put(6400,3919){\ellipse{336}{336}}
\put(3250,6169){\blacken\ellipse{336}{336}}
\put(3250,6169){\ellipse{336}{336}}
\thicklines
\path(62,3544)(337,3632)
\path(1387,494)(1212,407)
\path(200,3594)(1300,457)
\path(2250,3107)(2125,2944)
\path(2475,1957)(2662,1919)
\path(2187,3044)(4212,1694)(3275,4232)(2575,1944)
\path(6112,3969)(6187,3757)
\path(4462,1532)(4250,1394)
\path(6137,3844)(4650,3269)(2512,1657)
	(1387,194)(5325,194)(4350,1469)
\path(3137,4532)(3350,4532)
\path(5250,532)(5462,457)
\path(3237,4507)(3237,6182)(6375,3932)(5362,494)
\path(2925,6082)(3037,5944)
\path(4387,3307)(4387,3144)
\path(2962,6007)(200,3857)(2025,3207)(4387,3207)
\end{picture}
}}
\caption{Three compatible normal odd partitions  $\mathcal {T}_1$, $\mathcal {T}_2$, and $\mathcal {T}_3$, of the Petersen graph.}
\label{Figure:3PNICPourPetersen}
\end{center}
\end{figure}

In Figures \ref{Figure:PetersenPartitionT1}, \ref{Figure:PetersenPartitionT2}
and \ref{Figure:PetersenPartitionT3} we give \TroisPNIC of the
Petersen graph. It can be pointed out that these \TroisPNIC are
isomorphic. Indeed, we have in each partition, a path of length
five, three paths of lengths three, and one path of length unity. In some
sense this fact shows that Theorem \ref{Theorem:1Length3PlusTwoOdd} is sharp.

\subsection{Flower snarks}
For an odd $k$ such that $k\geq 3$, let $F_k$ be the cubic graph on $4k$ vertices $u_1,u_2,\ldots u_k$, $v_1,v_2,\ldots v_k$, $w_1,w_2,\ldots w_k$, $t_1,t_2,\ldots t_k$ such that $u_1u_2\ldots u_k$ 
is an induced cycle of length $k$, $w_1w_2\ldots w_kt_1t_2\ldots t_k$ is an induced  cycle of length $2k$ and for $1\leq i\leq k$ the vertex $v_i$ is adjacent to $u_i$, $w_i$ and $t_i$. 
For odd $k$ such that $k\geq 5$, the graph $F_k$ is known as a Flower snark  (see \cite{Isa75}) while $F_3$ is sometimes known as Tietze's graph (see \cite{BonMur08}).

\begin{Prop}\label{Proposition:FlowerSnarksOnt3PNIC}
If $k\geq 3$ is an odd integer, $F_k$ can be provided with \TroisPNICsv.
\end{Prop}
\begin{Prf}
In Figure \ref{Figure:3PNICdeFowerSnarkF3} we propose \TroisPNICsv, namely $\mathcal{T}_1$, $\mathcal{T}_2$ and $\mathcal{T}_3$, of Tietze's graph  $F_3$. 
Moreover, when considering the vertices $u_1$, $v_1$, $w_1$, $t_1$, $u_2$, $v_2$, $w_2$, $t_2$, we have the following situation~(recall that $k=3$):

\begin{eqnarray}
\label{Equation:FlowerSnarks:Claw1:1}e_{\mathcal{T}_1}(u_1)=u_1v_1, \; e_{\mathcal{T}_1}(v_1)=v_1w_1, \; e_{\mathcal{T}_1}(w_1)=w_1w_2, \; e_{\mathcal{T}_1}(t_1)=t_1t_2\\
\label{Equation:FlowerSnarks:Claw1:2}e_{\mathcal{T}_2}(u_1)=u_1u_k, \; e_{\mathcal{T}_2}(v_1)=v_1t_1, \; e_{\mathcal{T}_2}(w_1)=w_1v_1, \; e_{\mathcal{T}_2}(t_1)=t_1w_k\\
\label{Equation:FlowerSnarks:Claw1:3}e_{\mathcal{T}_3}(u_1)=u_1u_2, \; e_{\mathcal{T}_3}(v_1)=v_1u_1, \; e_{\mathcal{T}_3}(w_1)=w_1t_k, \; e_{\mathcal{T}_3}(t_1)=t_1v_1\\
\nonumber\\
\label{Equation:FlowerSnarks:Claw2:1}e_{\mathcal{T}_1}(u_2)=u_2u_3, \; e_{\mathcal{T}_1}(v_2)=v_2u_2, \; e_{\mathcal{T}_1}(w_2)=w_2v_2, \; e_{\mathcal{T}_1}(t_2)=t_2t_1\\
\label{Equation:FlowerSnarks:Claw2:2}e_{\mathcal{T}_2}(u_2)=u_2v_2, \; e_{\mathcal{T}_2}(v_2)=v_2t_2, \; e_{\mathcal{T}_2}(w_2)=w_2w_3, \; e_{\mathcal{T}_2}(t_2)=t_2t_3\\
\label{Equation:FlowerSnarks:Claw2:3}e_{\mathcal{T}_3}(u_2)=u_2u_1, \; e_{\mathcal{T}_3}(v_2)=v_2w_2, \; e_{\mathcal{T}_3}(w_2)=w_2w_1, \; e_{\mathcal{T}_3}(t_2)=t_2v_2.
\end{eqnarray}
Observe that among the edges $u_1u_2, w_1w_2\:$ and $t_1t_2\:$ we have

$u_1u_2\:$ is an odd edge in $\mathcal{T}_1$, $t_1t_2$ is an odd edge in $\mathcal{T}_2$ and in $\mathcal{T}_3.$
%\begin{eqnarray}
%\nonumber u_1u_2\: text{is an odd} edge in}\: \mathcal{T}_1,
%\nonumber t_1t_2\: \text{is an odd edge in}\: \mathcal{T}_2 \: \text{and in } \mathcal{T}_3
%\end{eqnarray}
\begin{figure}
\begin{center}
\subfigure[partition $\mathcal{T}_1$.]{\label{Figure:FlowerSnarkF3T1}\setlength{\unitlength}{0.00024996in}
\begingroup\makeatletter\ifx\SetFigFont\undefined%
\gdef\SetFigFont#1#2#3#4#5{%
  \reset@font\fontsize{#1}{#2pt}%
  \fontfamily{#3}\fontseries{#4}\fontshape{#5}%
  \selectfont}%
\fi\endgroup%
{\renewcommand{\dashlinestretch}{30}
\begin{picture}(6874,6450)(0,-10)
\put(6452,586){\blacken\ellipse{336}{336}}
\put(6452,586){\ellipse{336}{336}}
\put(5588,1242){\blacken\ellipse{336}{336}}
\put(5588,1242){\ellipse{336}{336}}
\put(5412,2106){\blacken\ellipse{336}{336}}
\put(5412,2106){\ellipse{336}{336}}
\put(4644,1338){\blacken\ellipse{336}{336}}
\put(4644,1338){\ellipse{336}{336}}
\put(244,618){\blacken\ellipse{336}{336}}
\put(244,618){\ellipse{336}{336}}
\put(1732,1418){\blacken\ellipse{336}{336}}
\put(1732,1418){\ellipse{336}{336}}
\put(948,1258){\blacken\ellipse{336}{336}}
\put(948,1258){\ellipse{336}{336}}
\put(1044,2090){\blacken\ellipse{336}{336}}
\put(1044,2090){\ellipse{336}{336}}
\put(2660,4266){\blacken\ellipse{336}{336}}
\put(2660,4266){\ellipse{336}{336}}
\put(3812,4170){\blacken\ellipse{336}{336}}
\put(3812,4170){\ellipse{336}{336}}
\put(3252,4906){\blacken\ellipse{336}{336}}
\put(3252,4906){\ellipse{336}{336}}
\put(3236,5850){\blacken\ellipse{336}{336}}
\put(3236,5850){\ellipse{336}{336}}
\thicklines
\path(2027,1287)(2027,1555)
\path(3437,5716)(3605,5984)
\path(567,742)(418,884)
\path(485,817)(955,1287)
\path(3505,4340)(3705,4508)
\path(3605,4408)(3236,4910)
\path(3236,4910)(3236,5884)
\path(2699,4240)(1089,2092)
\path(1089,2092)(951,1574)
\path(854,1555)(1063,1510)
\path(1390,2159)(1290,1891)
\path(2094,1420)(5115,2025)
\path(1759,1420)(3773,4240)
\path(3773,4240)(4712,1287)
\path(955,1287)(1826,1420)
\path(4712,1287)(1357,2025)
\path(5078,2166)(5148,1891)
\path(6457,615)(5852,1018)
\path(5718,902)(5894,1142)
\path(5383,2092)(5652,1220)
\path(5652,1152)(4980,1220)
\path(4947,1354)(4913,1085)
\path(2900,4105)(5383,2092)
\path(2833,3971)(3034,4240)
\path(3034,4710)(2699,4240)
\path(2900,4776)(3168,4642)
\path(6310,886)(6557,884)
\path(283,615)(284,615)(285,614)
	(287,614)(291,612)(297,611)
	(305,608)(316,605)(328,602)
	(344,597)(362,592)(383,586)
	(407,579)(434,572)(464,563)
	(497,554)(533,545)(571,535)
	(612,524)(655,513)(701,501)
	(749,489)(799,477)(851,464)
	(905,452)(960,439)(1017,426)
	(1076,414)(1136,401)(1198,389)
	(1262,376)(1327,364)(1394,352)
	(1463,341)(1534,329)(1607,318)
	(1682,308)(1760,297)(1840,287)
	(1923,278)(2008,269)(2097,260)
	(2189,252)(2285,245)(2384,238)
	(2486,232)(2593,226)(2702,222)
	(2815,218)(2931,215)(3048,214)
	(3168,213)(3276,213)(3383,215)
	(3490,217)(3596,220)(3700,224)
	(3802,228)(3902,233)(4000,239)
	(4096,245)(4189,252)(4280,259)
	(4369,267)(4456,275)(4542,283)
	(4625,292)(4706,301)(4786,310)
	(4864,320)(4941,330)(5017,340)
	(5091,350)(5164,361)(5235,372)
	(5306,383)(5375,394)(5444,405)
	(5511,416)(5576,428)(5641,439)
	(5704,450)(5766,462)(5826,473)
	(5884,484)(5941,495)(5995,505)
	(6047,516)(6098,526)(6145,535)
	(6190,544)(6233,553)(6272,561)
	(6309,569)(6342,576)(6373,582)
	(6400,588)(6424,593)(6445,598)
	(6464,602)(6479,605)(6492,608)
	(6502,610)(6510,612)(6516,613)
	(6520,614)(6522,615)(6523,615)(6524,615)
\path(3236,5850)(3235,5850)(3234,5849)
	(3231,5848)(3226,5846)(3219,5844)
	(3210,5840)(3198,5835)(3183,5830)
	(3165,5823)(3144,5814)(3119,5804)
	(3091,5793)(3060,5780)(3025,5766)
	(2987,5751)(2946,5733)(2902,5715)
	(2855,5694)(2805,5673)(2753,5650)
	(2698,5626)(2642,5600)(2584,5573)
	(2525,5545)(2464,5516)(2402,5485)
	(2339,5453)(2275,5420)(2210,5386)
	(2146,5351)(2081,5314)(2015,5277)
	(1950,5237)(1884,5197)(1818,5155)
	(1752,5111)(1687,5066)(1621,5019)
	(1555,4970)(1490,4920)(1424,4866)
	(1359,4811)(1294,4753)(1229,4693)
	(1163,4630)(1099,4564)(1034,4495)
	(970,4423)(907,4348)(844,4269)
	(782,4188)(722,4104)(663,4017)
	(606,3927)(552,3836)(501,3743)
	(453,3650)(408,3556)(367,3462)
	(330,3370)(295,3278)(264,3188)
	(236,3099)(211,3011)(189,2925)
	(170,2840)(153,2757)(138,2675)
	(126,2595)(116,2516)(108,2438)
	(102,2361)(97,2285)(94,2210)
	(93,2136)(93,2063)(94,1991)
	(97,1920)(100,1850)(105,1780)
	(110,1712)(117,1644)(124,1578)
	(132,1513)(140,1449)(148,1387)
	(157,1327)(166,1268)(176,1211)
	(185,1157)(194,1105)(204,1056)
	(212,1009)(221,966)(229,925)
	(237,888)(244,854)(251,823)
	(257,796)(262,772)(267,752)
	(271,734)(274,720)(277,708)
	(279,699)(281,693)(282,688)
	(282,685)(283,684)(283,683)
\path(3505,5850)(3506,5850)(3507,5849)
	(3509,5848)(3514,5846)(3519,5843)
	(3527,5839)(3538,5834)(3551,5828)
	(3566,5821)(3585,5812)(3606,5802)
	(3630,5790)(3657,5776)(3687,5761)
	(3720,5745)(3756,5726)(3794,5707)
	(3835,5686)(3878,5663)(3924,5639)
	(3971,5613)(4020,5587)(4070,5559)
	(4122,5529)(4175,5499)(4230,5467)
	(4285,5434)(4341,5400)(4397,5365)
	(4454,5329)(4512,5292)(4569,5253)
	(4627,5213)(4686,5172)(4744,5130)
	(4802,5086)(4861,5040)(4920,4993)
	(4979,4944)(5038,4894)(5097,4841)
	(5157,4786)(5217,4729)(5276,4669)
	(5336,4607)(5396,4542)(5456,4475)
	(5516,4404)(5576,4331)(5636,4255)
	(5695,4176)(5753,4094)(5810,4010)
	(5866,3924)(5920,3836)(5973,3744)
	(6024,3651)(6071,3558)(6116,3465)
	(6157,3374)(6195,3284)(6231,3195)
	(6263,3108)(6293,3023)(6320,2939)
	(6344,2856)(6366,2776)(6386,2697)
	(6404,2619)(6420,2543)(6434,2468)
	(6446,2394)(6457,2321)(6466,2249)
	(6474,2178)(6481,2108)(6486,2039)
	(6490,1972)(6494,1905)(6496,1839)
	(6498,1774)(6499,1710)(6499,1647)
	(6499,1586)(6498,1527)(6496,1469)
	(6494,1413)(6492,1360)(6490,1308)
	(6487,1259)(6484,1213)(6482,1169)
	(6479,1129)(6476,1091)(6473,1057)
	(6471,1026)(6468,999)(6466,975)
	(6464,954)(6462,936)(6461,921)
	(6460,910)(6459,900)(6458,894)
	(6458,889)(6457,886)(6457,885)(6457,884)
\put(2027,4373){\makebox(0,0)[lb]{\smash{{\SetFigFont{6}{7.2}{}{}{}$t_3$}}}}
\put(4309,4240){\makebox(0,0)[lb]{\smash{{\SetFigFont{6}{7.2}{}{}{}$w_3$}}}}
\put(3773,5045){\makebox(0,0)[lb]{\smash{{\SetFigFont{6}{7.2}{}{}{}$v_3$}}}}
\put(4712,750){\makebox(0,0)[lb]{\smash{{\SetFigFont{6}{7.2}{}{}{}$w_2$}}}}
\put(5920,2226){\makebox(0,0)[lb]{\smash{{\SetFigFont{6}{7.2}{}{}{}$t_2$}}}}
\put(6859,213){\makebox(0,0)[lb]{\smash{{\SetFigFont{6}{7.2}{}{}{}$u_2$}}}}
\put(418,2226){\makebox(0,0)[lb]{\smash{{\SetFigFont{6}{7.2}{}{}{}$w_1$}}}}
\put(1894,884){\makebox(0,0)[lb]{\smash{{\SetFigFont{6}{7.2}{}{}{}$t_1$}}}}
\put(1089,750){\makebox(0,0)[lb]{\smash{{\SetFigFont{6}{7.2}{}{}{}$v_1$}}}}
\put(15,78){\makebox(0,0)[lb]{\smash{{\SetFigFont{6}{7.2}{}{}{}$u_1$}}}}
\put(3236,6252){\makebox(0,0)[lb]{\smash{{\SetFigFont{6}{7.2}{}{}{}$u_3$}}}}
\put(5920,1420){\makebox(0,0)[lb]{\smash{{\SetFigFont{6}{7.2}{}{}{}$v_2$}}}}
\end{picture}
}}
\subfigure[partition $\mathcal{T}_2$.]{\label{Figure:FlowerSnarkF3T2}\setlength{\unitlength}{0.00024996in}
\begingroup\makeatletter\ifx\SetFigFont\undefined%
\gdef\SetFigFont#1#2#3#4#5{%
  \reset@font\fontsize{#1}{#2pt}%
  \fontfamily{#3}\fontseries{#4}\fontshape{#5}%
  \selectfont}%
\fi\endgroup%
{\renewcommand{\dashlinestretch}{30}
\begin{picture}(6874,6450)(0,-10)
\put(6452,586){\blacken\ellipse{336}{336}}
\put(6452,586){\ellipse{336}{336}}
\put(5588,1242){\blacken\ellipse{336}{336}}
\put(5588,1242){\ellipse{336}{336}}
\put(5412,2106){\blacken\ellipse{336}{336}}
\put(5412,2106){\ellipse{336}{336}}
\put(4644,1338){\blacken\ellipse{336}{336}}
\put(4644,1338){\ellipse{336}{336}}
\put(244,618){\blacken\ellipse{336}{336}}
\put(244,618){\ellipse{336}{336}}
\put(1732,1418){\blacken\ellipse{336}{336}}
\put(1732,1418){\ellipse{336}{336}}
\put(948,1258){\blacken\ellipse{336}{336}}
\put(948,1258){\ellipse{336}{336}}
\put(1044,2090){\blacken\ellipse{336}{336}}
\put(1044,2090){\ellipse{336}{336}}
\put(2660,4266){\blacken\ellipse{336}{336}}
\put(2660,4266){\ellipse{336}{336}}
\put(3812,4170){\blacken\ellipse{336}{336}}
\put(3812,4170){\ellipse{336}{336}}
\put(3252,4938){\blacken\ellipse{336}{336}}
\put(3252,4938){\ellipse{336}{336}}
\put(3236,5850){\blacken\ellipse{336}{336}}
\put(3236,5850){\ellipse{336}{336}}
\thicklines
\path(1750,1734)(2022,1590)
\path(3398,5286)(3078,5270)
\path(314,918)(42,854)
\path(375,742)(810,1174)
\path(3734,3846)(4054,3926)
\path(3686,4294)(3350,4774)
\path(3238,5286)(3236,5884)
\path(2699,4240)(1159,2246)
\path(1015,1814)(970,1446)
\path(1239,1174)(1207,1494)
\path(890,1830)(1146,1814)
\path(1894,1414)(5254,2054)
\path(1894,1654)(3670,4102)
\path(3894,3894)(4598,1622)
\path(1239,1334)(1559,1398)
\path(4712,1287)(1159,2054)
\path(5094,2198)(5238,2406)
\path(6182,806)(5686,1126)
\path(6134,694)(6262,902)
\path(5414,1926)(5558,1542)
\path(5652,1152)(4790,1302)
\path(5430,1494)(5734,1590)
\path(2774,4230)(5174,2294)
\path(2854,5894)(2950,5590)
\path(3142,4822)(2854,4454)
\path(2726,4534)(2982,4374)
\path(4730,1654)(4426,1574)
\path(283,615)(284,615)(285,614)
	(287,614)(291,612)(297,611)
	(305,608)(316,605)(328,602)
	(344,597)(362,592)(383,586)
	(407,579)(434,572)(464,563)
	(497,554)(533,545)(571,535)
	(612,524)(655,513)(701,501)
	(749,489)(799,477)(851,464)
	(905,452)(960,439)(1017,426)
	(1076,414)(1136,401)(1198,389)
	(1262,376)(1327,364)(1394,352)
	(1463,341)(1534,329)(1607,318)
	(1682,308)(1760,297)(1840,287)
	(1923,278)(2008,269)(2097,260)
	(2189,252)(2285,245)(2384,238)
	(2486,232)(2593,226)(2702,222)
	(2815,218)(2931,215)(3048,214)
	(3168,213)(3276,213)(3383,215)
	(3490,217)(3596,220)(3700,224)
	(3802,228)(3902,233)(4000,239)
	(4096,245)(4189,252)(4280,259)
	(4369,267)(4456,275)(4542,283)
	(4625,292)(4706,301)(4786,310)
	(4864,320)(4941,330)(5017,340)
	(5091,350)(5164,361)(5235,372)
	(5306,383)(5375,394)(5444,405)
	(5511,416)(5576,428)(5641,439)
	(5704,450)(5766,462)(5826,473)
	(5884,484)(5941,495)(5995,505)
	(6047,516)(6098,526)(6145,535)
	(6190,544)(6233,553)(6272,561)
	(6309,569)(6342,576)(6373,582)
	(6400,588)(6424,593)(6445,598)
	(6464,602)(6479,605)(6492,608)
	(6502,610)(6510,612)(6516,613)
	(6520,614)(6522,615)(6523,615)(6524,615)
\path(2902,5718)(2901,5718)(2900,5717)
	(2897,5716)(2893,5714)(2886,5711)
	(2877,5708)(2866,5703)(2852,5697)
	(2834,5690)(2814,5681)(2791,5671)
	(2765,5659)(2735,5646)(2702,5632)
	(2667,5616)(2628,5598)(2587,5579)
	(2543,5559)(2497,5537)(2449,5514)
	(2399,5489)(2347,5463)(2294,5436)
	(2239,5408)(2184,5378)(2127,5348)
	(2070,5316)(2012,5283)(1954,5248)
	(1895,5213)(1836,5176)(1777,5138)
	(1718,5098)(1658,5057)(1599,5014)
	(1539,4970)(1480,4924)(1420,4876)
	(1361,4825)(1301,4773)(1241,4718)
	(1182,4661)(1122,4600)(1062,4537)
	(1003,4471)(944,4402)(885,4330)
	(826,4255)(769,4177)(712,4095)
	(657,4011)(603,3925)(552,3836)
	(503,3746)(458,3655)(415,3564)
	(376,3474)(339,3384)(306,3295)
	(275,3208)(248,3122)(223,3037)
	(201,2954)(182,2873)(164,2793)
	(149,2714)(136,2637)(125,2560)
	(116,2485)(108,2411)(102,2338)
	(97,2267)(94,2195)(92,2125)
	(91,2056)(91,1988)(92,1920)
	(94,1853)(97,1788)(100,1724)
	(104,1661)(109,1599)(114,1539)
	(119,1480)(125,1424)(131,1369)
	(137,1317)(143,1267)(149,1220)
	(154,1176)(160,1135)(165,1097)
	(171,1062)(175,1031)(180,1003)
	(184,978)(187,957)(190,939)
	(192,924)(194,912)(196,903)
	(197,896)(198,891)(199,888)
	(199,887)(199,886)
\path(3398,5846)(3399,5846)(3400,5845)
	(3403,5844)(3407,5842)(3413,5840)
	(3422,5836)(3432,5831)(3446,5825)
	(3462,5818)(3481,5809)(3504,5800)
	(3529,5788)(3557,5775)(3589,5761)
	(3623,5745)(3661,5727)(3701,5708)
	(3743,5688)(3788,5666)(3836,5643)
	(3885,5618)(3936,5592)(3989,5565)
	(4043,5536)(4099,5507)(4156,5476)
	(4213,5444)(4272,5410)(4331,5376)
	(4390,5341)(4450,5304)(4511,5266)
	(4571,5226)(4632,5186)(4693,5144)
	(4754,5100)(4816,5055)(4877,5008)
	(4939,4959)(5001,4909)(5062,4856)
	(5125,4801)(5187,4743)(5249,4683)
	(5312,4620)(5374,4555)(5437,4487)
	(5500,4415)(5562,4341)(5624,4264)
	(5686,4183)(5746,4100)(5806,4014)
	(5864,3926)(5920,3836)(5975,3741)
	(6028,3646)(6077,3550)(6124,3455)
	(6167,3361)(6206,3268)(6243,3176)
	(6276,3086)(6307,2998)(6335,2911)
	(6360,2825)(6383,2742)(6404,2659)
	(6422,2579)(6438,2499)(6452,2421)
	(6465,2344)(6476,2268)(6485,2193)
	(6493,2119)(6500,2045)(6505,1973)
	(6510,1902)(6513,1832)(6515,1763)
	(6517,1695)(6517,1628)(6517,1562)
	(6517,1498)(6515,1435)(6514,1375)
	(6511,1316)(6509,1259)(6506,1205)
	(6503,1154)(6500,1105)(6497,1059)
	(6494,1016)(6491,977)(6488,941)
	(6485,908)(6482,879)(6480,854)
	(6478,832)(6476,813)(6474,797)
	(6473,785)(6472,775)(6471,768)
	(6471,763)(6470,760)(6470,759)(6470,758)
\put(2027,4373){\makebox(0,0)[lb]{\smash{{\SetFigFont{6}{7.2}{}{}{}$t_3$}}}}
\put(4309,4240){\makebox(0,0)[lb]{\smash{{\SetFigFont{6}{7.2}{}{}{}$w_3$}}}}
\put(3773,5045){\makebox(0,0)[lb]{\smash{{\SetFigFont{6}{7.2}{}{}{}$v_3$}}}}
\put(4712,750){\makebox(0,0)[lb]{\smash{{\SetFigFont{6}{7.2}{}{}{}$w_2$}}}}
\put(5920,2226){\makebox(0,0)[lb]{\smash{{\SetFigFont{6}{7.2}{}{}{}$t_2$}}}}
\put(6859,213){\makebox(0,0)[lb]{\smash{{\SetFigFont{6}{7.2}{}{}{}$u_2$}}}}
\put(418,2226){\makebox(0,0)[lb]{\smash{{\SetFigFont{6}{7.2}{}{}{}$w_1$}}}}
\put(1894,884){\makebox(0,0)[lb]{\smash{{\SetFigFont{6}{7.2}{}{}{}$t_1$}}}}
\put(1089,750){\makebox(0,0)[lb]{\smash{{\SetFigFont{6}{7.2}{}{}{}$v_1$}}}}
\put(15,78){\makebox(0,0)[lb]{\smash{{\SetFigFont{6}{7.2}{}{}{}$u_1$}}}}
\put(3236,6252){\makebox(0,0)[lb]{\smash{{\SetFigFont{6}{7.2}{}{}{}$u_3$}}}}
\put(5920,1420){\makebox(0,0)[lb]{\smash{{\SetFigFont{6}{7.2}{}{}{}$v_2$}}}}
\end{picture}
}}
\subfigure[partition $\mathcal{T}_3$.]{\label{Figure:FlowerSnarkF3T3}\setlength{\unitlength}{0.00024996in}
\begingroup\makeatletter\ifx\SetFigFont\undefined%
\gdef\SetFigFont#1#2#3#4#5{%
  \reset@font\fontsize{#1}{#2pt}%
  \fontfamily{#3}\fontseries{#4}\fontshape{#5}%
  \selectfont}%
\fi\endgroup%
{\renewcommand{\dashlinestretch}{30}
\begin{picture}(6874,6450)(0,-10)
\put(3236,5850){\blacken\ellipse{336}{336}}
\put(3236,5850){\ellipse{336}{336}}
\put(3252,4938){\blacken\ellipse{336}{336}}
\put(3252,4938){\ellipse{336}{336}}
\put(3812,4170){\blacken\ellipse{336}{336}}
\put(3812,4170){\ellipse{336}{336}}
\put(2643,4268){\blacken\ellipse{302}{338}}
\put(2643,4268){\ellipse{302}{338}}
\put(1044,2090){\blacken\ellipse{336}{336}}
\put(1044,2090){\ellipse{336}{336}}
\put(948,1258){\blacken\ellipse{336}{336}}
\put(948,1258){\ellipse{336}{336}}
\put(1732,1418){\blacken\ellipse{336}{336}}
\put(1732,1418){\ellipse{336}{336}}
\put(244,618){\blacken\ellipse{336}{336}}
\put(244,618){\ellipse{336}{336}}
\put(4644,1338){\blacken\ellipse{336}{336}}
\put(4644,1338){\ellipse{336}{336}}
\put(5412,2106){\blacken\ellipse{336}{336}}
\put(5412,2106){\ellipse{336}{336}}
\put(5588,1242){\blacken\ellipse{336}{336}}
\put(5588,1242){\ellipse{336}{336}}
\put(6452,586){\blacken\ellipse{336}{336}}
\put(6452,586){\ellipse{336}{336}}
\thicklines
\path(5526,1870)(5350,1819)
\path(2390,4072)(2566,3976)
\path(3142,4822)(2762,4358)
\path(3316,5624)(3135,5618)
\path(2774,4230)(5274,2214)
\path(5349,1316)(5338,1141)
\path(5346,1236)(4790,1302)
\path(5447,1842)(5562,1414)
\path(6154,451)(6125,609)
\path(6314,646)(5690,1142)
\path(4365,1265)(4410,1463)
\path(4390,1366)(1159,2054)
\path(1114,1286)(1488,1370)
\path(3850,4022)(4618,1510)
\path(1806,1552)(3630,3984)
\path(1894,1414)(5254,2054)
\path(1152,2402)(1304,2306)
\path(798,1045)(679,1158)
\path(1018,1942)(962,1424)
\path(2490,4022)(1226,2342)
\path(3242,5110)(3222,5618)
\path(3703,4308)(3390,4696)
\path(3726,3928)(3550,4056)
\path(375,742)(746,1094)
\path(443,456)(500,648)
\path(3478,4752)(3294,4632)
\path(1513,1271)(1484,1469)
\path(3398,5846)(3399,5846)(3400,5845)
	(3403,5844)(3407,5842)(3413,5840)
	(3422,5836)(3432,5831)(3446,5825)
	(3462,5818)(3481,5809)(3504,5800)
	(3529,5788)(3557,5775)(3589,5761)
	(3623,5745)(3661,5727)(3701,5708)
	(3743,5688)(3788,5666)(3836,5643)
	(3885,5618)(3936,5592)(3989,5565)
	(4043,5536)(4099,5507)(4156,5476)
	(4213,5444)(4272,5410)(4331,5376)
	(4390,5341)(4450,5304)(4511,5266)
	(4571,5226)(4632,5186)(4693,5144)
	(4754,5100)(4816,5055)(4877,5008)
	(4939,4959)(5001,4909)(5062,4856)
	(5125,4801)(5187,4743)(5249,4683)
	(5312,4620)(5374,4555)(5437,4487)
	(5500,4415)(5562,4341)(5624,4264)
	(5686,4183)(5746,4100)(5806,4014)
	(5864,3926)(5920,3836)(5975,3741)
	(6028,3646)(6077,3550)(6124,3455)
	(6167,3361)(6206,3268)(6243,3176)
	(6276,3086)(6307,2998)(6335,2911)
	(6360,2825)(6383,2742)(6404,2659)
	(6422,2579)(6438,2499)(6452,2421)
	(6465,2344)(6476,2268)(6485,2193)
	(6493,2119)(6500,2045)(6505,1973)
	(6510,1902)(6513,1832)(6515,1763)
	(6517,1695)(6517,1628)(6517,1562)
	(6517,1498)(6515,1435)(6514,1375)
	(6511,1316)(6509,1259)(6506,1205)
	(6503,1154)(6500,1105)(6497,1059)
	(6494,1016)(6491,977)(6488,941)
	(6485,908)(6482,879)(6480,854)
	(6478,832)(6476,813)(6474,797)
	(6473,785)(6472,775)(6471,768)
	(6471,763)(6470,760)(6470,759)(6470,758)
\path(3082,5814)(3081,5814)(3080,5813)
	(3077,5812)(3073,5810)(3066,5808)
	(3058,5804)(3047,5799)(3033,5794)
	(3016,5787)(2996,5778)(2973,5768)
	(2947,5757)(2918,5745)(2886,5730)
	(2850,5715)(2812,5698)(2771,5679)
	(2727,5659)(2680,5638)(2632,5615)
	(2581,5591)(2529,5565)(2474,5538)
	(2419,5510)(2362,5481)(2304,5451)
	(2245,5420)(2186,5387)(2125,5353)
	(2065,5318)(2004,5282)(1942,5245)
	(1881,5206)(1819,5166)(1758,5125)
	(1696,5082)(1634,5038)(1572,4992)
	(1510,4944)(1448,4894)(1387,4842)
	(1325,4787)(1263,4731)(1201,4672)
	(1139,4610)(1078,4545)(1016,4478)
	(955,4408)(894,4334)(834,4258)
	(775,4179)(717,4097)(660,4012)
	(605,3925)(552,3836)(500,3742)
	(452,3648)(406,3554)(364,3460)
	(326,3367)(291,3275)(259,3184)
	(230,3095)(204,3008)(181,2922)
	(161,2837)(143,2754)(127,2673)
	(114,2593)(103,2514)(94,2437)
	(86,2361)(80,2285)(76,2211)
	(73,2138)(72,2065)(71,1994)
	(72,1923)(74,1854)(77,1785)
	(81,1718)(85,1652)(90,1587)
	(96,1523)(102,1461)(109,1401)
	(115,1343)(122,1287)(129,1233)
	(136,1182)(143,1134)(150,1088)
	(157,1046)(163,1007)(169,971)
	(175,939)(180,910)(184,885)
	(188,863)(192,844)(194,829)
	(197,817)(199,807)(200,800)
	(201,795)(202,792)(202,791)(202,790)
\path(488,554)(489,554)(490,553)
	(493,553)(497,552)(503,550)
	(512,548)(523,545)(536,542)
	(553,538)(572,533)(594,527)
	(620,521)(648,514)(679,507)
	(714,499)(751,490)(790,481)
	(833,471)(877,461)(924,451)
	(973,440)(1024,430)(1076,419)
	(1131,408)(1187,397)(1245,386)
	(1304,375)(1365,364)(1427,353)
	(1491,343)(1557,332)(1624,322)
	(1693,312)(1765,303)(1838,293)
	(1914,284)(1993,276)(2074,267)
	(2158,259)(2245,252)(2335,245)
	(2428,239)(2525,233)(2625,228)
	(2729,223)(2835,219)(2944,216)
	(3055,214)(3168,213)(3273,213)
	(3377,214)(3480,215)(3582,217)
	(3682,220)(3779,224)(3875,228)
	(3968,233)(4058,238)(4146,244)
	(4232,250)(4315,256)(4397,263)
	(4476,270)(4553,278)(4629,286)
	(4703,294)(4775,302)(4846,311)
	(4916,319)(4984,328)(5051,338)
	(5117,347)(5182,356)(5245,366)
	(5307,376)(5368,385)(5427,395)
	(5485,404)(5541,414)(5595,423)
	(5647,432)(5697,441)(5745,450)
	(5791,458)(5834,466)(5874,473)
	(5912,480)(5946,487)(5978,493)
	(6006,498)(6032,503)(6054,508)
	(6073,511)(6089,515)(6103,517)
	(6114,519)(6122,521)(6128,522)
	(6132,523)(6135,524)(6136,524)(6137,524)
\put(3236,6252){\makebox(0,0)[lb]{\smash{{\SetFigFont{6}{7.2}{}{}{}$u_3$}}}}
\put(15,78){\makebox(0,0)[lb]{\smash{{\SetFigFont{6}{7.2}{}{}{}$u_1$}}}}
\put(1089,750){\makebox(0,0)[lb]{\smash{{\SetFigFont{6}{7.2}{}{}{}$v_1$}}}}
\put(1894,884){\makebox(0,0)[lb]{\smash{{\SetFigFont{6}{7.2}{}{}{}$t_1$}}}}
\put(418,2226){\makebox(0,0)[lb]{\smash{{\SetFigFont{6}{7.2}{}{}{}$w_1$}}}}
\put(6859,213){\makebox(0,0)[lb]{\smash{{\SetFigFont{6}{7.2}{}{}{}$u_2$}}}}
\put(5920,2226){\makebox(0,0)[lb]{\smash{{\SetFigFont{6}{7.2}{}{}{}$t_2$}}}}
\put(4712,750){\makebox(0,0)[lb]{\smash{{\SetFigFont{6}{7.2}{}{}{}$w_2$}}}}
\put(3773,5045){\makebox(0,0)[lb]{\smash{{\SetFigFont{6}{7.2}{}{}{}$v_3$}}}}
\put(4309,4240){\makebox(0,0)[lb]{\smash{{\SetFigFont{6}{7.2}{}{}{}$w_3$}}}}
\put(2027,4373){\makebox(0,0)[lb]{\smash{{\SetFigFont{6}{7.2}{}{}{}$t_3$}}}}
\put(5920,1420){\makebox(0,0)[lb]{\smash{{\SetFigFont{6}{7.2}{}{}{}$v_2$}}}}
\end{picture}
}}
\end{center}
\caption{Three compatible normal odd partitions of the Flower snark $F_3$.}
\label{Figure:3PNICdeFowerSnarkF3}
\end{figure}
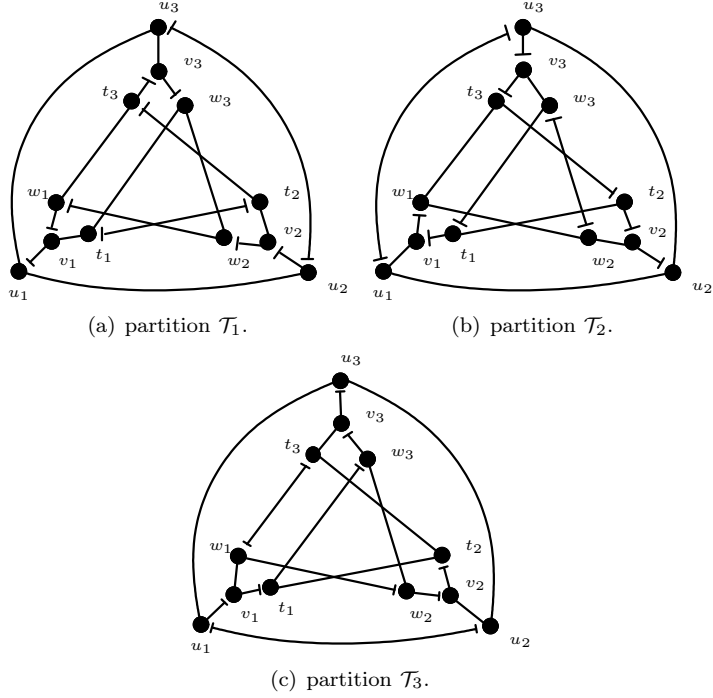
\newline Assume  that for an odd integer $k$, $k\geq 3$, $F_k$ is provided with \TroisPNICsv, namely $\mathcal{T}_1$, $\mathcal{T}_2$ and $\mathcal{T}_3$.  
Suppose further that the  Properties above (\ref{Equation:FlowerSnarks:Claw1:1})--(\ref{Equation:FlowerSnarks:Claw2:3}) are verified by $\mathcal{T}_1$, $\mathcal{T}_2$, and $\mathcal{T}_3$.

We derive $F_{k+2}$ from $F_k$ as follows:

$\begin{array}{ll}
V(F_{k+2})=V(F_k)\cup \{u'_1, v'_1, w'_1, t'_1, u'_2, v'_2, w'_2, t'_2 \}\\
E(F_{k+2})=E(F_k)-\!\{u_1u_2, w_1w_2, t_1t_2\}\\
\hspace{2.8cm}\cup \;\{u_1u'_2, w_1w'_2, t_1t'_2\}\\
\hspace{2.8cm}\cup\;\{v'_2u'_2, v'_2w'_2, v'_2t'_2\}\\
\hspace{2.8cm} \cup\;\{u'_2u'_1, w'_2w'_1, t'_2t'_1\}\\
\hspace{2.8cm} \cup\;\{v'_1u'_1, v'_1w'_1, v'_1t'_1\}\\
\hspace{2.8cm} \cup\;\{u'_1u_2, w'_1w_2, t'_1t_2\}
\end{array}
$

In other words, we insert eight new vertices into $F_k$, we delete the edges $u_1u_2$, $w_1w_2$, $t_1t_2$ and add new edges in order to obtain the Flower snark $F_{k+2}$.

\begin{figure}
\centering
\subfigure[partition $\mathcal{T}'_1$.] {\label{Figure:FlowerSnarkFkT1}\input{FlowerSnarkGeneralisePartitionT1.eepic}}
\subfigure[partition $\mathcal{T}'_2$.] {\label{Figure:FlowerSnarkFkT2}\input{FlowerSnarkGeneralisePartitionT2.eepic}}
\subfigure[partition $\mathcal{T}'_3$.] {\label{Figure:FlowerSnarkFkT3}\input{FlowerSnarkGeneralisePartitionT3.eepic}}
\caption{Extension of \TroisPNIC from the Flower snark $F_k$ to the Flower snark $F_{k+2}$.}
\label{Figure:3PNICdeFowerSnarkFk}
\end{figure}
Figures \ref{Figure:FlowerSnarkFkT1}, \ref{Figure:FlowerSnarkFkT2}, and \ref{Figure:FlowerSnarkFkT3} show three normal partitions of $F_{k+2}$, $\mathcal{T}'_1$, $\mathcal{T}'_2$ and $\mathcal{T}'_3$
 obtained, respectively, from $\mathcal{T}_1$, $\mathcal{T}_2$ and $\mathcal{T}_3$.

But now we rename some vertices of $F_{k+2}$ as follows:
For $i\geq 2$, the vertices $u_i$, $v_i$, $w_i$, and $t_i$, are renamed, respectively,  $u_{i+2}$, $v_{i+2}$, $w_{i+2}$, and $t_{i+2}$.
The vertices $u'_1$, $v'_1$, $w'_1$, and $t'_1$ are renamed, respectively, $u_3$, $v_3$, $w_3$, and $t_3$.
The vertices $u'_2$, $v'_2$, $w'_2$, and $t'_2$ are renamed, respectively, $u_2$, $v_2$, $w_2$, and $t_2$.

It is a routine matter to check that those partitions are odd, compatible, and satisfy Properties  (\ref{Equation:FlowerSnarks:Claw1:1})--(\ref{Equation:FlowerSnarks:Claw2:3}).
\end{Prf}
\subsection{Goldberg snarks}
\begin{figure}
\centering
\setlength{\unitlength}{0.00026247in}
\begingroup\makeatletter\ifx\SetFigFont\undefined%
\gdef\SetFigFont#1#2#3#4#5{%
  \reset@font\fontsize{#1}{#2pt}%
  \fontfamily{#3}\fontseries{#4}\fontshape{#5}%
  \selectfont}%
\fi\endgroup%
{\renewcommand{\dashlinestretch}{30}
\begin{picture}(6354,5750)(0,-10)
\put(6180,173){\blacken\ellipse{332}{332}}
\put(6180,173){\ellipse{332}{332}}
\put(4380,218){\blacken\ellipse{332}{332}}
\put(4380,218){\ellipse{332}{332}}
\put(2580,204){\blacken\ellipse{332}{332}}
\put(2580,204){\ellipse{332}{332}}
\put(780,204){\blacken\ellipse{332}{332}}
\put(780,204){\ellipse{332}{332}}
\put(825,5123){\blacken\ellipse{332}{332}}
\put(825,5123){\ellipse{332}{332}}
\put(6135,5123){\blacken\ellipse{332}{332}}
\put(6135,5123){\ellipse{332}{332}}
\put(6180,2918){\blacken\ellipse{332}{332}}
\put(6180,2918){\ellipse{332}{332}}
\put(2565,1517){\blacken\ellipse{332}{332}}
\put(2565,1517){\ellipse{332}{332}}
\put(2130,2873){\blacken\ellipse{332}{332}}
\put(2130,2873){\ellipse{332}{332}}
\put(3480,3773){\blacken\ellipse{332}{332}}
\put(3480,3773){\ellipse{332}{332}}
\put(4785,2873){\blacken\ellipse{332}{332}}
\put(4785,2873){\ellipse{332}{332}}
\put(4335,1568){\blacken\ellipse{332}{332}}
\put(4335,1568){\ellipse{332}{332}}
\put(3480,5123){\blacken\ellipse{332}{332}}
\put(3480,5123){\ellipse{332}{332}}
\put(735,2859){\blacken\ellipse{332}{332}}
\put(735,2859){\ellipse{332}{332}}
\path(780,173)(6180,173)
\path(780,2873)(6180,2873)
\path(780,5123)(6180,5123)
\path(4380,1523)(4380,263)
\path(2580,1523)(2580,173)
\path(3480,5123)(3480,3773)
\path(3480,3773)(2580,1523)(4830,2873)
	(2130,2873)(4380,1523)(3480,3773)
\put(2738,357){\makebox(0,0)[lb]{\smash{{\SetFigFont{5}{6.0}{\rmdefault}{}{}$v_8^j$}}}}
\put(4600,1456){\makebox(0,0)[lb]{\smash{{\SetFigFont{5}{6.0}{\rmdefault}{}{}$v_2^j$}}}}
\put(6105,456){\makebox(0,0)[lb]{\smash{{\SetFigFont{5}{6.0}{\rmdefault}{}{}$v_8^{j+1}$}}}}
\put(285,533){\makebox(0,0)[lb]{\smash{{\SetFigFont{5}{6.0}{\rmdefault}{}{}$v_7^{j-1}$}}}}
\put(1770,1433){\makebox(0,0)[lb]{\smash{{\SetFigFont{5}{6.0}{\rmdefault}{}{}$v_5^j$}}}}
\put(4560,488){\makebox(0,0)[lb]{\smash{{\SetFigFont{5}{6.0}{\rmdefault}{}{}$v_7^j$}}}}
\put(15,3233){\makebox(0,0)[lb]{\smash{{\SetFigFont{5}{6.0}{\rmdefault}{}{}$v_4^{j-1}$}}}}
\put(375,5528){\makebox(0,0)[lb]{\smash{{\SetFigFont{5}{6.0}{\rmdefault}{}{}$v_6^{j-1}$}}}}
\put(6045,3233){\makebox(0,0)[lb]{\smash{{\SetFigFont{5}{6.0}{\rmdefault}{}{}$v_3^{j+1}$}}}}
\put(6000,5438){\makebox(0,0)[lb]{\smash{{\SetFigFont{5}{6.0}{\rmdefault}{}{}$v_6^{j+1}$}}}}
\put(3210,5483){\makebox(0,0)[lb]{\smash{{\SetFigFont{5}{6.0}{\rmdefault}{}{}$v_6^j$}}}}
\put(3615,4133){\makebox(0,0)[lb]{\smash{{\SetFigFont{5}{6.0}{\rmdefault}{}{}$v_1^j$}}}}
\put(1770,3233){\makebox(0,0)[lb]{\smash{{\SetFigFont{5}{6.0}{\rmdefault}{}{}$v_3^j$}}}}
\put(4515,3323){\makebox(0,0)[lb]{\smash{{\SetFigFont{5}{6.0}{\rmdefault}{}{}$v_4^j$}}}}
\end{picture}
}
\caption{Adjacencies in a Goldberg Snark.}
\label{Figure:AdjacencesDansGoldbergSnark}
\end{figure}
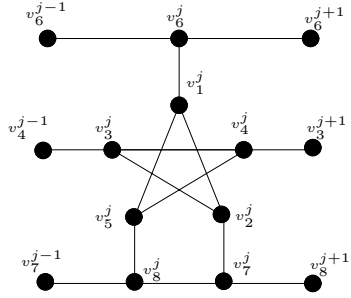

For every odd $k$ such that $k\geq 3$, the Goldberg snark $G_k$ is defined as follows: $V(G_k)=\{v_i^j: 1\leq i\leq 8, 0\leq j\leq k-1\}$ and adjacencies are defined as shown in Figure \ref{Figure:AdjacencesDansGoldbergSnark}. 
The superscript $j$ is always considered modulo $k$. Moreover, $v_6^k=v_6^0$, $v_3^k=v_4^0$, and $v_8^k=v_7^0$.

\begin{Prop}\label{Proposition:GoldbergSnarksOnt3PNIC}
If $k\geq 3$ is an odd integer, $G_k$ can be provided with \TroisPNICsv.
\end{Prop}
\begin{Prf}
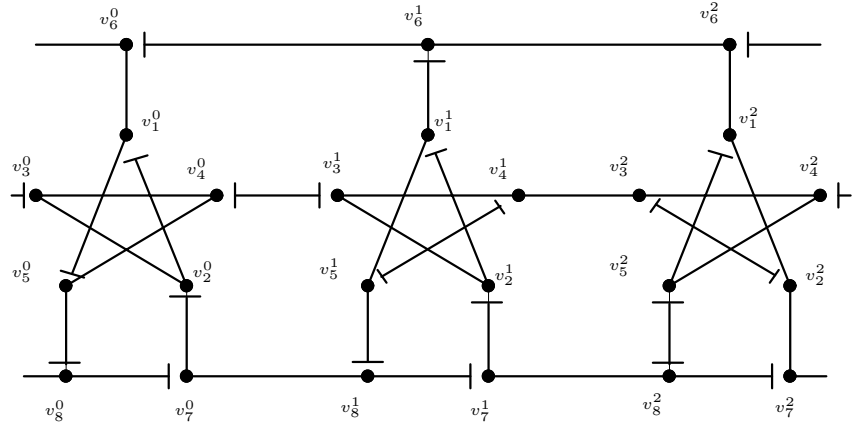
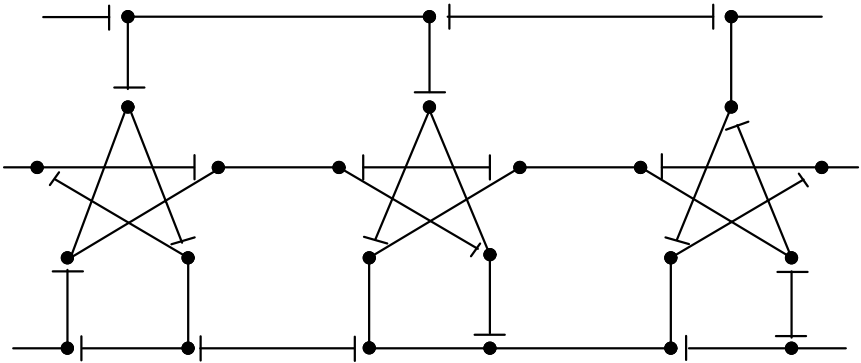
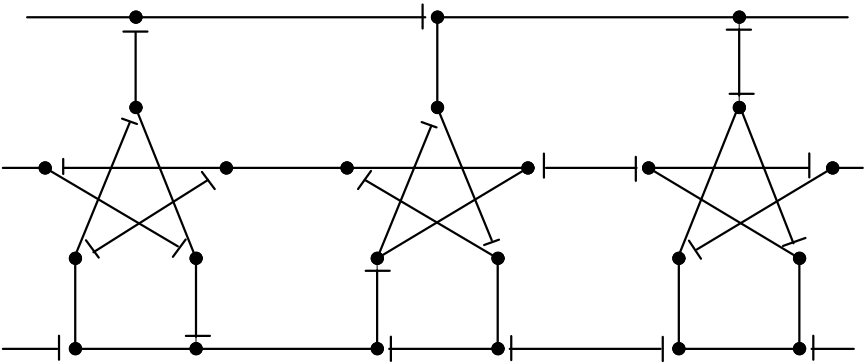
\begin{figure}
\centering
\subfigure[partition $\mathcal{T}_1$.]{\label{Figure:GoldbergSnarkG3T1}\setlength{\unitlength}{0.00034996in}
\begingroup\makeatletter\ifx\SetFigFont\undefined%
\gdef\SetFigFont#1#2#3#4#5{%
  \reset@font\fontsize{#1}{#2pt}%
  \fontfamily{#3}\fontseries{#4}\fontshape{#5}%
  \selectfont}%
\fi\endgroup%
{\renewcommand{\dashlinestretch}{30}
\begin{picture}(12598,6282)(0,-10)
\put(1754,5631){\blacken\ellipse{180}{180}}
\put(1754,5631){\ellipse{180}{180}}
\put(1754,4281){\blacken\ellipse{180}{180}}
\put(1754,4281){\ellipse{180}{180}}
\put(404,3381){\blacken\ellipse{180}{180}}
\put(404,3381){\ellipse{180}{180}}
\put(2654,2031){\blacken\ellipse{180}{180}}
\put(2654,2031){\ellipse{180}{180}}
\put(854,2031){\blacken\ellipse{180}{180}}
\put(854,2031){\ellipse{180}{180}}
\put(3104,3381){\blacken\ellipse{180}{180}}
\put(3104,3381){\ellipse{180}{180}}
\put(854,681){\blacken\ellipse{180}{180}}
\put(854,681){\ellipse{180}{180}}
\put(2654,681){\blacken\ellipse{180}{180}}
\put(2654,681){\ellipse{180}{180}}
\put(7154,681){\blacken\ellipse{180}{180}}
\put(7154,681){\ellipse{180}{180}}
\put(5354,681){\blacken\ellipse{180}{180}}
\put(5354,681){\ellipse{180}{180}}
\put(7604,3381){\blacken\ellipse{180}{180}}
\put(7604,3381){\ellipse{180}{180}}
\put(5354,2031){\blacken\ellipse{180}{180}}
\put(5354,2031){\ellipse{180}{180}}
\put(7154,2031){\blacken\ellipse{180}{180}}
\put(7154,2031){\ellipse{180}{180}}
\put(4904,3381){\blacken\ellipse{180}{180}}
\put(4904,3381){\ellipse{180}{180}}
\put(6254,4281){\blacken\ellipse{180}{180}}
\put(6254,4281){\ellipse{180}{180}}
\put(6254,5631){\blacken\ellipse{180}{180}}
\put(6254,5631){\ellipse{180}{180}}
\put(10754,5631){\blacken\ellipse{180}{180}}
\put(10754,5631){\ellipse{180}{180}}
\put(10754,4281){\blacken\ellipse{180}{180}}
\put(10754,4281){\ellipse{180}{180}}
\put(9404,3381){\blacken\ellipse{180}{180}}
\put(9404,3381){\ellipse{180}{180}}
\put(11654,2031){\blacken\ellipse{180}{180}}
\put(11654,2031){\ellipse{180}{180}}
\put(9854,2031){\blacken\ellipse{180}{180}}
\put(9854,2031){\ellipse{180}{180}}
\put(12104,3381){\blacken\ellipse{180}{180}}
\put(12104,3381){\ellipse{180}{180}}
\put(9854,681){\blacken\ellipse{180}{180}}
\put(9854,681){\ellipse{180}{180}}
\put(11654,681){\blacken\ellipse{180}{180}}
\put(11654,681){\ellipse{180}{180}}
\thicklines
\path(224,3381)(44,3381)
\path(2024,5631)(6254,5631)(10754,5631)
	(10754,4281)(11654,2031)(11654,681)(12194,681)
\path(2384,681)(854,681)(224,681)
\path(9854,1761)(9854,861)
\path(12374,3561)(12374,3201)
\path(12374,3381)(12554,3381)
\path(9606,883)(10056,883)
\path(11384,861)(11384,501)
\path(6884,861)(6884,501)
\path(2384,861)(2384,501)
\path(7154,1761)(7154,681)(9854,681)(11384,681)
\path(2654,1851)(2654,681)(5354,681)(6884,681)
\path(9606,1783)(10056,1783)
\path(6906,1783)(7356,1783)
\path(2406,1873)(2856,1873)
\path(3374,3381)(4634,3381)
\thinlines
\path(3374,3381)(4634,3381)
\thicklines
\path(4634,3561)(4634,3201)
\path(3374,3561)(3374,3201)
\path(224,3561)(224,3201)
\path(2024,5811)(2024,5451)
\thinlines
\path(1754,5631)(1754,4281)
\path(854,2031)(854,681)
\path(2654,2031)(2654,771)
\path(7154,2031)(7154,771)
\path(6254,5631)(6254,4281)
\path(10754,5631)(10754,4281)
\path(9854,2031)(9854,681)
\path(11654,2031)(11654,771)
\thicklines
\path(404,5631)(1754,5631)(1754,4281)(944,2211)
\path(1124,2155)(775,2256)
\path(11024,5811)(11024,5451)
\path(11024,5631)(12104,5631)
\path(2069,3977)(1720,3876)
\path(1889,3921)(2654,2031)(404,3381)
	(3104,3381)(854,2031)(854,861)
\path(606,884)(1056,884)
\path(6051,5384)(6501,5384)
\path(5129,895)(5579,895)
\path(6254,5361)(6254,4281)(5354,2031)(5354,895)
\path(6524,4068)(6220,3966)
\path(5635,2076)(5500,2268)
\path(7390,3111)(7255,3303)
\path(5556,2166)(7334,3224)
\path(6366,4000)(7154,2031)(4904,3381)
	(7604,3381)(9404,3381)(12104,3381)
	(9854,2031)(10630,3978)
\path(10788,3944)(10439,4045)
\path(9573,3134)(9708,3325)
\path(11350,2076)(11485,2268)
\path(9651,3246)(11429,2189)
\put(1349,5946){\makebox(0,0)[lb]{\smash{{\SetFigFont{6}{7.2}{\rmdefault}{}{}$v_6^0$}}}}
\put(5894,5991){\makebox(0,0)[lb]{\smash{{\SetFigFont{6}{7.2}{\rmdefault}{}{}$v_6^1$}}}}
\put(10304,6036){\makebox(0,0)[lb]{\smash{{\SetFigFont{6}{7.2}{\rmdefault}{}{}$v_6^2$}}}}
\put(1979,4416){\makebox(0,0)[lb]{\smash{{\SetFigFont{6}{7.2}{\rmdefault}{}{}$v_1^0$}}}}
\put(6344,4416){\makebox(0,0)[lb]{\smash{{\SetFigFont{6}{7.2}{\rmdefault}{}{}$v_1^1$}}}}
\put(10889,4461){\makebox(0,0)[lb]{\smash{{\SetFigFont{6}{7.2}{\rmdefault}{}{}$v_1^2$}}}}
\put(11789,3741){\makebox(0,0)[lb]{\smash{{\SetFigFont{6}{7.2}{\rmdefault}{}{}$v_4^2$}}}}
\put(8954,3741){\makebox(0,0)[lb]{\smash{{\SetFigFont{6}{7.2}{\rmdefault}{}{}$v_3^2$}}}}
\put(7154,3696){\makebox(0,0)[lb]{\smash{{\SetFigFont{6}{7.2}{\rmdefault}{}{}$v_4^1$}}}}
\put(4679,3786){\makebox(0,0)[lb]{\smash{{\SetFigFont{6}{7.2}{\rmdefault}{}{}$v_3^1$}}}}
\put(2654,3696){\makebox(0,0)[lb]{\smash{{\SetFigFont{6}{7.2}{\rmdefault}{}{}$v_4^0$}}}}
\put(44,3741){\makebox(0,0)[lb]{\smash{{\SetFigFont{6}{7.2}{\rmdefault}{}{}$v_3^0$}}}}
\put(44,2166){\makebox(0,0)[lb]{\smash{{\SetFigFont{6}{7.2}{\rmdefault}{}{}$v_5^0$}}}}
\put(2744,2166){\makebox(0,0)[lb]{\smash{{\SetFigFont{6}{7.2}{\rmdefault}{}{}$v_2^0$}}}}
\put(4634,2211){\makebox(0,0)[lb]{\smash{{\SetFigFont{6}{7.2}{\rmdefault}{}{}$v_5^1$}}}}
\put(7244,2121){\makebox(0,0)[lb]{\smash{{\SetFigFont{6}{7.2}{\rmdefault}{}{}$v_2^1$}}}}
\put(8954,2256){\makebox(0,0)[lb]{\smash{{\SetFigFont{6}{7.2}{\rmdefault}{}{}$v_5^2$}}}}
\put(11879,2121){\makebox(0,0)[lb]{\smash{{\SetFigFont{6}{7.2}{\rmdefault}{}{}$v_2^2$}}}}
\put(11429,141){\makebox(0,0)[lb]{\smash{{\SetFigFont{6}{7.2}{\rmdefault}{}{}$v_7^2$}}}}
\put(9449,186){\makebox(0,0)[lb]{\smash{{\SetFigFont{6}{7.2}{\rmdefault}{}{}$v_8^2$}}}}
\put(6884,96){\makebox(0,0)[lb]{\smash{{\SetFigFont{6}{7.2}{\rmdefault}{}{}$v_7^1$}}}}
\put(4949,141){\makebox(0,0)[lb]{\smash{{\SetFigFont{6}{7.2}{\rmdefault}{}{}$v_8^1$}}}}
\put(2474,96){\makebox(0,0)[lb]{\smash{{\SetFigFont{6}{7.2}{\rmdefault}{}{}$v_7^0$}}}}
\put(539,96){\makebox(0,0)[lb]{\smash{{\SetFigFont{6}{7.2}{\rmdefault}{}{}$v_8^0$}}}}
\end{picture}
}}
\subfigure[partition $\mathcal{T}_2$.]{\label{Figure:GoldbergSnarkG3T2}\setlength{\unitlength}{0.00034996in}
\begingroup\makeatletter\ifx\SetFigFont\undefined%
\gdef\SetFigFont#1#2#3#4#5{%
  \reset@font\fontsize{#1}{#2pt}%
  \fontfamily{#3}\fontseries{#4}\fontshape{#5}%
  \selectfont}%
\fi\endgroup%
{\renewcommand{\dashlinestretch}{30}
\begin{picture}(12823,5420)(0,-10)
\put(1889,5181){\blacken\ellipse{180}{180}}
\put(1889,5181){\ellipse{180}{180}}
\put(1889,3831){\blacken\ellipse{180}{180}}
\put(1889,3831){\ellipse{180}{180}}
\put(539,2931){\blacken\ellipse{180}{180}}
\put(539,2931){\ellipse{180}{180}}
\put(2789,1581){\blacken\ellipse{180}{180}}
\put(2789,1581){\ellipse{180}{180}}
\put(989,1581){\blacken\ellipse{180}{180}}
\put(989,1581){\ellipse{180}{180}}
\put(3239,2931){\blacken\ellipse{180}{180}}
\put(3239,2931){\ellipse{180}{180}}
\put(989,231){\blacken\ellipse{180}{180}}
\put(989,231){\ellipse{180}{180}}
\put(2789,231){\blacken\ellipse{180}{180}}
\put(2789,231){\ellipse{180}{180}}
\put(7289,231){\blacken\ellipse{180}{180}}
\put(7289,231){\ellipse{180}{180}}
\put(5489,237){\blacken\ellipse{180}{180}}
\put(5489,237){\ellipse{180}{180}}
\put(7739,2931){\blacken\ellipse{180}{180}}
\put(7739,2931){\ellipse{180}{180}}
\put(5489,1581){\blacken\ellipse{180}{180}}
\put(5489,1581){\ellipse{180}{180}}
\put(7289,1626){\blacken\ellipse{180}{180}}
\put(7289,1626){\ellipse{180}{180}}
\put(5039,2931){\blacken\ellipse{180}{180}}
\put(5039,2931){\ellipse{180}{180}}
\put(6389,3831){\blacken\ellipse{180}{180}}
\put(6389,3831){\ellipse{180}{180}}
\put(6389,5181){\blacken\ellipse{180}{180}}
\put(6389,5181){\ellipse{180}{180}}
\put(10889,5181){\blacken\ellipse{180}{180}}
\put(10889,5181){\ellipse{180}{180}}
\put(10889,3831){\blacken\ellipse{180}{180}}
\put(10889,3831){\ellipse{180}{180}}
\put(9539,2931){\blacken\ellipse{180}{180}}
\put(9539,2931){\ellipse{180}{180}}
\put(11789,1581){\blacken\ellipse{180}{180}}
\put(11789,1581){\ellipse{180}{180}}
\put(9989,1581){\blacken\ellipse{180}{180}}
\put(9989,1581){\ellipse{180}{180}}
\put(12239,2931){\blacken\ellipse{180}{180}}
\put(12239,2931){\ellipse{180}{180}}
\put(9989,231){\blacken\ellipse{180}{180}}
\put(9989,231){\ellipse{180}{180}}
\put(11789,231){\blacken\ellipse{180}{180}}
\put(11789,231){\ellipse{180}{180}}
\thicklines
\path(5579,1851)(6389,3786)(7289,1626)(7289,456)
\path(10259,231)(11789,231)(12599,231)
\path(11789,1379)(11789,389)
\path(11969,2751)(9989,1581)(9989,231)
	(5489,231)(5489,1581)(7739,2931)
	(9539,2931)(11789,1581)(10979,3561)
\path(9854,2931)(12284,2931)(12779,2931)
\path(989,1401)(989,231)(179,231)
\path(2879,2931)(539,2931)(44,2931)
\path(5399,2931)(7289,2931)
\path(2969,231)(5264,231)
\path(809,2751)(2789,1581)(2789,231)(1214,231)
\path(7109,1716)(5039,2931)(3284,2931)
	(1034,1581)(1889,3876)(2699,1806)
\path(10079,1851)(10889,3831)(10889,5181)(12239,5181)
\path(1889,4101)(1889,5181)(6389,5181)(6389,4056)
\path(6659,5181)(10619,5181)
\path(1613,5176)(627,5176)
\path(10215,417)(10215,57)
\path(2975,409)(2975,49)
\path(5274,404)(5274,44)
\path(7061,433)(7511,433)
\path(1611,5346)(1611,4986)
\thinlines
\path(2789,1581)(2789,321)
\path(5489,1581)(5489,231)
\path(10889,5181)(10889,3831)
\path(9989,1581)(9989,231)
\thicklines
\path(7289,3111)(7289,2751)
\path(11556,413)(12006,413)
\path(2888,1885)(2539,1784)
\path(770,1380)(1220,1380)
\path(1686,4124)(2136,4124)
\path(11578,1372)(12028,1372)
\path(12031,2646)(11896,2838)
\path(10258,1786)(9909,1887)
\path(7008,1603)(7143,1795)
\path(728,2666)(863,2857)
\path(1196,411)(1196,51)
\path(2883,3114)(2883,2754)
\path(5399,3111)(5399,2751)
\path(10619,5361)(10619,5001)
\path(6686,5361)(6686,5001)
\path(6169,4053)(6619,4053)
\path(9853,3128)(9853,2768)
\path(11144,3613)(10812,3495)
\path(5759,1795)(5410,1896)
\end{picture}
}}
\subfigure[partition $\mathcal{T}_3$.]{\label{Figure:GolbergSnarkG3T3}\setlength{\unitlength}{0.00034996in}
\begingroup\makeatletter\ifx\SetFigFont\undefined%
\gdef\SetFigFont#1#2#3#4#5{%
  \reset@font\fontsize{#1}{#2pt}%
  \fontfamily{#3}\fontseries{#4}\fontshape{#5}%
  \selectfont}%
\fi\endgroup%
{\renewcommand{\dashlinestretch}{30}
\begin{picture}(12913,5425)(0,-10)
\put(2024,5176){\blacken\ellipse{180}{180}}
\put(2024,5176){\ellipse{180}{180}}
\put(2024,3826){\blacken\ellipse{180}{180}}
\put(2024,3826){\ellipse{180}{180}}
\put(674,2926){\blacken\ellipse{180}{180}}
\put(674,2926){\ellipse{180}{180}}
\put(2924,1576){\blacken\ellipse{180}{180}}
\put(2924,1576){\ellipse{180}{180}}
\put(1124,1576){\blacken\ellipse{180}{180}}
\put(1124,1576){\ellipse{180}{180}}
\put(3374,2926){\blacken\ellipse{180}{180}}
\put(3374,2926){\ellipse{180}{180}}
\put(1124,226){\blacken\ellipse{180}{180}}
\put(1124,226){\ellipse{180}{180}}
\put(2924,226){\blacken\ellipse{180}{180}}
\put(2924,226){\ellipse{180}{180}}
\put(7424,226){\blacken\ellipse{180}{180}}
\put(7424,226){\ellipse{180}{180}}
\put(5624,226){\blacken\ellipse{180}{180}}
\put(5624,226){\ellipse{180}{180}}
\put(7874,2926){\blacken\ellipse{180}{180}}
\put(7874,2926){\ellipse{180}{180}}
\put(5624,1576){\blacken\ellipse{180}{180}}
\put(5624,1576){\ellipse{180}{180}}
\put(7424,1576){\blacken\ellipse{180}{180}}
\put(7424,1576){\ellipse{180}{180}}
\put(5174,2926){\blacken\ellipse{180}{180}}
\put(5174,2926){\ellipse{180}{180}}
\put(6524,3826){\blacken\ellipse{180}{180}}
\put(6524,3826){\ellipse{180}{180}}
\put(6524,5176){\blacken\ellipse{180}{180}}
\put(6524,5176){\ellipse{180}{180}}
\put(11024,5176){\blacken\ellipse{180}{180}}
\put(11024,5176){\ellipse{180}{180}}
\put(11024,3826){\blacken\ellipse{180}{180}}
\put(11024,3826){\ellipse{180}{180}}
\put(9674,2926){\blacken\ellipse{180}{180}}
\put(9674,2926){\ellipse{180}{180}}
\put(11924,1576){\blacken\ellipse{180}{180}}
\put(11924,1576){\ellipse{180}{180}}
\put(10124,1576){\blacken\ellipse{180}{180}}
\put(10124,1576){\ellipse{180}{180}}
\put(12419,2926){\blacken\ellipse{180}{180}}
\put(12419,2926){\ellipse{180}{180}}
\put(10124,226){\blacken\ellipse{180}{180}}
\put(10124,226){\ellipse{180}{180}}
\put(11924,226){\blacken\ellipse{180}{180}}
\put(11924,226){\ellipse{180}{180}}
\path(1124,1576)(1124,226)
\path(2924,1576)(2924,316)
\thicklines
\path(12015,1882)(11677,1760)
\thinlines
\path(7424,1576)(7424,316)
\path(5624,1576)(5624,226)
\path(6524,5176)(6524,3826)
\path(11024,5176)(11024,3826)
\path(10124,1576)(10124,226)
\thicklines
\path(6303,5366)(6303,5006)
\path(944,3061)(944,2836)
\path(8114,3140)(8114,2780)
\path(9484,3093)(9484,2733)
\path(12072,3144)(12072,2784)
\path(880,423)(880,63)
\path(5830,407)(5830,47)
\path(7626,417)(7626,57)
\path(9885,404)(9885,44)
\path(12130,423)(12130,63)
\path(1841,4960)(2201,4960)
\path(10841,4991)(11201,4991)
\path(10883,4032)(11243,4032)
\path(5453,1392)(5813,1392)
\path(2772,418)(3132,418)
\path(3202,2610)(3011,2868)
\path(1473,1587)(1282,1846)
\path(5339,2609)(5533,2882)
\path(10275,1842)(10454,1569)
\path(1821,3662)(2042,3583)
\path(6289,3609)(6510,3530)
\path(7221,1775)(7443,1854)
\path(2771,1859)(2579,1600)
\path(5444,2746)(7424,1576)(7424,226)(5804,226)
\path(5624,1396)(5624,226)(1124,226)
	(1124,1621)(1934,3601)
\path(2924,406)(2924,1576)(2024,3826)(2024,4951)
\path(7334,1846)(6524,3826)(6524,5176)
	(11024,5176)(12644,5176)
\path(11024,4996)(11024,4006)
\path(8144,2926)(9494,2926)
\path(7604,226)(9854,226)
\path(6344,5176)(404,5176)
\path(2654,1756)(674,2926)(44,2926)
\path(854,226)(44,226)
\path(12104,226)(12734,226)
\path(10394,1711)(12419,2926)(12869,2926)
\path(12059,2926)(9674,2926)(11924,1576)
	(11924,226)(10124,226)(10124,1621)
	(11024,3871)(11834,1801)
\path(944,2926)(7874,2926)(5624,1576)(6434,3556)
\path(1394,1666)(3104,2746)
\end{picture}
}}
\caption{Three compatible normal odd partitions of the Goldberg snark $G_3$.}
\label{Figure:3PNICdeGolbergSnarkF3}
\end{figure}
The proofs of Propositions \ref{Proposition:FlowerSnarksOnt3PNIC} and \ref{Proposition:GoldbergSnarksOnt3PNIC} are similar. 
Thus we do not give the details. We just mention that Figure \ref{Figure:3PNICdeGolbergSnarkF3} gives three compatible normal odd  partitions of $G_3$ 
while Figure \ref{Figure:3PNICdeGolbergSnarkGk} describes the construction of such partitions for $G_{k+2}$ from those of $G_k$.
\begin{figure}
\centering
\subfigure[partition $\mathcal{T}'_1$.] {\label{Figure:GoldbergSnarkGkT1}\setlength{\unitlength}{0.00030621in}
\begingroup\makeatletter\ifx\SetFigFont\undefined
% extract first six characters in \fmtname
\def\x#1#2#3#4#5#6#7\relax{\def\x{#1#2#3#4#5#6}}%
\expandafter\x\fmtname xxxxxx\relax \def\y{splain}%
\ifx\x\y   % LaTeX or SliTeX?
\gdef\SetFigFont#1#2#3{%
  \ifnum #1<17\tiny\else \ifnum #1<20\small\else
  \ifnum #1<24\normalsize\else \ifnum #1<29\large\else
  \ifnum #1<34\Large\else \ifnum #1<41\LARGE\else
     \huge\fi\fi\fi\fi\fi\fi
  \csname #3\endcsname}%
\else
\gdef\SetFigFont#1#2#3{\begingroup
  \count@#1\relax \ifnum 25<\count@\count@25\fi
  \def\x{\endgroup\@setsize\SetFigFont{#2pt}}%
  \expandafter\x
    \csname \romannumeral\the\count@ pt\expandafter\endcsname
    \csname @\romannumeral\the\count@ pt\endcsname
  \csname #3\endcsname}%
\fi
\fi\endgroup
{\renewcommand{\dashlinestretch}{30}
\begin{picture}(18503,5419)(0,-10)
\thicklines
\path(4098,2957)(5453,2957)
\thinlines
\put(7019,3853){\blacken\ellipse{180}{180}}
\put(7019,3853){\ellipse{180}{180}}
\put(5669,2953){\blacken\ellipse{180}{180}}
\put(5669,2953){\ellipse{180}{180}}
\put(7919,1603){\blacken\ellipse{180}{180}}
\put(7919,1603){\ellipse{180}{180}}
\put(6119,1603){\blacken\ellipse{180}{180}}
\put(6119,1603){\ellipse{180}{180}}
\put(8369,2953){\blacken\ellipse{180}{180}}
\put(8369,2953){\ellipse{180}{180}}
\put(6119,253){\blacken\ellipse{180}{180}}
\put(6119,253){\ellipse{180}{180}}
\put(7919,253){\blacken\ellipse{180}{180}}
\put(7919,253){\ellipse{180}{180}}
\put(12408,242){\blacken\ellipse{180}{180}}
\put(12408,242){\ellipse{180}{180}}
\put(10619,253){\blacken\ellipse{180}{180}}
\put(10619,253){\ellipse{180}{180}}
\put(12906,2953){\blacken\ellipse{180}{180}}
\put(12906,2953){\ellipse{180}{180}}
\put(10619,1603){\blacken\ellipse{180}{180}}
\put(10619,1603){\ellipse{180}{180}}
\put(12419,1603){\blacken\ellipse{180}{180}}
\put(12419,1603){\ellipse{180}{180}}
\put(10169,2953){\blacken\ellipse{180}{180}}
\put(10169,2953){\ellipse{180}{180}}
\put(11519,3853){\blacken\ellipse{180}{180}}
\put(11519,3853){\ellipse{180}{180}}
\put(11519,5203){\blacken\ellipse{180}{180}}
\put(11519,5203){\ellipse{180}{180}}
\put(3419,253){\blacken\ellipse{180}{180}}
\put(3419,253){\ellipse{180}{180}}
\put(1619,253){\blacken\ellipse{180}{180}}
\put(1619,253){\ellipse{180}{180}}
\put(3869,2953){\blacken\ellipse{180}{180}}
\put(3869,2953){\ellipse{180}{180}}
\put(1619,1603){\blacken\ellipse{180}{180}}
\put(1619,1603){\ellipse{180}{180}}
\put(3419,1603){\blacken\ellipse{180}{180}}
\put(3419,1603){\ellipse{180}{180}}
\put(1169,2953){\blacken\ellipse{180}{180}}
\put(1169,2953){\ellipse{180}{180}}
\put(2519,3853){\blacken\ellipse{180}{180}}
\put(2519,3853){\ellipse{180}{180}}
\put(2519,5203){\blacken\ellipse{180}{180}}
\put(2519,5203){\ellipse{180}{180}}
\put(16109,5202){\blacken\ellipse{180}{180}}
\put(16109,5202){\ellipse{180}{180}}
\put(16123,3833){\blacken\ellipse{180}{180}}
\put(16123,3833){\ellipse{180}{180}}
\put(14767,2935){\blacken\ellipse{180}{180}}
\put(14767,2935){\ellipse{180}{180}}
\put(17017,1585){\blacken\ellipse{180}{180}}
\put(17017,1585){\ellipse{180}{180}}
\put(15217,1585){\blacken\ellipse{180}{180}}
\put(15217,1585){\ellipse{180}{180}}
\put(17467,2935){\blacken\ellipse{180}{180}}
\put(17467,2935){\ellipse{180}{180}}
\put(15221,231){\blacken\ellipse{180}{180}}
\put(15221,231){\ellipse{180}{180}}
\put(17017,254){\blacken\ellipse{180}{180}}
\put(17017,254){\ellipse{180}{180}}
\thicklines
\path(13148,2951)(14544,2951)
\path(2744,5203)(7019,5203)(11519,5203)
	(16154,5203)(18044,5203)
\path(12419,1423)(12419,478)
\path(10619,1423)(10619,478)
\path(11384,3538)(10619,1603)(12914,2953)(10574,2953)
\path(7919,1378)(7919,478)
\path(7154,3538)(7919,1603)(5669,2953)
	(8369,2953)(10169,2953)(12419,1603)
	(11519,3853)(11519,4978)
\path(3059,253)(1619,253)(44,253)
\path(899,2953)(44,2953)
\path(44,5203)(2519,5203)(2519,3853)(1754,1963)
\path(6344,1738)(8144,2818)
\path(7019,4978)(7019,3808)(6119,1603)(6119,478)
\path(15522,1620)(15340,1844)
\path(17319,2669)(17137,2893)
\path(6440,1638)(6290,1822)
\path(16389,3616)(16088,3478)
\path(11543,3490)(11232,3605)
\path(8225,2693)(8043,2917)
\path(7318,3595)(7009,3455)
\path(1957,1903)(1601,2048)
\path(2487,3461)(2847,3608)
\path(10579,3157)(10579,2797)
\path(6831,4962)(7191,4962)
\path(11352,4972)(11712,4972)
\path(15974,5013)(16334,5013)
\path(16810,1384)(17170,1384)
\path(15054,454)(15414,454)
\path(12237,474)(12597,474)
\path(12265,1413)(12625,1413)
\path(10409,488)(10769,488)
\path(10440,1426)(10800,1426)
\path(7738,1384)(8098,1384)
\path(7744,475)(8104,475)
\path(5965,478)(6325,478)
\path(3261,1365)(3621,1365)
\path(1464,527)(1824,527)
\path(3095,430)(3095,70)
\path(16728,404)(16728,44)
\path(14536,3126)(14536,2766)
\path(13150,3126)(13150,2766)
\path(5465,3123)(5465,2763)
\path(4100,3135)(4100,2775)
\path(929,3133)(929,2773)
\path(2743,5360)(2743,5000)
\thinlines
\path(2519,5203)(2519,3853)
\thicklines
\path(2643,3517)(3419,1603)(1169,2953)
	(3824,2953)(1619,1558)(1619,523)
\path(16114,5006)(16114,3814)(15222,1606)(15222,469)
\path(15439,1736)(17230,2786)
\path(16239,3544)(17005,1586)(14780,2953)
	(17472,2936)(18447,2928)
\path(17005,1397)(17008,255)(18459,255)
\path(3419,1378)(3419,253)(6119,253)
	(7919,253)(10619,253)(12419,253)
	(15214,238)(16714,238)
\thinlines
\put(7019,5203){\blacken\ellipse{180}{180}}
\put(7019,5203){\ellipse{180}{180}}
\end{picture}
}}
\subfigure[partition $\mathcal{T}'_2$.] {\label{Figure:GoldbergSnarkGkT2}\setlength{\unitlength}{0.00030621in}
\begingroup\makeatletter\ifx\SetFigFont\undefined
% extract first six characters in \fmtname
\def\x#1#2#3#4#5#6#7\relax{\def\x{#1#2#3#4#5#6}}%
\expandafter\x\fmtname xxxxxx\relax \def\y{splain}%
\ifx\x\y   % LaTeX or SliTeX?
\gdef\SetFigFont#1#2#3{%
  \ifnum #1<17\tiny\else \ifnum #1<20\small\else
  \ifnum #1<24\normalsize\else \ifnum #1<29\large\else
  \ifnum #1<34\Large\else \ifnum #1<41\LARGE\else
     \huge\fi\fi\fi\fi\fi\fi
  \csname #3\endcsname}%
\else
\gdef\SetFigFont#1#2#3{\begingroup
  \count@#1\relax \ifnum 25<\count@\count@25\fi
  \def\x{\endgroup\@setsize\SetFigFont{#2pt}}%
  \expandafter\x
    \csname \romannumeral\the\count@ pt\expandafter\endcsname
    \csname @\romannumeral\the\count@ pt\endcsname
  \csname #3\endcsname}%
\fi
\fi\endgroup
{\renewcommand{\dashlinestretch}{30}
\begin{picture}(18403,5435)(0,-10)
\thicklines
\path(12578,3087)(12578,2812)
\thinlines
\put(16072,3810){\blacken\ellipse{180}{180}}
\put(16072,3810){\ellipse{180}{180}}
\put(14722,2910){\blacken\ellipse{180}{180}}
\put(14722,2910){\ellipse{180}{180}}
\put(16972,1560){\blacken\ellipse{180}{180}}
\put(16972,1560){\ellipse{180}{180}}
\put(15172,1560){\blacken\ellipse{180}{180}}
\put(15172,1560){\ellipse{180}{180}}
\put(17422,2910){\blacken\ellipse{180}{180}}
\put(17422,2910){\ellipse{180}{180}}
\put(15172,210){\blacken\ellipse{180}{180}}
\put(15172,210){\ellipse{180}{180}}
\put(16962,223){\blacken\ellipse{180}{180}}
\put(16962,223){\ellipse{180}{180}}
\put(6974,5187){\blacken\ellipse{180}{180}}
\put(6974,5187){\ellipse{180}{180}}
\put(6978,3860){\blacken\ellipse{180}{180}}
\put(6978,3860){\ellipse{180}{180}}
\put(5624,2928){\blacken\ellipse{180}{180}}
\put(5624,2928){\ellipse{180}{180}}
\put(7874,1578){\blacken\ellipse{180}{180}}
\put(7874,1578){\ellipse{180}{180}}
\put(6074,1578){\blacken\ellipse{180}{180}}
\put(6074,1578){\ellipse{180}{180}}
\put(8324,2928){\blacken\ellipse{180}{180}}
\put(8324,2928){\ellipse{180}{180}}
\put(6074,228){\blacken\ellipse{180}{180}}
\put(6074,228){\ellipse{180}{180}}
\put(7874,228){\blacken\ellipse{180}{180}}
\put(7874,228){\ellipse{180}{180}}
\put(12374,228){\blacken\ellipse{180}{180}}
\put(12374,228){\ellipse{180}{180}}
\put(10574,228){\blacken\ellipse{180}{180}}
\put(10574,228){\ellipse{180}{180}}
\put(12824,2928){\blacken\ellipse{180}{180}}
\put(12824,2928){\ellipse{180}{180}}
\put(10574,1578){\blacken\ellipse{180}{180}}
\put(10574,1578){\ellipse{180}{180}}
\put(12374,1578){\blacken\ellipse{180}{180}}
\put(12374,1578){\ellipse{180}{180}}
\put(10226,2922){\blacken\ellipse{180}{180}}
\put(10226,2922){\ellipse{180}{180}}
\put(11468,3840){\blacken\ellipse{180}{180}}
\put(11468,3840){\ellipse{180}{180}}
\put(11480,5178){\blacken\ellipse{180}{180}}
\put(11480,5178){\ellipse{180}{180}}
\put(3374,228){\blacken\ellipse{180}{180}}
\put(3374,228){\ellipse{180}{180}}
\put(1574,228){\blacken\ellipse{180}{180}}
\put(1574,228){\ellipse{180}{180}}
\put(3824,2928){\blacken\ellipse{180}{180}}
\put(3824,2928){\ellipse{180}{180}}
\put(1568,1572){\blacken\ellipse{180}{180}}
\put(1568,1572){\ellipse{180}{180}}
\put(3374,1578){\blacken\ellipse{180}{180}}
\put(3374,1578){\ellipse{180}{180}}
\put(1124,2928){\blacken\ellipse{180}{180}}
\put(1124,2928){\ellipse{180}{180}}
\put(2474,3828){\blacken\ellipse{180}{180}}
\put(2474,3828){\ellipse{180}{180}}
\put(2474,5178){\blacken\ellipse{180}{180}}
\put(2474,5178){\ellipse{180}{180}}
\thicklines
\path(12576,2928)(10236,2928)(8324,2928)
	(6074,1578)(6074,228)(7626,228)
\path(7970,2946)(5630,2946)(3830,2946)
	(1580,1596)(2480,3846)(3290,1866)
\path(10754,228)(12374,228)(12374,1578)
	(11474,3828)(10574,1578)(10574,228)(7877,230)
\path(1844,228)(3374,228)(3374,1578)(1394,2748)
\path(3644,228)(5804,228)
\path(1574,1398)(1574,228)(44,228)
\path(3554,2928)(1214,2928)(134,2928)
\path(2294,5178)(224,5178)
\path(7604,1758)(5894,2748)
\path(6977,4071)(6974,5178)(2474,5178)(2474,4008)
\path(11294,5178)(7244,5178)
\path(10439,2748)(12059,1758)
\path(12599,228)(14939,228)
\path(16694,1758)(14714,2928)(12824,2928)(10844,1713)
\path(16064,4098)(16064,5178)(11474,5178)(11474,4053)
\path(18134,228)(15164,228)(15164,1578)
	(17459,2928)(18359,2928)
\path(15299,1893)(16064,3828)(16964,1533)(16964,453)
\path(17099,2928)(15029,2928)
\path(16262,5181)(18120,5181)
\path(1394,1387)(1754,1387)
\path(1456,2874)(1315,2666)
\path(3120,1779)(3476,1924)
\path(2267,5358)(2267,4998)
\path(3599,488)(3599,128)
\path(5817,446)(5817,86)
\path(7624,415)(7624,55)
\path(6444,1879)(6088,2024)
\path(5977,2866)(5795,2642)
\path(7677,1882)(7495,1658)
\path(10767,421)(10767,61)
\path(10945,1596)(10763,1820)
\path(10511,2866)(10329,2642)
\path(12144,1882)(11962,1658)
\path(16761,1860)(16649,1668)
\path(17107,3135)(17107,2775)
\path(14950,404)(14950,44)
\path(15903,4087)(16263,4087)
\path(11286,4020)(11646,4020)
\path(6786,4070)(7146,4070)
\path(16264,5368)(16264,5008)
\path(11298,5376)(11298,5016)
\path(7248,5376)(7248,5016)
\thinlines
\path(15172,1560)(15172,210)
\path(16072,5160)(16072,3810)
\thicklines
\path(2278,4020)(2638,4020)
\path(16798,459)(17158,459)
\path(1833,455)(1833,95)
\path(12583,421)(12583,61)
\path(15041,3101)(15041,2741)
\path(7970,3132)(7970,2772)
\path(3549,3114)(3550,2798)
\thinlines
\path(7874,1578)(7874,318)
\path(12374,1578)(12374,318)
\path(10574,1578)(10574,228)
\path(3374,1578)(3374,318)
\thicklines
\path(15478,1829)(15122,1974)
\path(6242,1968)(6973,3815)(7892,1579)(7892,200)
\thinlines
\put(16072,5160){\blacken\ellipse{180}{180}}
\put(16072,5160){\ellipse{180}{180}}
\end{picture}
}}
\subfigure[partition $\mathcal{T}'_3$.] {\label{Figure:GoldbergSnarkGkT3}\setlength{\unitlength}{0.00030621in}
\begingroup\makeatletter\ifx\SetFigFont\undefined
% extract first six characters in \fmtname
\def\x#1#2#3#4#5#6#7\relax{\def\x{#1#2#3#4#5#6}}%
\expandafter\x\fmtname xxxxxx\relax \def\y{splain}%
\ifx\x\y   % LaTeX or SliTeX?
\gdef\SetFigFont#1#2#3{%
  \ifnum #1<17\tiny\else \ifnum #1<20\small\else
  \ifnum #1<24\normalsize\else \ifnum #1<29\large\else
  \ifnum #1<34\Large\else \ifnum #1<41\LARGE\else
     \huge\fi\fi\fi\fi\fi\fi
  \csname #3\endcsname}%
\else
\gdef\SetFigFont#1#2#3{\begingroup
  \count@#1\relax \ifnum 25<\count@\count@25\fi
  \def\x{\endgroup\@setsize\SetFigFont{#2pt}}%
  \expandafter\x
    \csname \romannumeral\the\count@ pt\expandafter\endcsname
    \csname @\romannumeral\the\count@ pt\endcsname
  \csname #3\endcsname}%
\fi
\fi\endgroup
{\renewcommand{\dashlinestretch}{30}
\begin{picture}(18736,5448)(0,-10)
\path(1745,1584)(1745,234)
\put(16251,3841){\blacken\ellipse{180}{180}}
\put(16251,3841){\ellipse{180}{180}}
\put(14893,2916){\blacken\ellipse{180}{180}}
\put(14893,2916){\ellipse{180}{180}}
\put(17143,1566){\blacken\ellipse{180}{180}}
\put(17143,1566){\ellipse{180}{180}}
\put(15367,1566){\blacken\ellipse{180}{180}}
\put(15367,1566){\ellipse{180}{180}}
\put(17593,2916){\blacken\ellipse{180}{180}}
\put(17593,2916){\ellipse{180}{180}}
\put(15343,216){\blacken\ellipse{180}{180}}
\put(15343,216){\ellipse{180}{180}}
\put(17143,216){\blacken\ellipse{180}{180}}
\put(17143,216){\ellipse{180}{180}}
\put(7145,5184){\blacken\ellipse{180}{180}}
\put(7145,5184){\ellipse{180}{180}}
\put(7145,3834){\blacken\ellipse{180}{180}}
\put(7145,3834){\ellipse{180}{180}}
\put(5786,2926){\blacken\ellipse{180}{180}}
\put(5786,2926){\ellipse{180}{180}}
\put(8045,1584){\blacken\ellipse{180}{180}}
\put(8045,1584){\ellipse{180}{180}}
\put(6245,1584){\blacken\ellipse{180}{180}}
\put(6245,1584){\ellipse{180}{180}}
\put(8495,2934){\blacken\ellipse{180}{180}}
\put(8495,2934){\ellipse{180}{180}}
\put(6245,234){\blacken\ellipse{180}{180}}
\put(6245,234){\ellipse{180}{180}}
\put(8045,234){\blacken\ellipse{180}{180}}
\put(8045,234){\ellipse{180}{180}}
\put(12545,234){\blacken\ellipse{180}{180}}
\put(12545,234){\ellipse{180}{180}}
\put(10745,234){\blacken\ellipse{180}{180}}
\put(10745,234){\ellipse{180}{180}}
\put(12977,2931){\blacken\ellipse{180}{180}}
\put(12977,2931){\ellipse{180}{180}}
\put(10745,1584){\blacken\ellipse{180}{180}}
\put(10745,1584){\ellipse{180}{180}}
\put(12545,1584){\blacken\ellipse{180}{180}}
\put(12545,1584){\ellipse{180}{180}}
\put(10295,2934){\blacken\ellipse{180}{180}}
\put(10295,2934){\ellipse{180}{180}}
\put(11645,3834){\blacken\ellipse{180}{180}}
\put(11645,3834){\ellipse{180}{180}}
\put(11645,5184){\blacken\ellipse{180}{180}}
\put(11645,5184){\ellipse{180}{180}}
\put(3545,234){\blacken\ellipse{180}{180}}
\put(3545,234){\ellipse{180}{180}}
\put(1745,234){\blacken\ellipse{180}{180}}
\put(1745,234){\ellipse{180}{180}}
\put(3995,2934){\blacken\ellipse{180}{180}}
\put(3995,2934){\ellipse{180}{180}}
\put(1757,1602){\blacken\ellipse{180}{180}}
\put(1757,1602){\ellipse{180}{180}}
\put(3545,1584){\blacken\ellipse{180}{180}}
\put(3545,1584){\ellipse{180}{180}}
\put(1295,2934){\blacken\ellipse{180}{180}}
\put(1295,2934){\ellipse{180}{180}}
\put(2645,3834){\blacken\ellipse{180}{180}}
\put(2645,3834){\ellipse{180}{180}}
\put(2645,5184){\blacken\ellipse{180}{180}}
\put(2645,5184){\ellipse{180}{180}}
\thicklines
\path(3540,438)(3543,1597)(2636,3833)(2636,4974)
\path(6935,5162)(2635,5162)(166,5162)
\path(1550,224)(159,224)
\path(3310,1727)(1277,2935)(44,2935)
\path(3752,2785)(2043,1752)
\path(6449,223)(8016,223)(8016,1590)
	(5748,2925)(3985,2922)(1589,2922)
\path(6244,1414)(6244,247)(3546,235)
	(1731,223)(1754,1567)(2532,3535)
\path(7035,3554)(6260,1579)(8534,2946)(6119,2946)
\path(12675,2744)(10765,1589)(10742,234)(12345,234)
\path(8703,2921)(10117,2921)
\path(10565,222)(8243,222)
\path(10836,1848)(11661,3828)(11638,5172)
	(7136,5172)(7136,3887)(7949,1836)
\path(11744,3545)(12474,1801)
\path(16135,3545)(15369,1565)(17596,2909)
	(14874,2921)(12988,2921)(10309,2921)
	(12536,1589)(12536,222)(15353,222)(15353,1341)
\path(16005,5172)(11845,5172)
\path(15121,2756)(17136,1565)(17125,234)(15557,234)
\path(17372,222)(18374,222)
\path(17785,2921)(18692,2921)
\path(18633,5172)(16241,5172)(16264,3805)(17042,1825)
\path(17375,404)(17375,44)
\path(15549,404)(15549,44)
\path(15169,1343)(15529,1343)
\path(17799,3103)(17799,2743)
\path(16902,1763)(17213,1878)
\path(15031,2639)(15213,2863)
\path(16289,3507)(15978,3622)
\path(16032,5389)(16032,5029)
\path(11836,5389)(11836,5029)
\path(11611,3495)(11922,3610)
\path(12294,1751)(12605,1866)
\path(12773,2639)(12618,2839)
\path(11021,1774)(10710,1889)
\path(12359,428)(12359,68)
\path(6934,5343)(6934,4983)
\path(10555,417)(10555,57)
\path(8234,405)(8234,45)
\path(6466,428)(6466,68)
\path(7816,1798)(8127,1913)
\path(8713,3100)(8713,2740)
\path(7229,3487)(6873,3632)
\path(6098,3104)(6097,2791)
\path(3215,1627)(3373,1802)
\path(2116,1626)(1934,1850)
\path(2689,3473)(2333,3618)
\path(3369,435)(3729,435)
\path(1540,440)(1540,80)
\path(1599,3139)(1599,2779)
\path(2461,4969)(2821,4969)
\thinlines
\path(7142,5196)(7136,3829)
\path(17143,1566)(17143,306)
\path(16243,5166)(16243,3816)
\thicklines
\path(3860,2651)(3678,2875)
\path(10115,3115)(10115,2755)
\path(6079,1402)(6439,1402)
\thinlines
\path(12545,1584)(12545,324)
\path(10745,1584)(10745,234)
\path(11645,5184)(11645,3834)
\put(16243,5166){\blacken\ellipse{180}{180}}
\put(16243,5166){\ellipse{180}{180}}
\end{picture}
}}
\caption{Extending \TroisPNIC from the Goldberg snark $G_k$ to the Goldberg snark $G_{k+2}$.}
\label{Figure:3PNICdeGolbergSnarkGk}
\end{figure}
\end{Prf}
\section{Open Problems}
Fan and Raspaud \cite{FanRas} conjectured that every bridgeless cubic
graph can be provided with three perfect matchings with empty
intersection. The following Conjecture \ref{Conjecture:Fulkerson} is due to Fulkerson: it appears first in
\cite{Ful71} and is known as the Berge--Fulkerson Conjecture.

\begin{Conj}\label{Conjecture:Fulkerson} If $G$ is a bridgeless
cubic graph, then there exist six perfect matchings $M_1,\ldots,M_6$
of $G$ with the property that every edge of $G$ is contained in
exactly two of $M_1,\ldots,M_6$.
\end{Conj}

\begin{Thm} \label{Theorem:3PerfectMatching}If $G$ is a cubic graph with \TroisPNIC, then there exist three perfect matchings $M$,
$M'$, and $M"$ such that $M \cap M' \cap M" = \emptyset$.
\end{Thm}

\begin{Prf}Let $M$, $M'$, and $M"$ be the associated perfect
matchings of $\mathcal T$, $\mathcal T'$, and $\mathcal T''$,
respectively. Let $v$ be a vertex and $u_1, u_2$ and $u_3$ its
neighbors. $\mathcal T$, $\mathcal T'$ and $\mathcal T''$ being
compatible, we can suppose  $ e_{\mathcal T}(v)=vu_1$,
$e_{\mathcal T'}(v)=vu_2$, and $ e_{\mathcal T''}(v)=vu_3$. The edge $vu_1$ is
an end edge of a trail in $\mathcal T$. This edge is not an odd edge
in $\mathcal T$ and thus $vu_1 \not \in M$. In the same way,  $vu_2
\not \in M'$ and  $vu_3 \not \in M"$. Hence every edge incident to $v$ is contained in at most two perfect matchings among $M, M'$, and $M"$. This means that $M \cap M' \cap M" = \emptyset$.
\end{Prf}

Theorem  \ref{Theorem:3PerfectMatching} above implies that the Fan and Raspaud
Conjecture is true for graphs with \TroisPNIC and we propose
the following new conjecture.

\begin{Conj} \label{Conjecture:3OddnormalCompatibleBridgeless}Any bridgeless cubic
graph can be provided with \TroisPNICsv.
\end{Conj}

We do not know whether this new conjecture is equivalent to the Fan and Raspaud
conjecture or not, or whether it is implied by the Berge--Fulkerson Conjecture.

Let $\mathcal S=\{\mathcal T_1, \mathcal T_2, \ldots \mathcal T_k\}$ ($k\geq 3$) be a set of odd normal partitions of a cubic graph $G$. The set $\mathcal S$ will be said to be a {\em complete system of odd normal partitions of order $k$} 
whenever for any vertex $v$ of $G$ there are three  partitions in $\mathcal S$ which are compatible on $v$, that is, there are $\mathcal T, \mathcal T'$, and $\mathcal  T''$ 
(depending on $v$) in $\mathcal S$ such that $e_{\mathcal T}(v)$, $e_{\mathcal T'}(v)$, and $e_{\mathcal  T''}(v)$ are three distinct edges.

\begin{Pb}\label{pb:SystemeComplet}
 Is it true that there exists  $k\geq 3$ such that every bridgeless cubic graph has a complete system of odd normal partitions of order at most $k$?
\end{Pb}

If a cubic graph has a complete system of normal odd partitions of order $k$, then it has $k$ perfect matchings with empty intersection. 
This conjecture would imply that the conclusions of Conjecture \ref{Conjecture:Jackson} below would hold for bridgeless cubic graphs.

\begin{Conj}\cite{God} \label{Conjecture:Jackson}There exists $k \geq 2$ such
that every $r$-graph contains $k+1$ perfect matchings with empty intersection.
\end{Conj}

As a matter of fact, in this Conjecture, the integer $k$ depends on $r$.

{\bf Acknowledgements:} The authors wish to thank the anonymous reviewers: their remarks and suggestions greatly contributed to many improvements. Let us thank also Robert Thorn for his helpful proofreading and correcting.

\bibliographystyle{plain}

\bibliography{Bibliographie}

\end{document}